\documentclass[fleqn,usenatbib]{mnras}

\usepackage{graphicx}
\usepackage{ifthen}
\usepackage{url}
\usepackage{amsmath}
\usepackage{bm}
\usepackage{lscape}
\usepackage{rotating}
\usepackage[graphicx]{realboxes}
\usepackage{supertabular}
\usepackage{pdfpages}
\usepackage{footnote}
\usepackage{hyperref}
\usepackage{amssymb}
\makesavenoteenv{tabular}
\makesavenoteenv{table}
\usepackage[T1]{fontenc}
\usepackage{tipa}
\usepackage{accents}
\newcommand{\ubar}[1]{\underaccent{\bar}{#1}}
\usepackage{tablefootnote}

\hypersetup{
  colorlinks   = true,    
  urlcolor     = blue,    
  linkcolor    = blue,    
  citecolor    = blue      
}

\usepackage{ulem}

\def\ltsima{$\; \buildrel < \over \sim \;$}
\def\lta{\lower.5ex\hbox{\ltsima}}
\def\gtsima{$\; \buildrel > \over \sim \;$}
\def\simgt{\lower.5ex\hbox{\gtsima}}
\def\kms{{\rm\,km \  s^{-1}}}
\def\mas{{\rm\,mas}}

\def\kpc{{\rm\,kpc}}

\def\msun{{\rm\,M_\odot}}

\usepackage{amsmath}

\def\s{\ifmmode \widetilde \else \~\fi}
\def\={\overline}

\def\spose#1{\hbox to 0pt{#1\hss}}

\def\eg{{e.g.,\ }}
\def\ie{{i.e.,\ }}
\def\lta{\mathrel{\spose{\lower 3pt\hbox{$\mathchar"218$}}
     \raise 2.0pt\hbox{$\mathchar"13C$}}}
\def\gta{\mathrel{\spose{\lower 3pt\hbox{$\mathchar"218$}}
     \raise 2.0pt\hbox{$\mathchar"13E$}}}
\def\Dt{\spose{\raise 1.5ex\hbox{\hskip3pt$\mathchar"201$}}}    
\def\dt{\spose{\raise 1.0ex\hbox{\hskip2pt$\mathchar"201$}}}    

\def\dotsfill{\leaders\hbox to 1em{\hss.\hss}\hfill}

\def\Gyr{{\rm\,Gyr}}
\def\FeH{{\rm[Fe/H]}}

\def\ione{{~\sc i}}
\def\ii{{~\sc ii}}

\loadboldmathitalic 
\title[The oldest and most metal-poor stars in the MW bulge]{The Pristine Inner Galaxy Survey (PIGS) V: a chemo-dynamical investigation of the early assembly of the Milky Way with the most metal-poor stars in the bulge} 
\author[F. Sestito et al.] {Federico Sestito$^{1}$\thanks{Email: \url{sestitof@uvic.ca}}, Kim A. Venn$^{1}$, Anke Arentsen$^{2,3}$,
David Aguado$^{4,5}$,
Collin L. Kielty$^{1}$,
\newauthor  
Carmela Lardo$^{6}$,
Nicolas F. Martin$^{3,7}$,
Julio F. Navarro$^{1}$, 
Else Starkenburg$^{8}$,
\newauthor 
Fletcher Waller$^{1}$,
Raymond G. Carlberg$^{9}$,
Patrick Fran\c{c}ois$^{10}$,
\newauthor
Jonay I. Gonz\'alez Hern\'andez $^{11,12}$,
Georges Kordopatis$^{13}$,
Sara Vitali$^{14}$, and
Zhen Yuan$^{3}$\\
$^{1}$ Department of Physics and Astronomy, University of Victoria, PO Box 3055, STN CSC, Victoria BC V8W 3P6, Canada
\\
$^{2}$ Institute of Astronomy, University of Cambridge, Madingley Road, Cambridge CB3 0HA, UK \\
 $^{3}$ Observatoire astronomique de Strasbourg, CNRS, UMR 7550, F-67000 Strasbourg, France\\
$^{4}$ Dipartimento di Fisica e Astrofisica, Univerisit\`a degli Studi di Firenze, via G. Sansone 1, I-50019 Sesto Fiorentino, Italy\\
$^{5}$ INAF/Osservatorio Astrofisico di Arcetri, Largo E. Fermi 5, I-50125 Firenze, Italy\\
$^{6}$ Dipartimento di Fisica e Astronomia, Universit\`a degli Studi di Bologna, Via Gobetti 93/2, I-40129 Bologna, Italy\\
$^{7}$ Max-Planck-Institut f\"ur Astronomie, K\"onigstuhl 17, D-69117, Heidelberg, Germany\\
$^{8}$ Kapteyn Astronomical Institute, University of Groningen, Landleven 12, NL-9747AD Groningen, the Netherlands\\
$^{9}$ Department of Astronomy \& Astrophysics, University of Toronto, Toronto, ON M5S 3H4, Canada\\
$^{10}$ GEPI, Observatoire de Paris, Universit\'e PSL, CNRS, Place Jules Janssen, F-92190 Meudon, France\\
$^{11}$ Instituto de Astrof\'isica de Canarias (IAC), V\'ia L\'actea, E-38200 La Laguna, Tenerife, Spain\\
$^{12}$ Universidad de La Laguna, Dept. Astrof\'isica, E-38200 La Laguna, Tenerife, Spain\\
$^{13}$ Universit\'e C\^{o}te d'Azur, Observatoire de la C\^{o}te d’Azur, CNRS, Laboratoire Lagrange, Nice, France\\
$^{14}$ N\'ucleo de Astronom\'a \& Millenium Nucleus ERIS, Facultad de Ingenier\'ia y Ciencias Universidad Diego Portales, Ej\'ercito 441, Santiago, Chile\\
}

\pubyear{2022}
\date{Accepted XXX. Received YYY; in original form ZZZ}

\begin{document} 
\label{firstpage}
\pagerange{\pageref{firstpage}--\pageref{lastpage}}
\maketitle 
\begin{abstract} 
The investigation of the  metal-poor tail in the Galactic bulge provides unique information on the early Milky Way assembly and evolution. A chemo-dynamical analysis of 17 very metal-poor stars (VMP, \FeH$<-2.0$) selected from the Pristine Inner Galaxy Survey was carried out  based on  Gemini/GRACES spectra. 
 The chemistry suggests that the majority of our stars are very similar to metal-poor stars in the Galactic halo.  Orbits calculated from {\it Gaia} EDR3 imply these stars are brought into the bulge during the earliest Galactic assembly. Most of our stars have large [Na,Ca/Mg] abundances, and thus show little evidence of enrichment by pair-instability supernovae.
Two of our stars (P171457, P184700) have chemical abundances compatible with second-generation globular cluster stars, suggestive of the presence of ancient and now dissolved globular clusters in the inner Galaxy. One of them (P171457) is extremely metal-poor (\FeH$<-3.0$) and well below the metallicity floor of globular clusters, which  supports the growing evidence for the existence of lower-metallicity globular clusters in the early Universe.
A third star (P180956, [Fe/H]$\sim-2$) has low [Na,Ca/Mg] and very low [Ba/Fe] for its metallicity, which are consistent with formation in a system polluted by only one or a few low-mass supernovae. Interestingly, its orbit is confined to the Galactic plane, like other very metal-poor stars found in the literature, which have been associated with the earliest building blocks of the Milky Way.
\end{abstract}
 

\section{Introduction}

The oldest and most chemically pristine stars in the Galaxy are expected to have been enriched by only one or a few individual supernovae or hypernovae events.  This means that studies of their chemical abundance patterns and orbital dynamics are invaluable for learning about the lives and deaths of the first stars, and the assembly history of the Galaxy \citep{Freeman02, Tumlinson10, Wise12, Karlsson13}.  Successive generations of stars  enriched the interstellar medium, while gas inflows dilute it, contributing to a complex star formation history that depends on location within the Milky Way Galaxy.
In cosmological simulations, low-metallicity stars (\FeH\footnote{\FeH $= \log(\rm{N_{Fe}/N_{H}})_{\star}-\log(\rm{N_{Fe}/N_{H}})_{\odot} $, in which $\rm{N_X}$ is the number density of element X.}$\leq-2.5$) form in the first 2--3 $\Gyr$ after the Big Bang, and mostly in low-mass systems, the so-called ``building blocks", \citep{Starkenburg17a, ElBadry18, Sestito21}.   
These building blocks gradually merged to form the proto-Milky Way.  These stars are expected to occupy the deepest parts of the gravitational potential, \ie near the bulge, while late accretion of dwarf satellites are expected to  deposit metal-poor stars primarily in the halo \citep{Bullock2005, Johnston2008, Tissera12}, or even in the disc for planar accretions \citep[\eg][]{Abadi03,Sestito21,Santistevan21}. 

While the metal-poor stars in the Galactic bulge are important tracers of the earliest stages in the formation of the Milky Way, they are extremely difficult to find \citep[e.g.,][]{Schlaufman14}.  Firstly, the region of the bulge is dominated by a metal-rich population of both young and old stars, disrupted globular clusters, and ongoing star formation \citep{Ness13a, Ness2014, Bensby13,Bensby17,Schiavon17, Schultheis19}.
Secondly, the heavy and variable interstellar extinction, extreme stellar crowding, and presence of complex foreground disc stellar populations have made photometric surveys of metal-poor stars extremely challenging.
The ARGOS spectroscopic survey found that fewer than 1\% (84) of the stars in their sample have $\FeH<-1.5$ \citep{Ness13b}. 
The Extremely Metal-poor BuLge stars with AAOmega  \citep[EMBLA,][]{Howes14,Howes15,Howes16} survey selected VMP targets with a metallicity-sensitive photometric filter from the SkyMapper Southern Survey \citep{Bessell11,Wolf18} for low-resolution spectroscopy with the Anglo-Australian Telescope.
EMBLA analysed with high-resolution spectroscopy 63 stars with [Fe/H] $< -2.0$, where the majority resemble chemically metal-poor stars in the Galactic halo.  The only noticeable differences were a lack of carbon-rich stars, and possibly a larger scatter in [$\alpha$/Fe] abundances. 
A detailed kinematics analysis of their sample also raised questions about what it means to be a ``bulge star", \ie a star that formed in the bulge versus one passing through the bulge on a radial orbit.  Reducing their sample to stars with apocentric distances $\le$5 kpc (36 stars), however, did not alter their conclusions \citep{Howes16}. 

The Pristine Inner Galaxy Survey \citep[PIGS,][]{Arentsen20a,Arentsen20b} is similar to the EMBLA survey in that metal-poor targets have been selected from the narrow-band photomery, in this case the Pristine survey \citep{Starkenburg17b}.  The Pristine survey is a narrow-band imaging survey carried out at the Canada-France-Hawaii Telescope (CFHT), where the Ca\ii{} HK filter, in combination with broad band photometry, has been shown to find low-metallicity stars ([Fe/H]$<-2.5$) with $\sim$56\% efficiency in the Galactic halo \citep{Youakim17, Aguado19, Venn20}. 
The power of the Pristine survey has been demonstrated by the discovery of two new ultra metal-poor stars \citep[``UMP", $\FeH<-4.0$,][]{Starkenburg19,Lardo21}, or $\sim$5\% of the total known UMP stars so far \citep[see the compilation in ][]{Sestito19}. 
The Pristine-selected metal-poor targets in the bulge were examined with low-/medium-resolution spectroscopic observations obtained with the the AAOmega spectrograph on the Anglo Australian Telescope (AAT), from which stellar parameters, metallicities, and carbon abundances were derived for $\sim$12,000 stars.
\citet{Arentsen20b} report an efficiency of $\sim$80\% in finding very metal-poor stars (``VMP"\footnote{This nomenclature, \eg VMP, EMP, UMP, has been introduced in \citet{Beers05}.}, $\FeH<-2.0$) in the bulge avoiding the most highly extincted regions. \citet{Arentsen20b} used the PIGS/AAT observations to study the kinematics of metal-poor stars in the inner Galaxy, finding that the rotation around the Galactic centre decreases with decreasing metallicity, \ie lower metallicity stars are more dispersion-dominated as the Galactic halo. 

To chemically examine the low-metallicity tail of the Galactic bulge also requires consideration of the contributions from disrupted globular clusters and later accretions of dwarf galaxies.
Some studies have suggested that up to  $\sim$25\% of the stellar mass of the inner region of the Milky Way is made of dissolved ancient globular clusters
\citep[\eg][]{Shapiro10, Kruijssen15, Schiavon17}. 
One study \citep{Schiavon17} based this claim on
the large number of nitrogen-rich stars 
that resemble the chemistry of second-generation stars in globular clusters \citep[\eg][]{Gratton04,Bastian18}. More recently, a few bulge stars with  chemistry similar to  second-generation globular cluster stars were found in the 
Chemical Origins of Metal-poor Bulge Stars (COMBS) survey \citep{Lucey19, Lucey21, Lucey22}.  The COMBS survey is based on VLT/UVES+GIRAFFE+FLAMES spectra of red giants in the bulge, and also reported that the number of halo stars passing through the bulge (``interlopers") increases with decreasing metallicity. 

In this paper, we report on a chemo-dynamical investigation of 17 VMP candidates selected from the PIGS survey and observed with the high-resolution GRACES spectrograph at Gemini North. This work aims to confirm the low metallicity of these targets, investigate the orbits of these stars within the inner Galaxy, and examine their connections with the early assembly of the Milky Way through studies of their detailed chemical abundance ratios. In particular, we explore the chemical signatures of our 17 metal-poor stars ($-3.3 <$ [Fe/H]$< -2.0$) in comparison to other metal-poor systems, such as ultra-faint dwarf galaxies and globular clusters, as well as the dispersed Galactic halo.

In Section~\ref{datasec}, we discuss the target selection process, the observations, and the data reduction of our spectra. Section~\ref{kinesec} describes the kinematical analysis, including estimates of the radial velocity, the heliocentric distance, and the orbital parameters using an appropriate Galactic potential. In Section~\ref{stellarparamssec}, a description of the stellar parameters (effective temperature and surface gravity) determinations and the effects of high extinction are reported. In Section~\ref{models}, the model atmospheres, radiative transfer, line lists, and metallicity determinations are discussed. In Section~\ref{chemsec}, the method used to measure the chemical abundances from the observed spectra is described, focusing on iron-group elements, $\alpha$-, odd-Z, and neutron-capture elements. The scientific implications of our chemo-dynamical analyses are described in Section~\ref{discussionsec}, particularly in the context of Galaxy formation and evolution.

\section{Data}\label{datasec}
\subsection{Target selection}
The targets in this work were selected from the larger sample of PIGS \citep{Arentsen20a}. All the PIGS stars were selected from MegaCam photometry observed at the CFHT\footnote{The authors wish to recognize and acknowledge the very significant cultural role and reverence that the summit of Maunakea has always had within the Native Hawaiian community. We are very fortunate to have had the opportunity to conduct observations from this mountain.}. Then,  AAT/AAOmega low/medium-resolution spectra were taken and analysed \citep{Arentsen20b} with both \textsc{FERRE} \citep{Allende06} and ULySS \citep{Koleva09} yielding estimates of metallicity, effective temperature, surface gravity, radial velocity, and carbonicity.
Throughout this paper, when comparing stellar parameters and metallicities to this sample, the \textsc{FERRE} output are considered and denoted as AAT (see Sections~\ref{gracesferreRV}~and~\ref{metsec}).

\begin{figure}
\centering
\includegraphics[width=0.5\textwidth]{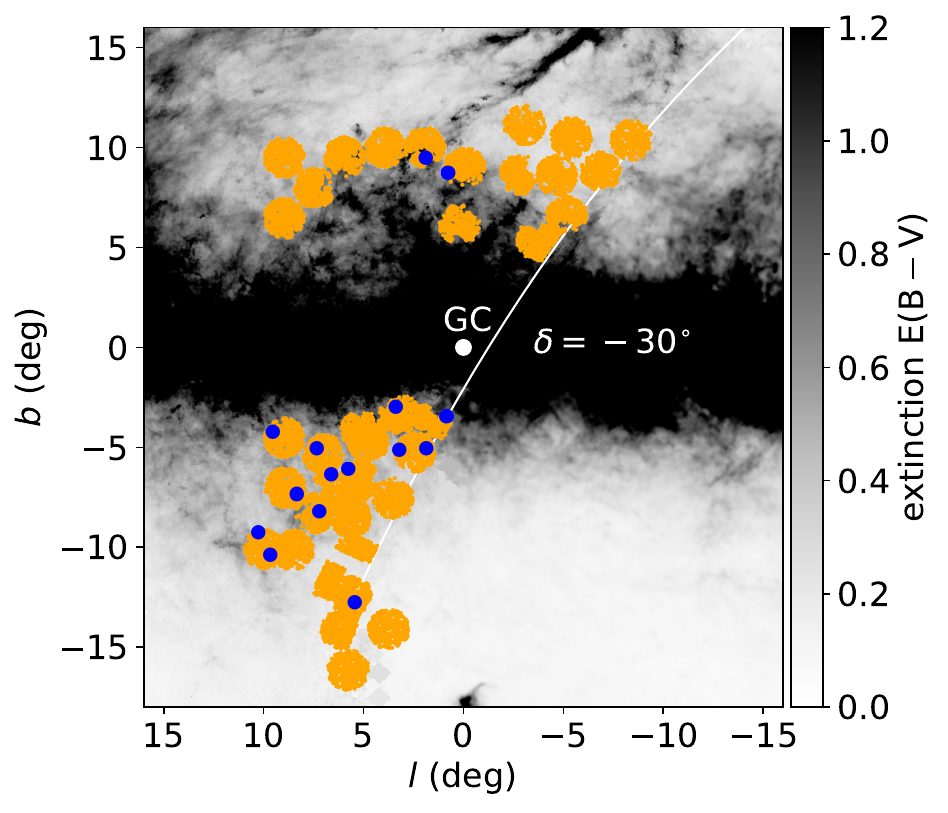}
\caption{The Pristine inner Galaxy survey spectroscopic footprint in the Galactic coordinates. All the $\sim12000$ PIGS stars observed at low/medium resolution \citep{Arentsen20b} are marked with orange, while the 17 PIGS/GRACES stars are shown with larger blue dots. The E\,(B-V) at a distance of $8\kpc$ from \citet{Green19} is overplotted with a grey colour map.}
\label{longlat}
\end{figure}

From the largest dataset of PIGS/AAT, a sample of 20 stars were observed at higher resolution with GRACES as part of the Gemini Large and Long Program LLP$-102$ started in 2019A and ended in 2021A.
We selected the targets to have $\FeH_{\mathrm{AAT}}\leq-2.5$, V $\leq15$ mag, $T_{\mathrm{eff,AAT}}\leq 5500$ K. The cut on magnitudes allows us to achieve a  SNR$\sim30$ within a reasonable exposure time, see Table~\ref{tab:obs}. The selection in effective temperature permits us to observe spectral lines that are too weak at higher temperatures, \eg Ba lines. While the metallicity limit is necessary to explore the most metal-poor tail of the inner galaxy. Of the full sample we observed only less than a third of the potential targets. At the epoch of the proposal submission, the astrometric  {\it Gaia} DR2 data \citep{Gaia16,Gaia18,Lindegren18} poorly constrained the orbits at the distance of the bulge, and hence no selection based on orbital properties was not done. Due to numerous unexpected events, such as the protests at  Mauna Kea, the global Covid-19 pandemic, and the  bad weather, we  were only able to collect a sample composed of $17$ VMP bulge stars and 2 VMP standard stars, HD~122563 and HD~84937 \citep[\eg][]{Amarsi16,Sneden16,Karovicova18}. Figure~\ref{longlat} shows the PIGS footprint in Galactic coordinates, with the GRACES stars highlighted.

\subsection{GRACES observations and reduction}
Observations were conducted with the Gemini Remote Access to CFHT ESPaDOnS Spectrograph \citep[GRACES,][]{Chene14,Pazder14} in the 2-fibre (object+sky) mode with a resolution of R$\sim40000$. GRACES consists in a 270-m optical fibre that links the Gemini North telescope to the Canada–France–Hawaii Telescope ESPaDOnS spectrograph \citep{Donati06}, which is a cross-dispersed high resolution \'echelle spectrograph. The spectral coverage of GRACES is from 4500 \AA{} to 10000 \AA{} \citep{Chene14}. However, our spectra are dominated by noise below the spectral region of 4900-5000 \AA{}.

\begin{figure*}
\includegraphics[width=1\textwidth]{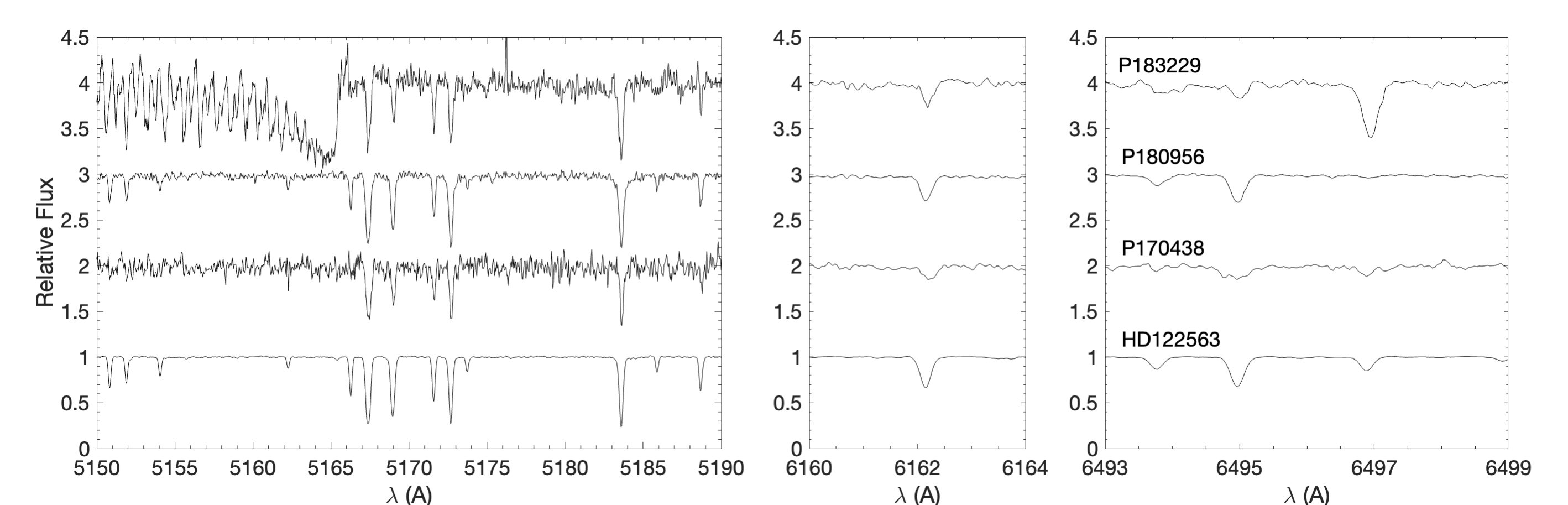}
\caption{Example of spectra of the PIGS/GRACES sample.
Left panel: Mg\ione{} T region. The strong absorption band on the left part of the spectrum of P183229 is the C$_2$ Swan band. This is one of the two C-enhanced stars in the sample. Central panel: Ca\ione{}  line at  6162.173 \AA{}. Right panel:  Ba\ii{} line region (6496.910 \AA{}). The high-resolution and high SNR GRACES spectrum of the very metal-poor star HD122563 is added in each panel as a comparison.}
\label{spectraex}
\end{figure*}

The GRACES spectra were first reduced using the Open source Pipeline for ESPaDOnS Reduction and Analysis \citep[OPERA,][]{Martioli12} tool, which also corrects for heliocentric motion. Then the reduced spectra were reprocessed following the procedure described in \citet{Kielty21}. The latter pipeline allows us to measure the radial velocity of the observed star, to co-add  multiple observations for a given target to check for possible radial velocity variations, to correct for the motion of the star, and to eventually renormalise the flux. This procedure also improves the signal-to-noise ratio in the overlapping spectral order regions. Figure~\ref{spectraex} shows the spectra of three stars (P170438, P180956, and P183229) with the very metal-poor standard star HD122563 in the Mg\ione{} b, the Ca\ione{}, and Ba\ii{} regions.

The observed stars with their Pristine name,  {\it Gaia} EDR3 ID, their  {\it Gaia} photometry, and the log of the observations are reported in Table~\ref{tab:obs}.

\begin{table*}
\caption{Log of the observations. The short name, the Pristine name, the  {\it Gaia} EDR3 source ID, the G and BP-RP from  {\it Gaia} EDR3, the reddening from the 3D map of \citet{Green19}, the total exposure time, the number of exposures, and the SNR are reported. The  SNR is measured as the ratio between the median flux and its standard deviation in two spectral regions, the  5175$-$5182 \AA{} (@Mg\ione{} b) and the 5950$-$6000 \AA{}  (@6000\AA{}) ranges. P180503 is the star for which the 3D extinction map from \citet{Green19} does not provide a value, therefore we report the \textsc{StarHorse} extinction \citep{Anders19}.}
\label{tab:obs}
\resizebox{\textwidth}{!}{
\begin{tabular}{ccccccccc}
\hline
Short name & Pristine name & source ID   & G & BP-RP & E(B-V) & T$_{\rm exp}$& N$_{\rm exp}$ & SNR   \\
& & & (mag) & (mag) &  (mag) &  (s) & & @Mg\ione{} b,@6000 \AA{}  \\
\hline
P170438     & Pristine\_170438.40-261742.8 & 4111904146257749504 & 15.52 & 1.33 & 0.41         & 4800 & 2  & 13, 37 \\
P170610     & Pristine\_170610.81-290322.3 & 6029916002325425536 & 14.30  & 1.47 & 0.34         & 1800 & 1  & 22, 51 \\
P171457     & Pristine\_171457.25-232718.6 & 4114176871140428416 & 14.12 & 1.68 & 0.52         & 1800 & 1  & 22, 44 \\
P171458     & Pristine\_171458.77-220807.2 & 4114599427192094592 & 14.83 & 1.63 & 0.46   & 1800 & 1  & 18, 43 \\
P180118     & Pristine\_180118.30-295346.9 & 4050241536946876928 & 14.96 & 1.75 & 0.85   & 3600 & 2  & 11, 37 \\
P180503     & Pristine\_180503.40-272725.4 & 4062947566467257472 & 13.80  & 1.93 &       0.64       & 1800 & 3  & 8, 34 \\
P180956     & Pristine\_180956.78-294759.8 & 4050071013878221696 & 13.50  & 1.55 & 0.53         & 2700 & 3  & 36, 78 \\
P181306     & Pristine\_181306.64-283901.5 & 4050649838085533184 & 14.98 & 1.33 & 0.47         & 7200 & 4  & 19, 53 \\
P182129     & Pristine\_182129.69-245815.3 & 4053226917177000448 & 13.60  & 1.77 & 0.51         & 1800 & 3  & 30, 63 \\
P182221     & Pristine\_182221.12-265025.3 & 4052704649149713536 & 15.78 & 1.22 & 0.44         & 4800 & 2 & 11, 34 \\
P182244     & Pristine\_182244.51-223836.4 & 4089749772871612416 & 14.12 & 1.85 & 0.58         & 1800 & 1  & 25, 47 \\
P182505     & Pristine\_182505.97-261308.2 & 4052845008684897408 & 14.86 & 1.39 & 0.38         & 2400 & 1  & 15, 37 \\
P183229     & Pristine\_183229.69-250729.1 & 4076280038267332736 & 15.13 & 1.41 & 0.35         & 3600 & 2  & 10, 26 \\
P183335     & Pristine\_183335.04-263056.1 & 4075864732108631808 & 14.38 & 1.37 & 0.36   & 2700 & 3 & 9, 31 \\
P184338     & Pristine\_184338.08-241508.3 & 4078080041827258112 & 14.56 & 1.43 & 0.35         & 1800 & 1  & 11, 29 \\
P184700     & Pristine\_184700.56-251720.9 & 4073414191180594944 & 16.19 & 1.55 & 0.33   & 7200 & 3  & 16, 16 \\
P184855     & Pristine\_184855.88-300124.9 & 6761365842153603968 & 13.67 & 1.42 & 0.22         & 3600 & 3  & 38, 70\\
\hline
\end{tabular}}
\end{table*}

\section{Kinematical analysis}\label{kinesec}

\subsection{Radial velocity: GRACES vs. medium-resolution spectroscopy}\label{gracesferreRV}
\begin{figure}

\includegraphics[width=0.5\textwidth]{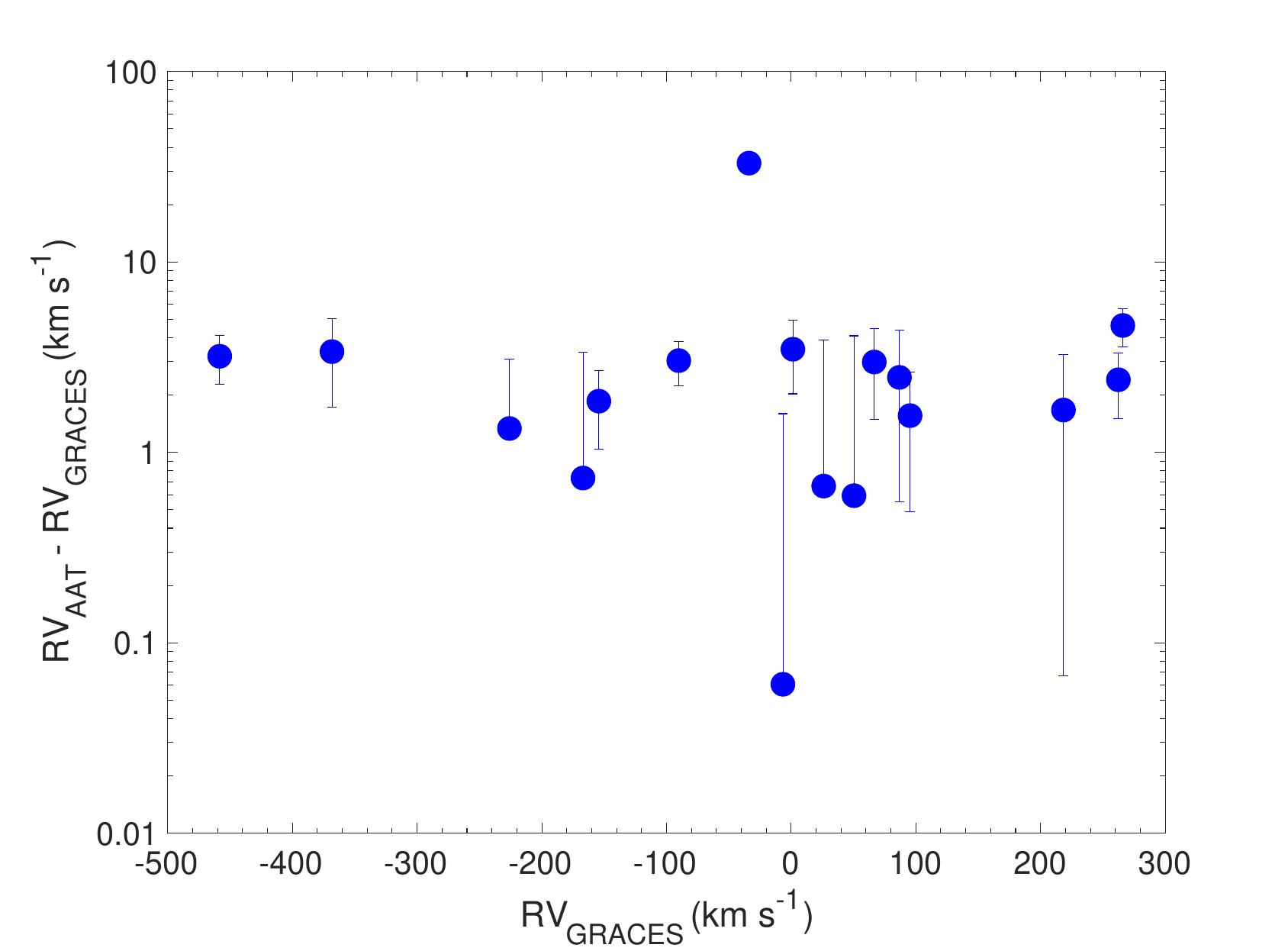}
\caption{Radial velocity differences between the high-resolution measurements from GRACES and the medium-resolution from AAT. Error bars in the y-axis are the sum in quadrature of the uncertainties on GRACES and AAT RV measurements.}
\label{rvgracesferre}

\end{figure}

A necessary step for the orbital parameters inference is to measure the radial velocity, RV. A first measurement for the RV is performed by the reduction pipeline. The spectra are cross-correlated with well known reference metal-poor star spectra (HD 122563). The cross-correlation is made selecting H$\alpha$, H$\beta$, and the Mg\ione{} b lines. Then the corrected spectra are  cross-correlated using the more precise \textsc{fxcor} routine from \textsc{IRAF}\footnote{\url{ https://github.com/iraf-community}} \citep{Tody86,Tody93}. The RV of our targets were previously determined by medium-resolution observations \citep{Arentsen20a}, and Figure~\ref{rvgracesferre} displays a comparison between  our measurements and the former, which shows in general good agreement.
The difference in velocity between the two sets of spectra at different resolution is always positive which might indicate a possible offset between the two instruments. The median difference in radial velocity between the two measurements is $2.13\kms$ with a dispersion of $1.28\kms$, removing the outlier star (P182221, $\sim33 \kms$). A discussion on P182221 is reported in Section~\ref{cempgc}. Due to the much higher resolution and the SNR of the GRACES spectra, the determination of RV and its uncertainty from  our observations are preferred and considered for the orbital parameters inference.

\subsection{Distances and spatial distribution}
Another essential ingredient for the recipe of the kinematical analysis is the determination of the heliocentric distances. The  improvements that the  {\it Gaia} early data release 3 \citep[hereafter  {\it Gaia} EDR3,][]{GaiaEDR3,Lindegren21}  provided for  astrometric measurements, \ie parallaxes and proper motions, play a crucial role for  stars in our sample.

It is now well-known that it is ill advised to simply invert the parallax for inferring the distance \citep[\eg ][]{Bailer15,Bailer18}. This is especially true when the parallax and its uncertainty are poorly constrained, \eg $\varpi\leq 0 \mas$ and/or $\sigma_{\varpi}/\varpi\geq 10$ percent. In this work, we infer the distances  with a Bayesian approach similar to that of \citet{Bailer15}, and following  Equations  8 to 11 of \citet{Sestito19}.  Briefly, this consists in a Gaussian likelihood for the parallax distribution and a  prior on the stellar density distribution that takes into account the Galactic disc and halo \citep[for more details see][]{Sestito19}. The zero point offset has been applied to the  {\it Gaia} EDR3 parallaxes \citep{Lindegren21} using the python \textsc{gaiadr3\_zeropoint}\footnote{\url{https://gitlab.com/icc-ub/public/gaiadr3\_zeropoint}} package.  The distance is provided with a probability distribution function (PDF or posterior), which, in some cases, is far from being a Gaussian-like distribution \citep{Bailer15}.
 Figure~\ref{spacedist} displays the spatial distribution in Galactic Cartesian coordinates of this PIGS sample. For each star, the median and the standard deviation of the Galactocentric coordinates PDF are shown.

\subsection{Orbital parameters}\label{orbsec}
The final step for the orbital inference is to feed \textsc{Galpy}\footnote{\url{http://github.com/jobovy/galpy}}  \citep{Bovy15} with the inferred distances and RV, and the proper motions and coordinates from \textit{Gaia} EDR3. Since we deal here with objects that are in the inner region of the Milky Way, we need to account for the presence of a rotating bar in the Galactic gravitational potential. Therefore, the potential we use is composed by  a Navarro-Frenk-White dark matter halo \citep[][\textsc{NFWPotential}]{NavarroFrenkWhite97}, a Miyamoto-Nagai potential disc \citep[][\textsc{MiyamotoNagaiPotential}]{MiyamotoNagai}, an exponentially cut-off bulge (\textsc{PowerSphericalPotentialwCutoff}), and a rotating bar potential (\textsc{DehnenBarPotential}). All of the aforementioned potentials, with the exclusion of the bar, are usually summoned by the \textsc{MWPotential14} package. However, we adopt a more massive and up-to-date halo \citep{BlandHawthorn16}, with a mass of $1.2\times10^{12}\msun$ (vs. $0.8\times10^{12}\msun$ for  \textsc{MWPotential14}). The bar is invoked from the \textsc{DehnenBarPotential} package, which consists in a Dehen bar potential \citep{Dehnen00} and generalised to 3D following \citet{Monari16}. This choice of the Galactic potential, especially the settings of the rotating bar (\eg pattern speed and scale length of the bar), is in line with the recent dynamical analysis of bulge stars by the COMBS survey \citep{Lucey21}.

\begin{figure*}
\begin{center}
\includegraphics[width=\textwidth]{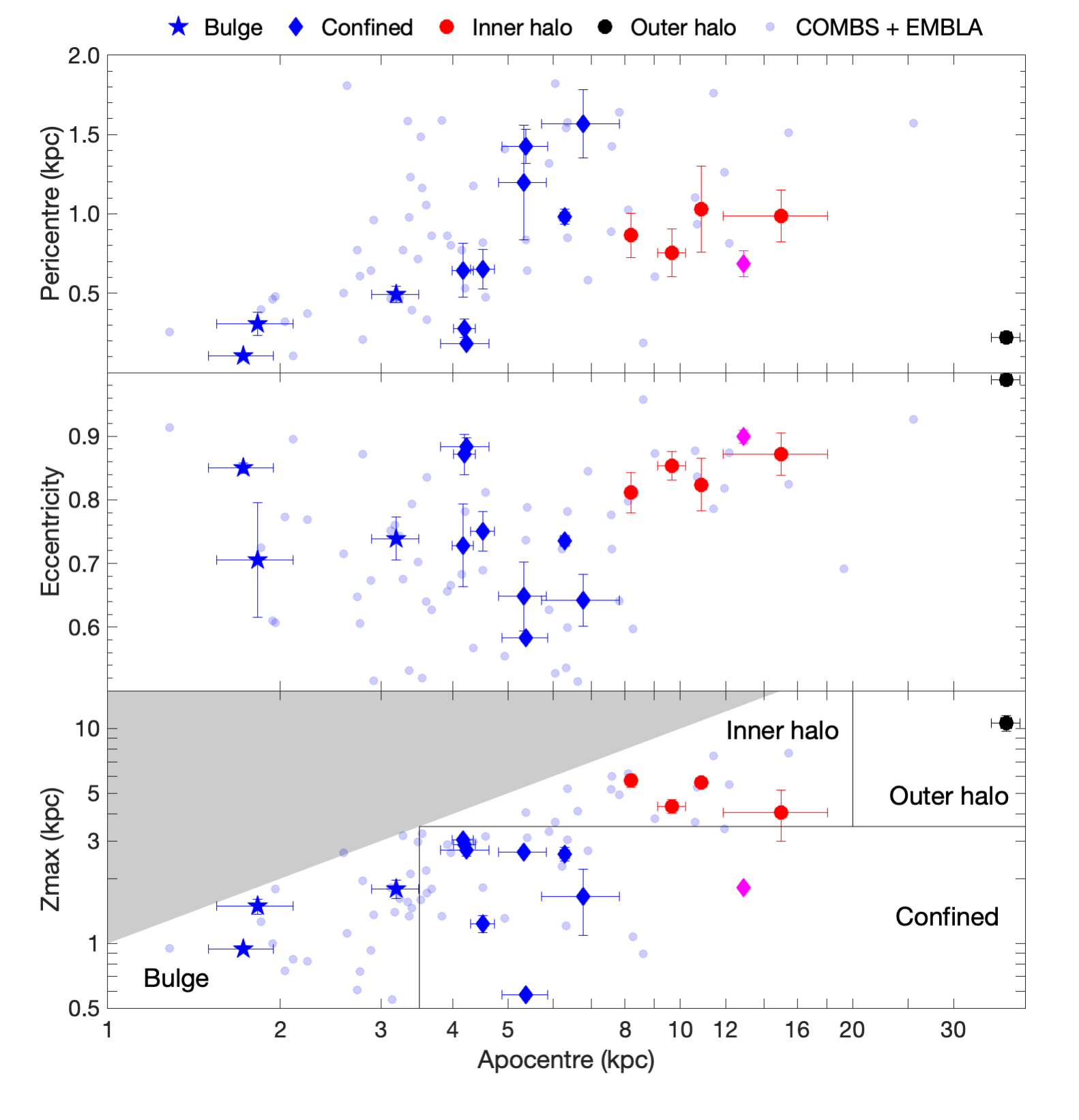}
\caption{Orbital parameters of the targets observed with GRACES. The GRACES sample is divided into 4 dynamical groups according to their $r_{\rm apo}$ $Z_{\rm max}$. The bulge group is marked with blue stars, the confined group is denoted with blue rhombuses, except for the one with largest $r_{\rm apo}$ (magenta rhombus), the inner halo grop with red circles, and the outer halo star with black square. Stars with metallicities below [Fe/H]=$-1.7$  from \citet{Howes14, Howes15, Howes16,Lucey22} are shown as light blue circles. The grey shaded area denotes the forbidden region in which  $Z_{\rm max}$>$r_{\rm apo}$. Vertical and horizontal lines separate the groups as defined in Section~\ref{orbsec}.  }
\label{kinefig}
\end{center}
\end{figure*}

From the distance inference the majority of the stars are placed very close to the inner region of the bulge ($<5\kpc$, see Figure~\ref{spacedist} in Appendix~\ref{apporb}), and due to the uncertainties on the distance, it is hard to discern if they are in front or just beyond the Galactic centre. Depending on their location either in front or behind, the orbital parameters might drastically change, resulting in a change of their Galactic rotation direction (retrograde vs. prograde). Therefore for each star, we create a grid of distances with a step of $0.1\kpc$ within $\pm 1\sigma$ from the maximum of the distance PDF. For each point of the grid (\ie at a fixed distance), we perform a Monte Carlo with 1000 random draws on the other parameters (\eg RV, coordinates) to infer the orbital parameters and their uncertainties. In case of the proper motion components, we consider their correlation given the coefficients from  {\it Gaia} EDR3, drawing randomly with a multivariate Gaussian function. The RV and coordinates are treated as a Gaussian. The integration time is set to 1 Gyr. Figure~\ref{kinefig} shows the main median orbital parameters inferred from \textsc{Galpy}\footnote{Note that \textsc{Galpy} cannot infer the energy, the angular momentum, and the action variables for the potential we have adopted, given the presence of a rotating bar.}  and considered for this analysis, namely the maximum height from the plane $Z_{\rm max}$, the apocentre $r_{\rm apo}$, the pericentre $r_{\rm peri}$, and the eccentricity $\epsilon$. 
The variation of the orbital parameters as a function of the distance grid steps is shown in Figure~\ref{kinefig_grid}. The sample is catalogued into 4 groups according to their median $Z_{\rm max}$ and their median $r_{\rm apo}$. In this space, the variation of the orbital parameters does not strongly impact the classification of the stars. When a star could be classified as belonging to more than one group, the final choice mirrors the classification according to the maximum of the distance PDF.

We want to emphasise that at this stage of the narrative, we are not discriminating between halo, bulge, or disc stars. This will be discussed in Section~\ref{discussionsec}. The 4 categories are described as follow:
\begin{itemize}
\item \textbf{Bulge group}. The stars in this group do not venture outside a sphere of 3.5 $\kpc$ from the Galactic centre, \ie $Z_{\rm max}\leq3.5 \kpc$ and $r_{\rm apo}\leq3.5 \kpc$. In this category there are 3 stars, P171458, P180118 and P180503. The targets within this group are marked with a blue star marker in the Figures of this work.
\item \textbf{Confined group}. These stars have $Z_{\rm max}\leq3.5 \kpc$ and $r_{\rm apo}>3.5 \kpc$. They are confined close to the Milky Way disc and this group is composed by 9 objects, P170438, P170610, P180956, P181306, P182244, P182505, P183229, P183335 and P184338.
\item \textbf{Inner halo group}. This group is composed by 4 stars with $Z_{\rm max}>3.5 \kpc$ and $r_{\rm apo}\leq20 \kpc$. P171457, P182129, P182221, and P184700 belong to this group.
\item \textbf{Outer halo group}. Only P184855 is catalogued in this group and it has a $Z_{\rm max}>3.5 \kpc$ and $r_{\rm apo}>20 \kpc$. 
\end{itemize}

As it appears from Figure~\ref{kinefig}, all the stars, independently of their group, display a high eccentricity $\epsilon >0.55$ and their pericentre is located in the inner Galactic region $r_{\rm peri}<2 \kpc$. To be noted, stars in the Confined group have not necessarily a planar orbit. This is because for the majority of them the $Z_{\rm max}$ to $r_{\rm apo}$ ratio is not small ($0.25<Z_{\rm max}/r_{\rm apo}<0.85$). Two of them, P170610 and P180956, have a very small ratio (\ie $Z_{\rm max}/r_{\rm apo}<0.15$), which indicates their planar orbit. Table~\ref{tablekine} contains all the kinematical parameters used in this work.

\begin{table*}
\caption{Orbital parameters for the stars in this sample. The heliocentric distance, the maximum height from the MW plane, the apocentre and pericentre distances, the eccentricity, the Galactic cartesian coordinates (X, Y, Z), and the dynamical group classification are reported for each star denoted by their short name. For the dynamical group classification, B = bulge, C = confined, I = inner halo, and O = outer halo.}
\label{tablekine}
\resizebox{\textwidth}{!}{
\begin{tabular}{cccccccccccccccccc}
\hline
name short & D  & $\sigma_{\rm D}$    & $Z_{\rm max}$     & $\sigma_{Z_{\rm max}}$        & $r_{\rm apo}$     & $\sigma_{r_{\rm apo}}$       & $r_{\rm peri}$       & $\sigma_{r_{\rm peri}}$         & $\epsilon$      & $\sigma_{\epsilon}$      & X                    & $\sigma_X$              & Y                   & $\sigma_Y$              & Z                    & $\sigma_Z$      &  Group    \\
& $(\kpc)$ &$(\kpc)$ &$(\kpc)$&$(\kpc)$&$(\kpc)$&$(\kpc)$&$(\kpc)$&$(\kpc)$& & &$(\kpc)$&$(\kpc)$&$(\kpc)$&$(\kpc)$&$(\kpc)$&$(\kpc)$ & \\
\hline
P170438     & 5.40 & 1.74      & 1.66    & 0.56    & 6.77   & 1.05   & 1.57    & 0.21    & 0.64   & 0.04   & 2.65  & 1.63   & -0.27 & 0.08   & 0.86  & 0.25  & C\\
P170610     & 4.09 & 0.48      & 0.58    & 0.01    & 5.38   & 0.49   & 1.43    & 0.11    & 0.58   & 0.01   & 3.97  & 0.48   & -0.35 & 0.04   & 0.52  & 0.06  & C \\
P171457     & 7.38 & 1.07      & 4.34    & 0.31    & 9.67   & 0.55   & 0.75    & 0.15    & 0.85   & 0.02   & 0.72  & 1.01   & 0.11  & 0.01   & 1.12  & 0.15  & I \\
P171458     & 7.59 & 1.38      & 1.79    & 0.17    & 3.19   & 0.30   & 0.49    & 0.05    & 0.74   & 0.03   & 0.65  & 1.38   & 0.25  & 0.05   & 1.23  & 0.23  & B \\
P180118     & 8.24 & 0.37      & 0.94    & 0.03    & 1.72   & 0.22   & 0.11    & 0.01    & 0.85   & 0.01   & -0.20 & 0.38   & 0.13  & 0.01   & -0.49 & 0.02  & B \\
P180503     & 8.22 & 0.47      & 1.49    & 0.12    & 1.83   & 0.28   & 0.31    & 0.07    & 0.71   & 0.09   & -0.18 & 0.45   & 0.49  & 0.03   & -0.42 & 0.03  & B \\
P180956     & 3.30 & 0.27      & 1.81    & 0.04    & 12.91  & 0.19   & 0.69    & 0.08    & 0.90   & 0.01   & 4.71  & 0.27   & 0.11  & 0.01   & -0.27 & 0.02  & C \\
P181306     & 5.24 & 1.22      & 2.87    & 0.08    & 4.20   & 0.19   & 0.28    & 0.06    & 0.87   & 0.03   & 2.84  & 1.18   & 0.29  & 0.07   & -0.45 & 0.11  & C \\
P182129     & 6.12 & 0.98      & 5.73    & 0.41    & 8.20   & 0.16   & 0.86    & 0.14    & 0.81   & 0.03   & 1.91  & 0.96   & 0.79  & 0.12   & -0.53 & 0.09  & I \\
P182221     & 7.94 & 1.43      & 4.07    & 1.09    & 14.98  & 3.09   & 0.99    & 0.16    & 0.87   & 0.03   & 0.06  & 1.39   & 0.81  & 0.14   & -0.84 & 0.15  & I \\
P182244     & 7.99 & 0.92      & 3.03    & 0.12    & 4.17   & 0.18   & 0.64    & 0.17    & 0.73   & 0.06   & 0.08  & 0.95   & 1.34  & 0.16   & -0.59 & 0.07  & C \\
P182505     & 4.80 & 0.86      & 2.65    & 0.03    & 5.32   & 0.51   & 1.20    & 0.36    & 0.65   & 0.05   & 3.31  & 0.86   & 0.55  & 0.10   & -0.51 & 0.10  & C \\
P183229     & 4.23 & 0.61      & 1.23    & 0.12    & 4.52   & 0.22   & 0.65    & 0.13    & 0.75   & 0.03   & 3.86  & 0.59   & 0.61  & 0.09   & -0.52 & 0.08  & C \\
P183335     & 6.82 & 1.13      & 2.72    & 0.17    & 4.23   & 0.41   & 0.18    & 0.02    & 0.88   & 0.01   & 1.32  & 1.16   & 0.85  & 0.15   & -0.96 & 0.17  & C \\
P184338     & 8.00 & 1.07      & 2.61    & 0.19    & 6.28   & 0.12   & 0.98    & 0.05    & 0.74   & 0.01   & 0.19  & 1.04   & 1.42  & 0.19   & -1.28 & 0.17  & C \\
P184700     & 8.25 & 1.39      & 5.60    & 0.16    & 10.88  & 0.25   & 1.03    & 0.27    & 0.82   & 0.04   & -0.04 & 1.40   & 1.37  & 0.24   & -1.49 & 0.26  & I \\
P184855     & 5.71 & 0.67      & 10.62   & 0.89    & 37.06  & 2.15   & 0.22    & 0.03    & 0.99   & 0.00   & 2.46  & 0.63   & 0.53  & 0.06   & -1.25 & 0.14  & O \\
\hline

\end{tabular}}
\end{table*}

\subsection{The kinematical sample from the literature}
We compare the orbital parameters of the PIGS/GRACES sample with two datasets from the literature. The first is a compilation made of 36 stars from the EMBLA survey \citep{Howes14,Howes15,Howes16}. Since the orbital parameters in that sample were inferred before  {\it Gaia} DR2, we re-calculate their orbit using the most up-to-date  {\it Gaia} EDR3 astrometric solutions, the method describe in this Section, and the RV from the \citet{Howes15,Howes16}. In the case of  stars from \citet{Howes16}, the RV has been inferred from the Galactocentric velocity inverting their Equation~4. The second dataset is from the recent work of \citet{Lucey22}, which is part of the COMBS survey \citep{Lucey19}. The complete sample is composed of 319 stars, however we restrict the kinematical comparison to the 27 stars with \FeH$<-1.7$. Both  datasets are marked with blue circles in Figure~\ref{kinefig}. All the panels in Figure~\ref{kinefig} display that the distribution of the stars in our sample kinematically matches the literature's compilation. All the displayed VMPs  towards the inner region of the MW have high eccentricity ($\epsilon>0.5$), small pericentre ($\lesssim 2\kpc$), and a combination of $Z_{\rm max}$ and $r_{\rm apo}$ that makes them inhabit various regions of the Galaxy.

\section{The stellar parameters and the effects of the high extinction}\label{stellarparamssec}
\begin{figure}
\begin{center}
\includegraphics[width=0.5\textwidth]{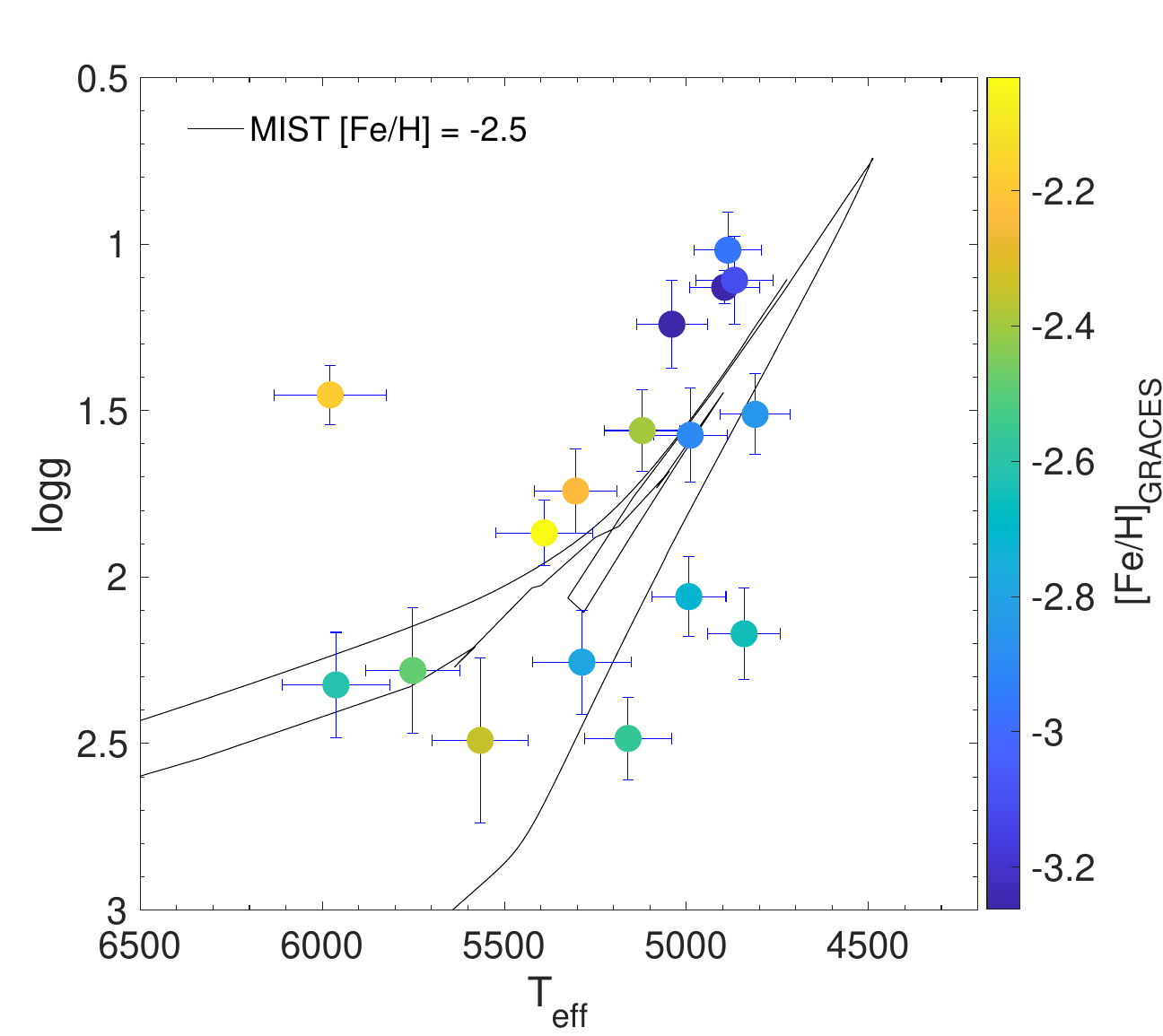}
\caption{Kiel diagram of the PIGS/GRACES sample. The markers are colour-coded by their GRACES metallicities. A MESA/MIST \citep{Choi16,Dotter16} isochrone at \FeH$=-2.5$ is overplotted as a comparison.}
\label{kielfig}
\end{center}
\end{figure}

The effective temperature is measured using the \citet[][hereafter MBM21]{Mucciarelli21} colour-temperature relation which combines the InfraRed flux method from \citet{Gonzalez09} with the photometry from  {\it Gaia} EDR3. The input parameters for this inference are the  {\it Gaia} EDR3 (BP-RP) colour, the reddening in this colour, a metallicity estimate, and  knowledge of whether a star is in the dwarf or giant phase. The 3D extinction map from \citet{Green19} was used to correct the photometry for  extinction\footnote{For P180503, the \citet{Green19} 3D extinction map does not provide a value, therefore the \textsc{StarHorse} extinction \citep{Anders19} was adopted for this star.}. This map provides the reddening E(B-V) that has to be converted to  {\it Gaia} filters. Then, the  {\it Gaia} extinction coefficients were derived using  $\rm A_V/E(B-V)= 3.1$ \citep{Schultz75} and the $\rm A_G/A_V = 0.85926$, $\rm A_{BP} /A_V = 1.06794$, $\rm A_{RP} /A_V = 0.65199$ relations \citep{Marigo08,Evans18}. As input metallicities, we adopt the values from the AAT analysis \citep{Arentsen20a}. Since the MBM21 relation needs the knowledge on the nature of the star (\ie dwarf or giant), we infer a first effective temperature that is the average of both the dwarf and giant solutions. With this first guess, a first estimate on the surface gravity using the Stefan-Boltzmann law\footnote{$L_{\star} = 4\pi R_{\star}^2 \sigma T_{\star}^4$; the radius of the star can be calculated from this equation, then the surface gravity is inferred assuming the mass.} is derived. This latter step requires as input the previously inferred effective temperature, the distance of the object, the  {\it Gaia} EDR3 G photometry, the extinction, and the bolometric corrections on the flux \citep{Andrae18}.  Then, with this first estimate of the surface gravity, we iterate the process to find the effective temperature and subsequently a new inference on the surface gravity. These steps have been iterated 1000 times in order to converge to a final estimate of surface gravity and effective temperature. The final values do not depend on the initial inference, especially on the dwarf-giant average made in the  first step. For each step, we perform a Monte Carlo on all the input parameters to estimate the uncertainties on the effective temperature and surface gravity. 
The input parameters are randomised within $1\sigma$ using a  Gaussian distribution, except for the stellar mass and the extinction.  The stellar mass is treated with a flat prior from 0.5 to 0.8 $\msun$, which is consistent with the mass of  very metal-poor stars. The extinction is also described with a flat prior with a width of $30$ percent from the assumed value from \citet{Green19}. The mean uncertainties on the effective temperature is $\sim113$ K, while on the surface gravity it is $\sim0.14$ dex.
This method has been shown to provide reliable stellar parameters that are in agreement, within the uncertainties, with  values obtained with spectroscopic methods \citep[\eg][]{Kielty21,Lardo21}.
Figure~\ref{kielfig} shows the Kiel Diagram of the inferred stellar parameters colour-coded by $\FeH_{\rm GRACES}$ with a VMP MESA/MIST\footnote{\url{https://waps.cfa.harvard.edu/MIST/}} \citep{Dotter16,Choi16} isochrone as a reference. All the stars appear to be giants. The inferred stellar parameters are reported in Table~\ref{tab:stellarparams}.

The photons of these targets travelled across multiple clouds of interstellar medium (ISM) before they got collected by the CCD of the telescope. Since the determination of the effective temperature and surface gravity highly depends on the de-reddened photometry, we investigate how much a different set of extinction coefficients can affect the final values of $\rm A_{G}$, $\rm A_{BP}$, and $\rm A_{RP}$. For this purpose, the extinction coefficients from \citet[][]{Casagrande21} have been tested. Using GALAH data \citep{Buder21}, they calibrate the extinction coefficients for  {\it Gaia} filters as a function of the colour, BP$-$RP, \citep[see Figure~1 and Appendix~B in ][]{Casagrande21}.
Therefore, $\rm A_{G}$, $\rm A_{BP}- A_{RP}$ have been computed with the new relation and tested against the ones adopted in this work. Considering a 30 percent  relative uncertainty on the E(B-V) from the \citet{Green19} 3D dust map towards the bulge region, we find that the $\rm A_{G}$ from \citet{Casagrande21} and the one adopted in this work differs by less than 0.6$\sigma$. The difference between the two methods for   $\rm A_{BP}- A_{RP}$ is less than 0.15$\sigma$. Therefore, the two methods agree within the uncertainties. Nonetheless, the derived stellar parameters  used for this work have been checked against the excitation potential as described in Section~\ref{chemsec} (see Figure~\ref{fitfig}) before being adopted as the final values. We stressed that, since the  {\it Gaia} filters are very broad, the extinction coefficient of these filters depends on the effective temperature of a star, and is not a fixed number. This  effect would be very important for photometric calibrations, while is negligible for spectroscopic analyses, as this work.

\section{Model Atmospheres Analysis} \label{models}

\subsection{Model atmospheres}
The first step to measure the chemical abundances in the stellar spectra is to have an ad hoc model atmosphere. The most up-to-date \textsc{MARCS}\footnote{\url{https://marcs.astro.uu.se}} models \citep{Gustafsson08,Plez12} are generated. In particular, for  stars with log(g)$<3.5$, \textsc{OSMARCS} spherical models are used.

\subsection{The lines list and the atomic data}
A lines list very similar to the one adopted by \citet{Kielty21} is employed for this work. It contains Fe lines from \cite{Norris17} and \cite{Monty20}, while the other species are generated with \textsc{linemake}\footnote{\url{https://github.com/vmplacco/linemake}} \citep{Placco21}. In addition, K\ione{} lines are from the National Institute of Standards and Technology \citep[NIST,][]{NIST_ASD}\footnote{NIST database at \url{https://physics.nist.gov/asd}}.

\subsection{Lines measurements}
Some of the observed spectra suffer from  poor flux normalization due to the suboptimal signal-to-noise ratio across the wavelength range. This translates into a failure of automatic line fits and procedures to measure chemical abundances. To obviate to this problem, the automatic procedure described in \citet{Kielty21} is run to identify the spectral lines and to create a common line list between the stars. Then, their equivalent widths (EW) are measured with \textsc{IRAF} \citep{Tody86,Tody93} using the \textsc{splot} routine. Multiple profiles (Gaussian, Voigt, integral of the flux) have been adopted to measure the EW, then the median is taken as final value. We discard strong lines (EW $>140$ m\AA, where differences in our measurement methods exceeded $\sim$15\%) and very weak lines at the level of the noise (EW $<20-25$ m\AA), depending on the location in our spectra.
Then, the \textsc{autoMOOG} code is used to infer the chemical abundances from the input EW and atmosphere models. This code is an automated version of the more popular \textsc{MOOG}\footnote{\url{https://www.as.utexas.edu/~chris/moog.html}} code \citep{Sneden73,Sobeck11}.

A table containing the EW measurements is provided as a machine readable table in the Supplementary materials.

\subsection{Metallicity: GRACES vs. AAT}\label{metsec}
Given the SNR of the observed spectra, the A(Fe\ione{})  is measured using from 8 to 63 lines in the $[4871,6678]$ \AA{} spectral range, while A(Fe\ii{}) only from 2 lines ($\lambda\lambda 4923.922, 5018.435$ \AA). The final [Fe/H] is calculated as the mean of A(Fe\ione{}) and A(Fe\ii{}), weighted by the number of lines and then scaled by the solar Fe content. Solar abundances are from \citet{Asplund09}.

\begin{figure}
\includegraphics[width=0.5\textwidth]{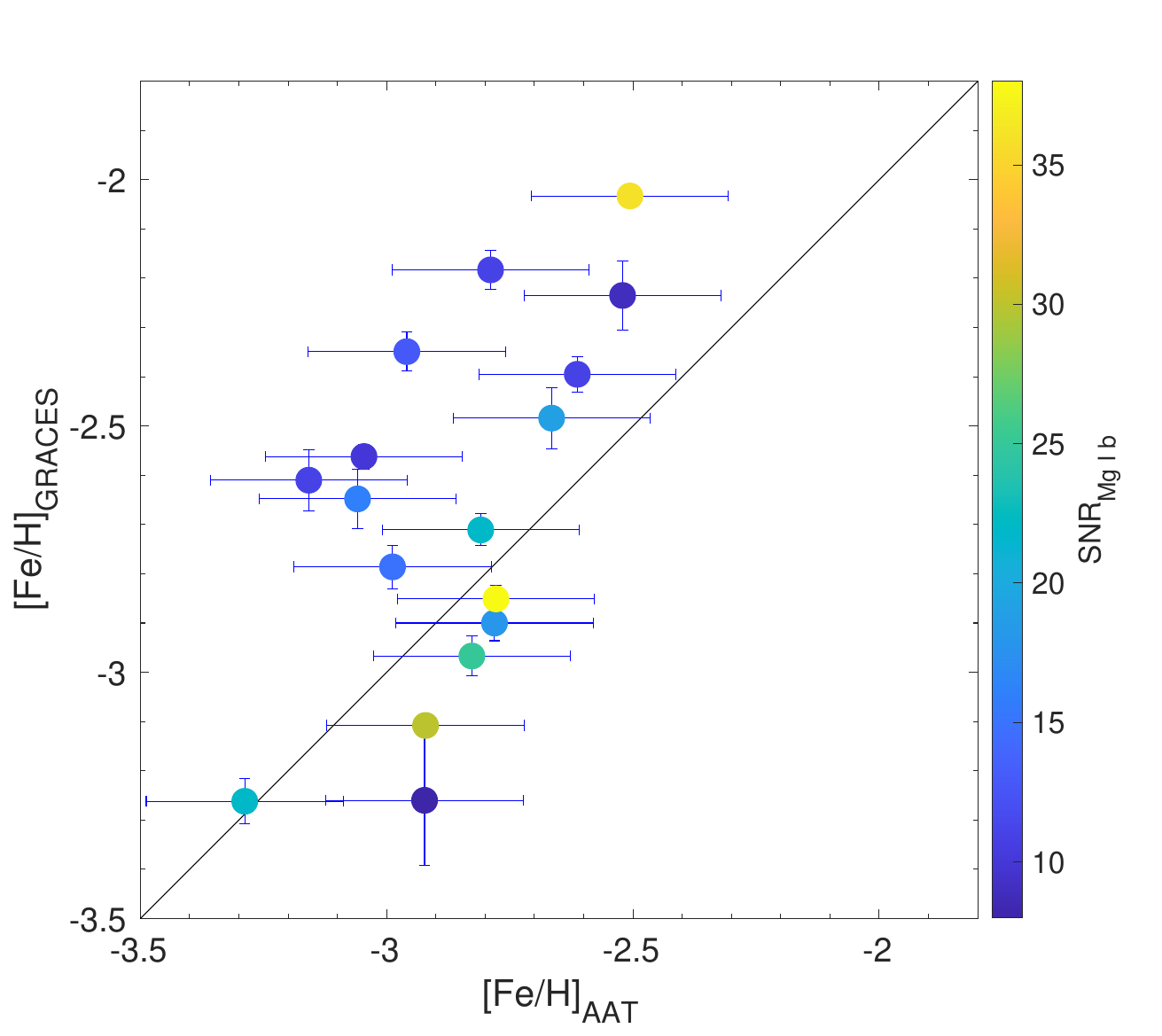}
\caption{Comparison of the metallicities between  GRACES high-resolution vs. AAT low/medium resolution. Both the \FeH{} are from  LTE analysis. Markers are colour-coded by the  SNR of the GRACES observations.}
\label{gracesferre}
\end{figure}

Figure~\ref{gracesferre} shows the comparison between the LTE $\FeH$ inferred in this work vs. the ones previously determined with FERRE at low/medium-resolution from AAT \citep{Arentsen20a}. The markers are colour-coded by the SNR measured at the Mg\ione{} b region (see also Table~\ref{tab:obs}). All the stars have low/medium $\FeH_{\rm AAT}\leq-2.5$, while only 11 have also $\FeH_{\rm GRACES}\leq-2.5$ from high-resolution. The remaining 6 stars are still VMP. Our method has also been calibrated on two VMP standard stars (HD122563 and HD84937), reproducing the literature values from the stellar parameters to the metallicity. Therefore, the deviation of the two measurements is thought to originate from a difference in the spectral resolution, and the high-resolution results are preferred in this work.

Table~\ref{tab:stellarparams} reports the RV of the targets, the inferred stellar parameters (T$_{\rm eff}$, logg, $\xi$), and the \FeH{} from LTE analysis.

\begin{table*}
\caption{The radial velocities, the effective temperature, surface gravity, microturbulence velocities, and LTE metallicities are reported with their uncertainties.}
\label{tab:stellarparams}
\begin{tabular}{ccccccccccc}
\hline
name short & RV   & $\sigma_{\rm RV}$ & T$_{\rm eff}$  & $\sigma_{\rm T_{eff}}$& logg & $\sigma_{\rm logg}$ & $\xi$ & $\sigma_{\xi}$ & \FeH{} & $\sigma_{\rm [Fe/H]}$ \\
& $(\kms)$ & $(\kms)$& (K) & (K) & & &$(\kms)$ &$(\kms)$ &  & \\
\hline
P170438     & -166.93 & 1.00        & 5566 & 132     & 2.49 & 0.25      & 2.33       & 0.10           & -2.35  & 0.04        \\
P170610     & -6.60    & 1.31       & 4994& 102    & 2.06 & 0.12      & 2.24       & 0.10           & -2.71  & 0.03        \\
P171457     & 265.99  & 0.59       & 5040& 98     & 1.24 & 0.13      & 2.29       & 0.10           & -3.26  & 0.05        \\
P171458     & 95.36   & 0.45       & 4990& 101    & 1.57 & 0.14      & 2.27       & 0.10           & -2.9   & 0.04        \\
P180118     & 86.84   & 0.84       & 5978& 154    & 1.45 & 0.09      & 2.61       & 0.10           & -2.18  & 0.04        \\
P180503     & -90.12  & 0.47       & 4895& 97     & 1.13 & 0.05      & 2.18       & 0.10           & -3.26  & 0.13        \\
P180956     & 262.45  & 0.54       & 5391& 133    & 1.87 & 0.10       & 2.50       & 0.10           & -2.03  & 0.02        \\
P181306     & 66.65   & 1.12       & 5752& 129    & 2.28 & 0.19      & 2.30       & 0.10           & -2.48  & 0.06        \\
P182129     & -154.23 & 0.56       & 4868& 107    & 1.11 & 0.13      & 2.44       & 0.10           & -3.11  & 0.02        \\
P182221     & -33.79  & 2.48       & 5962& 148    & 2.32 & 0.16      & 2.42       & 0.10           & -2.61  & 0.06        \\
P182244     & -225.93 & 1.58       & 4886& 92     & 1.02 & 0.12      & 2.48       & 0.10           & -2.97  & 0.04        \\
P182505     & 26.10    & 2.97       & 5287& 135    & 2.26 & 0.16      & 1.91       & 0.10           & -2.79  & 0.04        \\
P183229     & 1.41    & 1.02       & 5160& 119    & 2.48 & 0.12      & 2.14       & 0.10           & -2.56  & 0.03        \\
P183335     & 50.41   & 1.85       & 5304& 113    & 1.74 & 0.13      & 2.52       & 0.10           & -2.23  & 0.07        \\
P184338     & 218.38  & 0.35       & 5121& 103    & 1.56 & 0.12      & 2.63       & 0.10           & -2.39  & 0.04        \\
P184700     & -368.27 & 1.17       & 4842& 100    & 2.17 & 0.14      & 1.94       & 0.10           & -2.65  & 0.06        \\
P184855     & -458.3  & 0.78       & 4811& 95      & 1.51 & 0.12      & 2.41       & 0.10           & -2.85  & 0.03       \\
\hline
\end{tabular}
\end{table*}

\subsection{Checking the stellar parameters}
With high-resolution spectroscopy, it is possible to test if the input stellar parameters, \ie effective temperature, surface gravity and microturbulence velocity, are correct. Wrong estimates of the microturbulence velocity will produce a slope in the A(Fe\ione) vs. reduced EW, log(EW/$\lambda$), relation. We adopted the \citet{Mashonkina17} and \citet{Sitnova19} relations as  starting values for the microturbulence velocities and then refined to flatten the slope of the A(Fe\ione) $-$ log(EW/$\lambda$) curve.

\begin{figure}
\begin{center}
\includegraphics[width=0.5\textwidth]{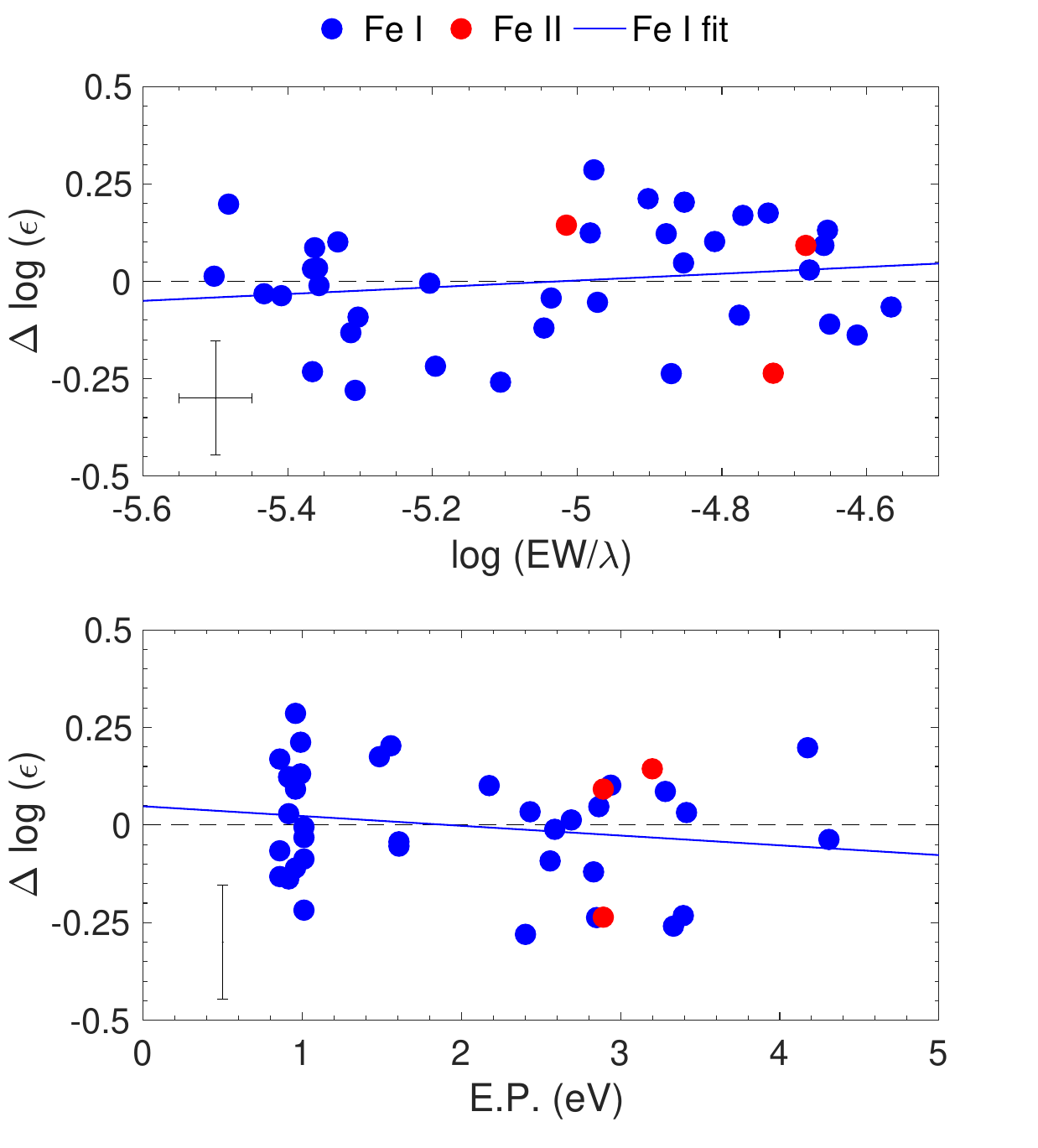}
\caption{Output from \textsc{autoMOOG} for P182244. Top: the deviation from the mean Fe chemical abundance as a function of the reduced equivalent width. Bottom: the deviation from the mean Fe chemical abundance as a function of the excitation potential. Values from Fe\ione{} and Fe\ii{} are denoted by blue and red markers, respectively. The fit for Fe\ione{} is shown with the solid blue line, while the dashed black line is centred on the null abundance difference. For P182244, the slope in the top panel is  0.087, while in the bottom panel the value is $-0.025$ dex eV$^{-1}$. A typical uncertainty on the reduced EW is reported in the top panel, while the standard deviation from \textsc{AutoMOOG} is reported in both the panels. The uncertainty on the excitation potential is not reported.}
\label{fitfig}
\end{center}
\end{figure}

In a similar way, a wrong estimate of the effective temperature would produce a slope in the A(Fe\ione) $-$ Excitation potential (EP) space. The slope from the linear fit has an absolute value $< 0.1$ dex eV$^{-1}$ using the  effective temperatures from MBM21. These values are smaller than the dispersion in the measurements of the chemical abundances, therefore there is no need to further tune the  effective temperatures. Figure~\ref{fitfig} summarises the aforementioned relations for P182244 using the output from \textsc{autoMOOG}.

Historically, the Fe\ione{} $-$ Fe\ii{} ionisation balance has been widely used as a sanity check on the surface gravity. However, the recent work from \citet{Karovicova20} showed that tuning the surface gravity to achieve the balance between Fe\ione{} and Fe\ii{} can lead to wrong  chemical abundances and gravity estimations. They used extremely precise interferometric observations of metal-poor stars, producing the most precise and accurate stellar parameters for a set of metal-poor benchmark stars. \citet{Karovicova20} found that the deviation from  the Fe\ione{} $-$ Fe\ii{} ionisation balance can reach up to $\sim0.8$ dex. This effect is very important when dealing with (very) metal-poor cold giants \citep[see Figure~7 of][\eg \FeH$<-2.0$, log(g)$<3$, and T$_{\rm eff}\sim5500$ K]{Karovicova20}. This is exactly the same range in stellar parameters of the PIGS/GRACES stars. For this reason, we refrain from tuning the surface gravity to reach this balance.

Appendix~\ref{appdr3} and Figure~\ref{gaiadr3_params} discuss the comparison between our set of stellar parameters and the one released from the most recent  {\it Gaia} DR3 \citep[][]{Andrae22,Gaia16}.

\subsection{Uncertainties on the chemical abundances}
\textsc{AutoMOOG} provides estimates of the chemical abundances A(X) and their uncertainties $\sigma_{\rm A(X)}$. The abundance uncertainties are calculated by adding the line-to-line scatter ($\sigma_{\rm EW}$) in quadrature with the uncertainties imposed by the stellar parameter uncertainties ($\sigma_{\rm T_{eff}}$, $\sigma_{\rm logg}$, $\sigma_{\rm \FeH}$, see Table~\ref{tab:stellarparams}). The final uncertainty  on element X is given by $\delta_{\rm A(X)}=\sigma_{\rm A(X)}/\sqrt{{\rm N_X}}$ if the number of the measured spectral lines is ${\rm N_X}>5$, or  $\delta_{\rm A(X)}=\sigma_{\rm A(Fe\ione)}/\sqrt{{\rm N_X}}$. The uncertainties on the chemical abundances are provided as a machine readable table in the Supplementary material.

\section{Chemical abundances analysis}\label{chemsec}
The spectral coverage of the GRACES spectrograph  enables estimates of the abundances of Fe-peak  (Fe, Cr, Ni), $\alpha-$ (Mg, Ca, Ti), odd-Z (Na, K, Sc), and neutron-capture process (Ba) elements. In addition, some intrinsically weak lines (\eg O\ione{} 7770 \AA{} and Eu\ii{} 6645 \AA) are in the spectral region covered by our GRACES spectra; however, these lines are detectable only if the stars are highly enhanced in the element \citep[][]{Kielty21}.

Chemical abundances in both LTE and NLTE analysis are provided as a machine-readable table in the Supplementary material.

\subsection{The Swan band and the Carbon from low-resolution spectroscopy}\label{cempsec}
According to the  low/medium resolution PIGS campaign  \citep[][]{Arentsen21}, P183229 and P184700 are C-enhanced with [C/Fe]$=2.10\pm 0.22$ and [C/Fe]$=2.85\pm 3.12$, respectively. 
In general, Carbon is visible in the wavelength range covered by GRACES spectra  through the Swan band ($\lambda5100-5200$\AA) but only when the star is C-enhanced and relatively cold \citep[with the exception for SDSS J081554.26$+$472947.5][]{Ganzalez20}. This is the case for P183229 and P184700, for which the Swan band is very pronounced. The large uncertainty on the [C/Fe] for P184700 might originate from a combination of low SNR and such extreme carbon-enhancement that the models don't fit well.   The left panel of Figure~\ref{spectraex} displays the Swan band for P183229 in comparison to other Carbon-normal stars.  We thus qualitatively confirm the C-enhanced nature of these two stars.  A third star, P182221 has [C/Fe] $ =0.64 \pm0.73$ from  PIGS/AAT. This value is slightly below the CEMP threshold ([C/Fe]$=0.7$), although its high uncertainty. The GRACES spectrum does not display an evident Swan band as for the other two CEMP stars.

\subsection{$\alpha$-elements}
\begin{figure*}
\begin{center}
\includegraphics[width=\textwidth]{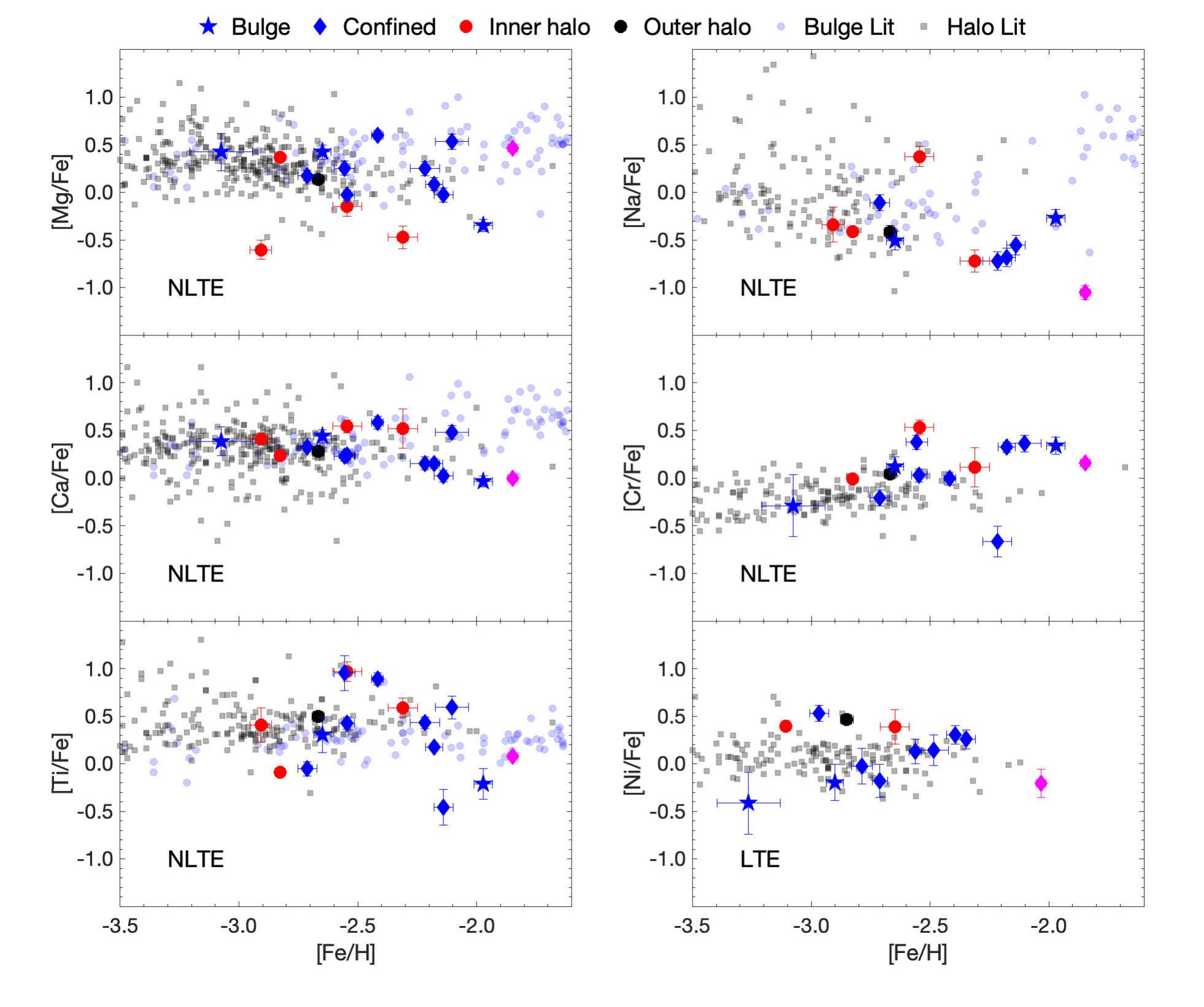}
\caption{Chemical abundances ratios as a function of \FeH. $\alpha-$elements Mg, Ca and Ti are displayed on the left panels, Odd-Z Na is on the top right, and the Fe-peak elements Cr and Ni are on the central and bottom right panels. All the panels, except [Ni/Fe] vs. \FeH{}, are corrected by NLTE effect as discussed in Section~\ref{nltesec}. The stars in the PIGS/GRACES sample are marked with the same symbols and colours as Figure~\ref{kinefig}. The bulge literature sample (blue circles) is composed by stars from \citet{Howes14,Howes15,Howes16,Koch16,Reggiani20,Lucey22}. Halo literature compilation (grey squares) are from \citet{Aoki13,Yong13,Kielty21}.}
\label{chemfig}
\end{center}
\end{figure*}

$\alpha$-elements are primarily formed in massive stars before being ejected by core-collapse supernovae and during the supernovae event \citep[\eg][]{Timmes95,Kobayashi20}. There are only three $\alpha$-elements which produce lines in the GRACES spectra, Mg, Ca and Ti. The A(Mg\ione{}) is from two lines of the Mg\ione{} Triplet ($\lambda\lambda 5172.684, 5183.604$\AA), the weaker $5528.405$\AA{} line, and the $4702.991$ \AA{} line for which the SNR is high. The A(Ca\ione{}) is inferred from up to 10 spectral lines, from 5588 \AA{} to 6500 \AA. The Ca Triplet has been excluded since it shows strong lines ($>140$ m\AA). In these GRACES spectra, up to 4 and 6 lines of Ti\ione{} and Ti\ii{} are present \citep{Lawler13,Wood13}, respectively. The three left panels of Figure~\ref{chemfig} display the [Mg,Ca,Ti/Fe] ratios as a function of the \FeH{}, corrected for NLTE effects (see Section~\ref{nltesec}). When present both Ti\ione{} and Ti\ii{} lines, the [Ti/Fe] is the average of [Ti\ione{}/Fe] and [Ti\ii{}/Fe].

\subsection{Odd-Z elements}
Odd-Z elements are tracers of core-collapse supernovae. In particular, the difference in energy between the neutron capture and the $\alpha-$ particle capture produce the so called odd-even effect in the chemical yields  \citep[\eg][]{Heger10,Takahashi18}.

\begin{figure}
\hspace{-0.5cm}
\includegraphics[width=0.5\textwidth]{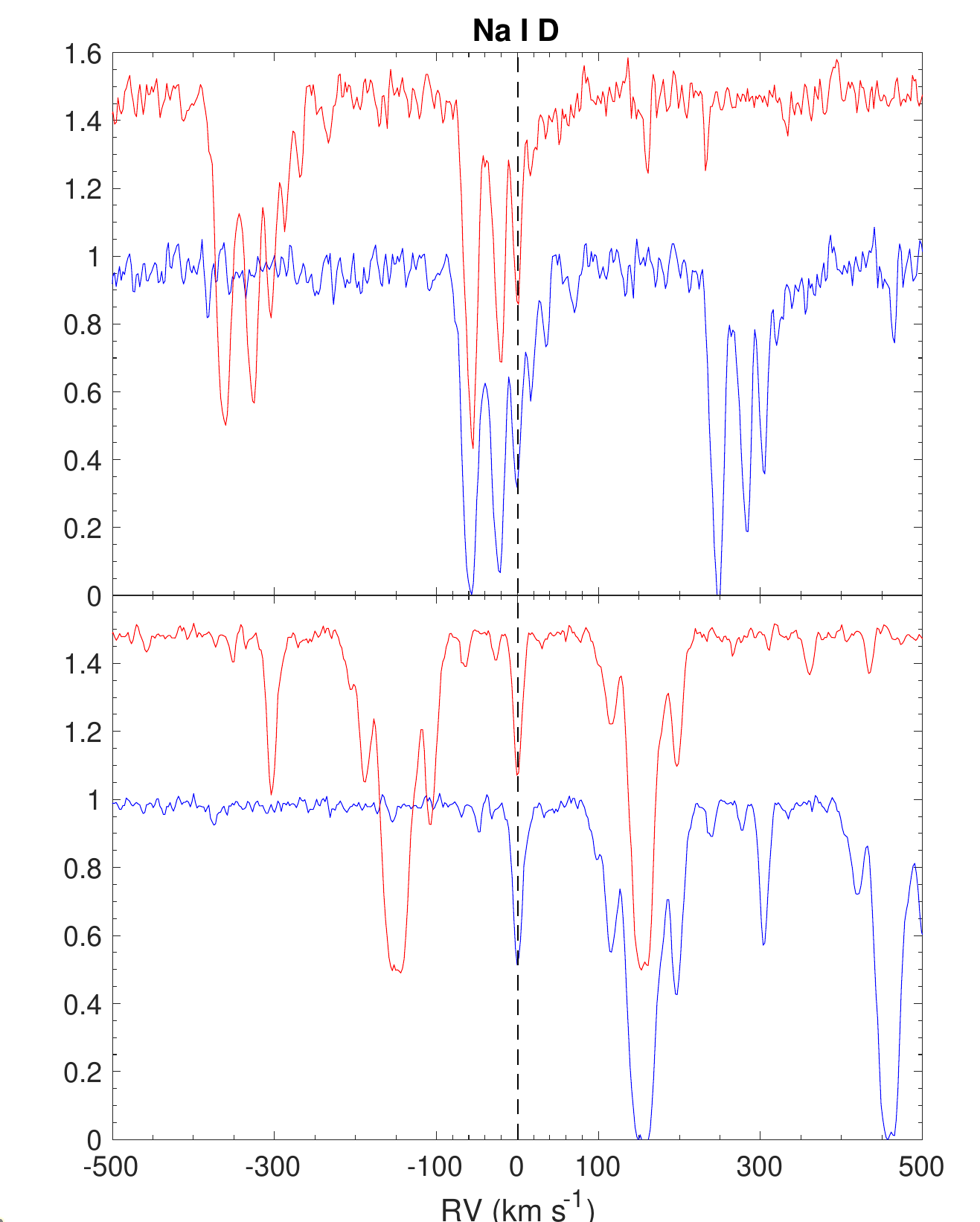}
\caption{Na\ione{} D in the spectral velocity space. The velocity for the blue lines is null at the Na\ione{} D 5890 \AA{} line, while for the red lines is null at the  Na\ione{} D 5896 \AA{} line. The vertical dashed line is positioned at the velocity at rest of the targets. Top: P183335, example of a star in which the ISM Na\ione{} D is blended with the stellar component for both the lines. Bottom: P182129, example of star in which the ISM Na\ione{} D is not blended to the stellar component. In both panels, it clear that the presence of multiple clouds of ISM with different velocities between us and the targets. }
\label{Nafig}
\end{figure}

Three odd-Z elements are observable in the spectra, Na, K and Sc. The Na abundance is measurable from the Na\ione{} Doublet ($\lambda\lambda 5889.951,5895.924$ \AA).
 The ISM Na\ione{} D lines are present with multiple components and might have formed from clouds at a similar RV of our targets. Therefore, the ISM and stellar component could be blended. In 5 stars analysed out of 17, it is not possible to measure Na\ione{} D EW due to the blending. The two panels of Figure~\ref{Nafig} show  two cases in which the ISM Na\ione{} D is blended (upper panel for P183335) and not blended (bottom panel for P182129). 
 
K\ione{} is observable with two lines at $\lambda\lambda7664.899, 7698.965$ \AA{} \citep{Falke06,Trubko17}. These lines are very close to water vapour lines of the Earth's atmosphere. A(K\ione{}) is measurable only when at least one line is not blended with the atmospheric lines and when the SNR is sufficiently high. A(K\ione{}) is measurable in 1 star from the $\lambda\lambda7664.899$\AA{} line and from $\lambda\lambda7698.965$ \AA{} in 6 other stars. Only in one star are both K\ione{} visible and  yield the same estimate of [K/Fe]. Therefore, A(K\ione{}) is measured in 8 stars out of 17. 

Sc is present with one Sc\ii{} line at $\lambda\lambda5526.785$ \AA{} \citep{Lawler19}.
For 10 stars, the SNR in the Sc\ii{} region is very low; therefore, it is not possible to  fit the line and measure A(Sc\ii). The abundances of K and Sc have been measured with the  \textsc{synth} configuration in \textsc{MOOG} and taking the hyperfine structure into account for Sc.

Figure~\ref{chemfig} shows only [Na/Fe] (NLTE corrected) ratio among the odd-Z elements, since Sc and K are measurable only for a few stars.

\subsection{Fe-peak elements}
 Fe-peak elements are important tracers of stellar evolution. In the early Universe, when the very metal-poor stars were forming, Fe-peak elements were produced primarily in core collapse supernovae \citep[\eg][]{Tolstoy09,Heger10}. While at much higher metallicities, hence later in cosmic time, Fe-peak elements were formed in supernovae type Ia \citep{Nomoto13}.

The Fe-peak elements that are observable in the GRACES spectra are Fe (see Section~\ref{metsec}), Cr and Ni.
A(Cr\ione) is measured with up to 5 spectral lines \citep[$\lambda\lambda 5206.023, 5208.409, 5345.796, 5348.314, 5409.783  $\AA,][]{Sobeck07}, while Ni\ione{} is present with up to 4 lines \citep[$\lambda\lambda 5115.389, 5476.904, 5754.656, 6482.796 $ \AA,][]{Wood14}. 
Figure~\ref{chemfig} shows [Cr/Fe] (NLTE) and [Ni/Fe] (LTE) as a function of \FeH{} for the stars in this sample.

\subsection{Neutron-capture process elements}
Neutron-capture elements can be formed through two main channels, the rapid and the slow neutron captures. Rapid-process elements are formed if their nuclear production timescale is much shorter than the time needed by the $\beta^{-}$ decay. This is the case for core collapse supernovae and neutron-star mergers. Otherwise, if the timescale for their synthesis is longer as in the stellar atmospheres of  AGB stars, then these elements are named slow-process.

\begin{figure}

\includegraphics[width=0.5\textwidth]{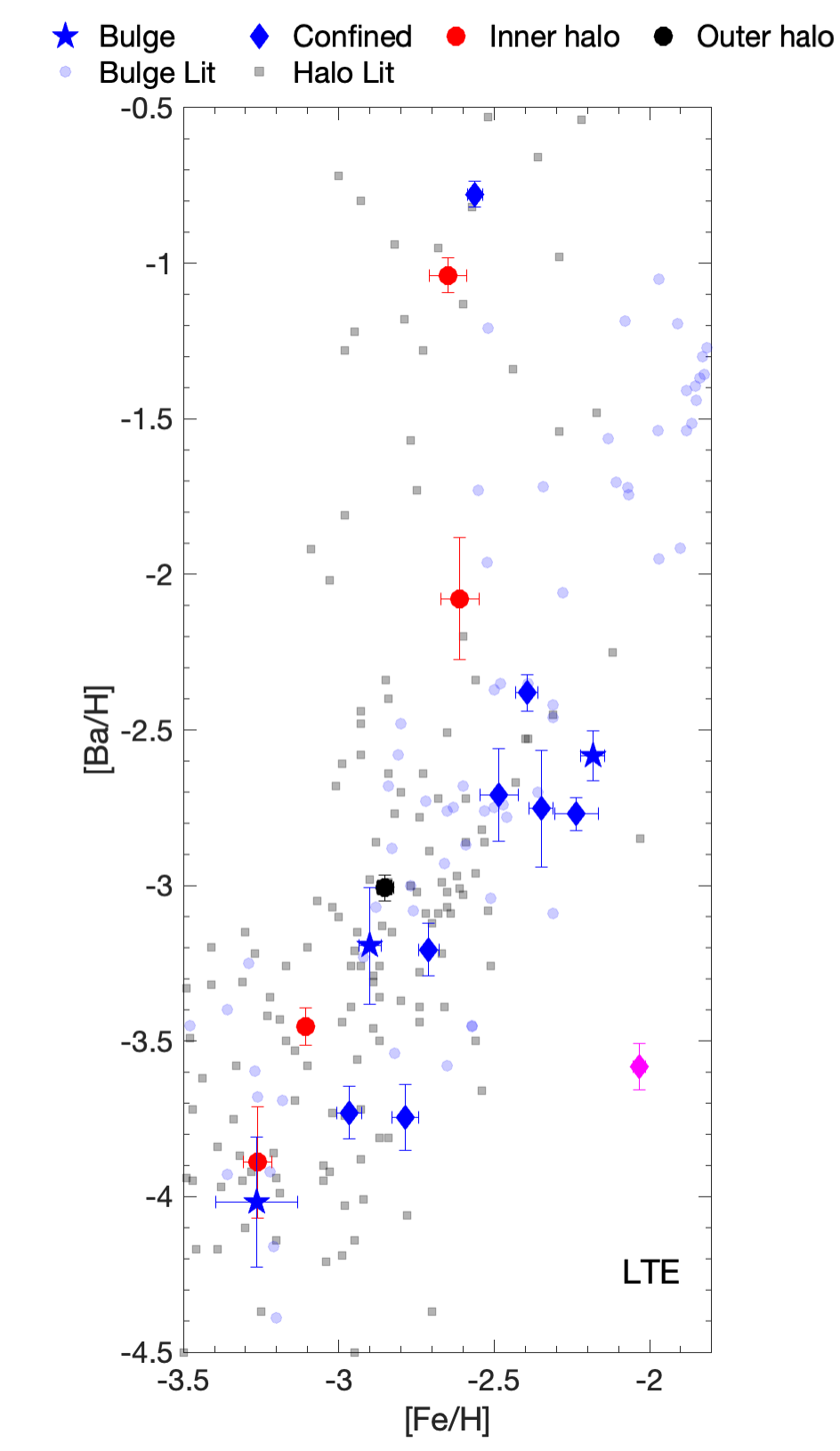}
\caption{[Ba/H] vs. \FeH. Ba abundances are not corrected by NLTE effect (see Section~\ref{nltesec}). The stars in the PIGS/GRACES sample are marked with the same symbols and colours as Figure~\ref{kinefig}. The bulge literature sample (blue circles) is composed by stars from \citet{Howes14,Howes15,Howes16,Koch16,Reggiani20,Lucey22}. Halo literature compilation (grey squares) are from \citet{Aoki13,Yong13,Kielty21}. The two Ba-rich stars are C-enhanced stars, therefore they high Ba is produced by the accretion of material from an AGB companion (see Section~\ref{cempsec}~and~\ref{cempgc}. The low-Ba star (magenta diamond marker) is likely accreted from a dwarf galaxy (see Section~\ref{planarsec} and Figure~\ref{badwarf}). The majority of the sample possesses a [Ba/H] ratio very similar to the MW halo stars at the same \FeH{} (see Section~\ref{nonpeculiar}).}
\label{bafig}
\end{figure}

The neutron-capture process elements present are Ba, with up to three Ba\ii{} lines ($\lambda\lambda 5853.69, 6141.73, 6496.91 $ \AA, e.g., see Fig.~\ref{spectraex}), and Y, with only one Y\ii{} line \citep[$\lambda\lambda 5200.413$ \AA,][]{Hannaford82,Biemont11}. To infer the A(Ba\ii{}), \textsc{MOOG} has been run with the synthetic configuration to take the hyperfine structure and corrections into account. The Y\ii{} line is measurable in only 6 stars. [Ba/H] (LTE) as a  function of \FeH{} is shown in Figure~\ref{bafig}, while [Ba/Fe] is displayed in Figure~\ref{badwarf} and discussed in Section~\ref{planarsec}.

The GRACES spectra cover only one very weak Eu\ii{} line at $\lambda\lambda 6645.11$\AA{}. This line is visible only when a star is Eu-rich \citep[see Figure~8 in][]{Kielty21}, which is not the case for 
any of our stars. Upper limits to A(Eu\ii) do not constrain if these stars were polluted by r- or s-processes.

Figure~\ref{bafig} shows that two stars, P183229 and P184700, are strongly Ba-rich. This enhancement is explained by their CEMP nature as discussed in Sections~\ref{cempsec}~and~\ref{cempgc}. P182221, seems to be slightly enriched in Ba too ([Ba/H]$\sim-2$ at $\FeH\sim-2.6$). As for the other two Ba-rich stars, the Eu line is not detectable, therefore we can exclude that this star is r-process enhanced. The likely CEMP nature of this star is discussed in Section~\ref{cempgc}.

\subsection{NLTE corrections}\label{nltesec}
The elemental abundances in the atmospheres of very metal-poor stars are affected by  departures from Local Thermodynamic Equilibrium (LTE). Thus, the statistical equilibrium solutions need to  correct for radiative effects (non-LTE, or NLTE effects), which can be large for some species. 
To correct for  NLTE effects in Fe\ione{} and Fe\ii{} \citep{Bergemann2012}, Mg\ione{} \citep{Bergemann2017}, Ca\ione{} \citep{Mashonkina17}, Ti\ione{} and Ti\ii{} \citep{Bergemann2011}, and Cr\ione{} \citep{Bergemann2010b} the MPIA webtool database\footnote{\url{http://nlte.mpia.de}} has been used. On the other hand, \textsc{INSPECT}\footnote{\url{http://inspect-stars.com}} has been adopted to correct for NLTE effect in Na\ione{} \citep{Lind2012}. In this sample, NLTE corrections are on  order of $0.1-0.3$ dex for lines of Fe\ione{} and $<0.1$ dex for lines of Fe\ii{}.  Similarly small NLTE corrections are found for lines of Mg\ione{} ($<0.1$ dex) and Ti\ione{} ($<0.05$ dex).  Larger corrections were found for lines of Ca\ione{} ($\sim0.25$), Ti\ii{} ($\sim0.5$ dex), Na\ione{} ($0.2-0.4$ dex for the Na\ione{} D resonance lines), and Cr\ione{} ($\sim0.5$ dex). NLTE chemical abundances are provided as a machine-readable table in the Supplementary material.

\section{Discussion} \label{discussionsec}
For this discussion, a literature selection of stars in the Milky Way halo and bulge, globular clusters, and dwarf galaxies was put together to compare with the chemical results from this work. 
Metal-poor MW halo and bulge stars  have been selected with $-4<$\FeH$<-1.7$ to match the PIGS/GRACES sample. 
The compilation includes the high-resolution spectral analysis results from \citet{Aoki13,Yong13}, and particularly stars in \citet{Kielty21}, which have been observed with GRACES at SNR comparable to targets in this work, and analysed with a similar methodology (line lists, stellar parameters, model atmospheres).  The bulge compilation  is from the high-resolution optical spectral analyses by \citet{Howes14,Howes15,Howes16,Koch16,Reggiani20,Lucey22}. 
A compilation of stars with high resolution spectral analyses in globular clusters was collected from \citet{Pancino17,Pritzl05,Larsen22,Martin22}, and for stars in a selection of dwarf galaxies, including Hercules \citep{Koch08,Koch13,Francois16} and Segue 1 \citep{Frebel14}.

\subsection{The Inner Galaxy}\label{nonpeculiar}
According to various cosmological simulations \citep[\eg][]{Salvadori10,Tumlinson10,ElBadry18,Starkenburg17a,Sestito21}, during the first $2-3$\Gyr{} many low-mass systems ($\sim10^8\msun$) merge somewhat chaotically to form the proto-Galaxy. Each of these building blocks brought in the oldest and most pristine stars, the ISM, and the dark matter. During this phase, because the mass ratio of the merging clumps was not extreme, stars from different building blocks were able to populate may regions of the Galaxy, including its very central regions. As the Galaxy grew more massive, later accretions were able to provide their stars mainly to the outer  halo and possibly to the disc. Therefore, the inner galaxy should include some of the oldest stars accreted during the early assembly, while the halo would be a mixture of stars brought in across the early and later accretion history. 
In addition, a VMP star ($\FeH\leq-2.0$) that forms in an evolved dwarf galaxy and is accreted at a later time, will show a different chemical signature than a \textit{normal} halo star. 
This has been shown for stars in the halo, \eg  r-process-weak stars \citep[\eg][]{Kielty21,Lucchesi22}, r-process-enhanced stars \citep[\eg][]{Hansen18}, or stars with low [Ca/Mg] ratio \citep[\eg][]{Sitnova19, Venn20}.
In combination with the kinematics, this is the basis for the discovery of accreted structures in the Galaxy, \eg the Gaia-Enceladus/Sausage system \citep[\eg][]{Belokurov18,Helmi18}, Sequoia \citep[\eg][]{Myeong19, Monty20}, the Inner Galactic Structure \citep{Horta21}, and  Thamnos \citep{Koppelman19}.
In particular, the [$\alpha$/Fe],  Fe-peak, and the odd-Z element abundance ratios can differ from those of the metal-poor MW halo stars.
Thus, the chemistry of stars in the inner Galaxy can be used to test scenarios for  the formation and evolution of the Milky Way. 

When we examine our chemical abundances for stars that appear constrained to the bulge, we do not find significant differences for most stars from that of the MW halo metal-poor stars. This includes the distribution of the $[\alpha/{\rm Fe}]$ ratio around $\sim0.4$, as in Figure~\ref{chemfig}, and the typical [Ba/H] ratio trend with the increase of the \FeH{}, as in Figure~\ref{bafig}. These findings are in agreement with the  high-resolution analysis of bulge VMPs from \citet{Howes14,Howes15,Howes16}. They discuss that the majority of the stars in their sample are indistinguishable from MW halo stars at the same metallicities.
Generally, we observe the same and we conclude that the majority of stars with high resolution spectral analyses in the bulge resemble those of the majority of stars in the halo.

\subsection{No evidence for PISNe}\label{pisnesec}

During the early Universe and amongst the earliest generations of stars, some of the low-metallicity stars are predicted to have formed from gas enriched only in the yields from Pair-Instability Supenovae \citep[PISNe,][]{Ji15}.  This is when the highly energetic thermonuclear explosions of very massive stars ($150<{\rm M}_{\mathrel{prog}} < 260 \msun$) are carried by the pair production of electrons and positrons formed in the massive CO cores ($ >65 \msun$). The question of the initial mass function for the first stars is still unresolved; however, many theoretical models suggest that first stars were very massive, up to a few hundreds of $\msun$, given that molecular hydrogen would have been the only available coolant \citep[\eg][]{Omukai01,Bromm02,Stacy10}. 

\begin{figure}

\includegraphics[width=0.5\textwidth]{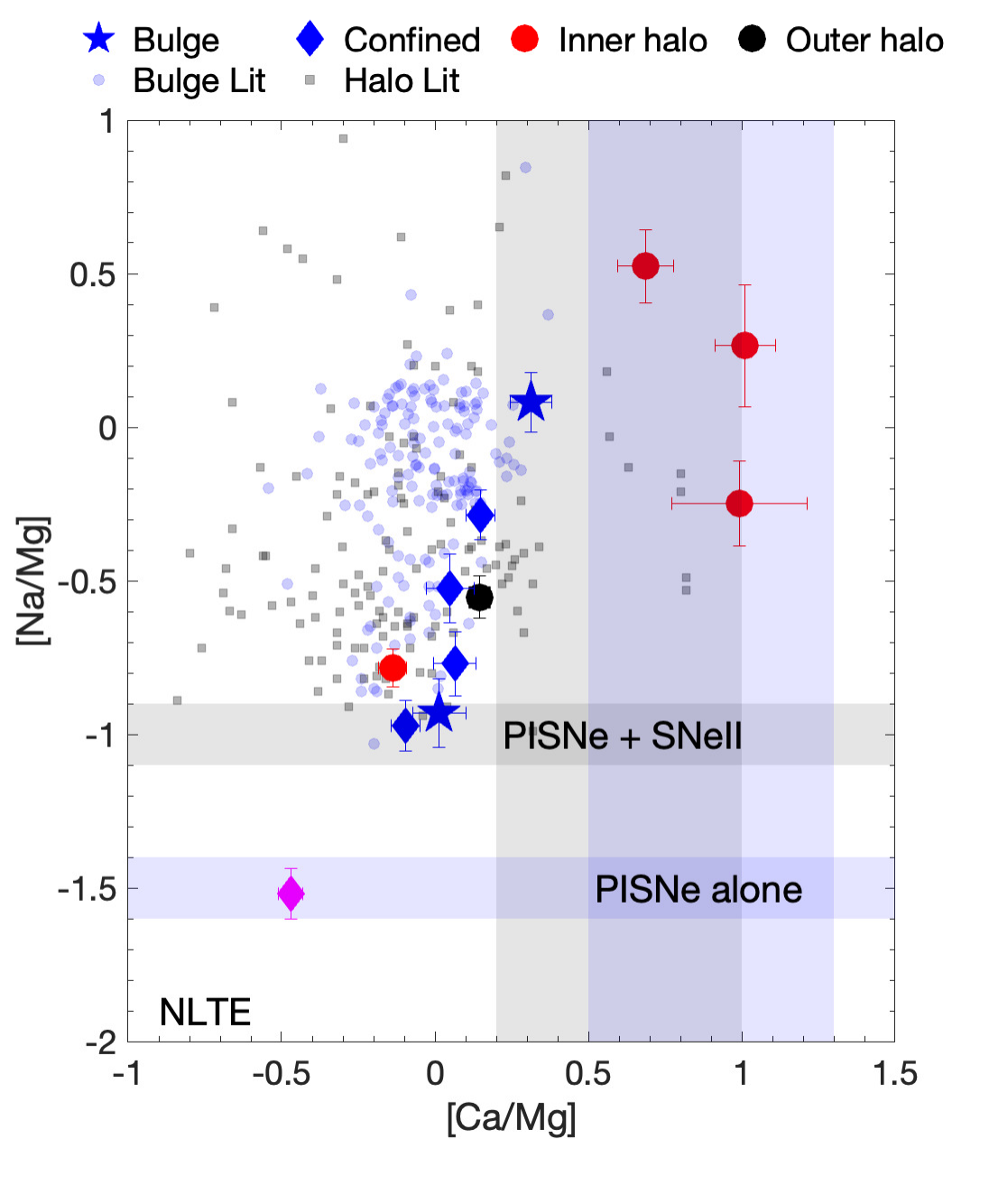}
\caption{Pair-Instability Supernovae yields space. The chemical abundances ratios are corrected for NLTE effects. The compilations from the bulge \citep[blue circles,][]{Howes14,Howes15,Howes16,Koch16,Reggiani20,Lucey22} and from the literature \citep[grey squares,][]{Aoki13,Yong13,Kielty21} are displayed. Stars in the overlapping region between the two blue bands would possess the signature of the PISNe yields alone scenario as shown in \citet{Takahashi18}. While the overlapping region of the grey bands is the locus in which the  stars would have been polluted by a PISNe and a SN II as in \citet{Salvadori19}. For the latter case, we show the yields relative to a PISNe to SN II ratio between 0.5 and 0.9, following Figure~6 from \citet{Salvadori19}.
}
\label{pisnefig}

\end{figure}

PISNe produce a strong odd-even effect \citep[ratios of odd-Z to even-Z elements, such as $\rm{[Al/Mg]}$,][]{Heger02, Aoki14}, however the typical metal-poor halo stars (with [Fe/H] $< -2.5$) have chemical abundances that resemble the predicted yields from lower mass, metal-poor core-collapse supernovae \citep[CCSNe, \eg][]{Joggerst10, Ishigaki18}.  As the ejecta from multiple CCSNe mix together, any unique Pop III abundance patterns which may have been preserved in metal-poor stars would be erased.
Nevertheless, \citet{Takahashi18} predict PISNe yields as a function of progenitor mass (and with few differences between rotating and non-rotating models), which we compare to the results for the inner Galaxy stars from our GRACES spectral analysis.
The chemical abundances for Mg, Ca, and Na, observable in our GRACES spectra, do not vary significantly with PISNe progenitor mass, i.e., a PISN produces a high ratio of Ca to Mg ($0.5<$[Ca/Mg]$<1.3$) and very low amounts of Na ([Na/Mg]$\sim-1.5$). In Figure~\ref{pisnefig}, the [Na/Mg] vs. [Ca/Mg] for our PIGS/GRACES sample are shown, together with bulge stars from the literature. For this latter comparison, stars with $\FeH >-2$ have not been removed to show the dearth of stars with detectable PISNe signatures.
Three stars with inner halo-like orbits (P171457, P182221, and P184700) have high [Ca/Mg] compatible with PISNe yields (blue shaded area), yet their observed [Na/Mg] are too high.  This is similar to the results from the COMBS survey \citep[Figure~13 of][]{Lucey22}, which made use of Al (not covered by GRACES) instead of Na. Thus, in both studies, no signature of  PISNe yields is found.

Alternatively, \citet{Salvadori19} investigate the combination of PISNe and SN II yields, which produce different yields  (see grey shaded area in Figure~\ref{pisnefig}). 
Again, none of our stars occupy the overlapping PISNe + SNe II regions.
\citet{Salvadori19} suggest that the smoking gun for the detection of PISNe yields would include low chemical abundances of N, Zn, and Cu; unfortunately, none are measurable in our GRACES spectra. 
They also suggest that PISNe would pollute the ISM up to $\FeH\sim-2.0$,
making it  unlikely to ever detect PISNe signatures in the most metal-poor stars.

\subsection{Comparisons with Globular Clusters}
\subsubsection{Second-generation globular cluster stars}\label{secgen}

It has been proposed that some of the building blocks of the Galactic bulge could also have been from ancient globular clusters disrupted at early times \citep[\eg][]{Shapiro10,Kruijssen15,Bournaud16}. Quantitatively, \citet{Schiavon17} and \citet{Horta21b} have estimated that up to $\sim25$\% of the stellar mass of the inner halo within 2 \kpc{} from the Galactic centre is made of disrupted GCs, where those clusters were more massive (by $10-100$) than those observed today. Similarly, \citet{Martell11} suggested that a minimum of 17 percent of the present-day mass of the stellar halo was originally formed in globular clusters. Some research has confirmed the presence of stars in the bulge with peculiar chemical signatures similar to those in present-day GCs \citep[\eg][]{Trincado17,Schiavon17,Lucey19,Lucey22}, \eg the Na-O and/or Al-Mg anticorrelation found only in globular cluster red giants \citep[\eg][]{Gratton04, Martell11, Carretta12, Pancino17}. This is typical of the so-called second-generation globular cluster stars (or enriched stars), which their chemical imprint is thought to be governed by CNO cycle processing at high temperatures \citep[\eg][]{Gratton04,Bastian18}.

\begin{figure}
\hspace{-0.9cm}
\includegraphics[width=0.55\textwidth]{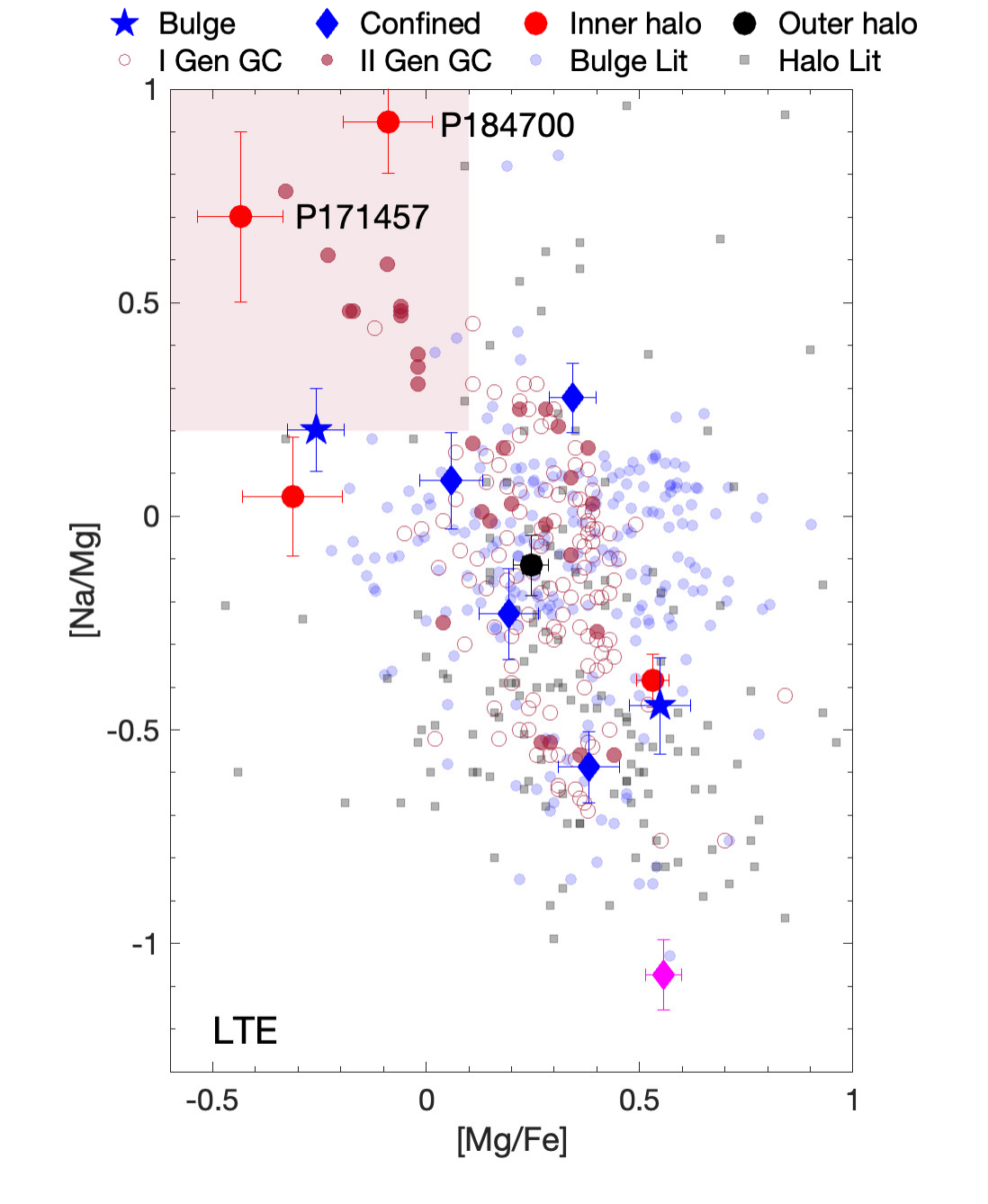}
\caption{Second-generation globular cluster stars space. Together with the PIGS/GRACES sample, three compilations for the bulge \citep[blue circles,][]{Howes14,Howes15,Howes16,Koch16,Reggiani20,Lucey22}, for the literature \citep[grey squares,][]{Aoki13,Yong13,Kielty21}, and for GC \citep[open red circles for second-generation stars and solid red circles  for first generation stars from GCs,][]{Pancino17} are displayed. Chemical ratios are LTE since the GC comparison stars from \citet{Pancino17} are in  LTE.
}
\label{gcsecfig}
\end{figure}

The spectral coverage of GRACES contains four Al\ione{} lines (at $\AA6 696.015,6 698.673,8 772.866,8 773.896$\AA), however, no lines were detected.  Upper limit estimates suggest [Al/Fe]$\leq+2$, which does not provide a meaningfully discriminating constraint. Therefore, we substitute Al with Na to study the odd-even effect.  \citet{Pancino17} suggests that [Na/Mg] is smaller than [Al/Mg] in displaying the anti-correlation effects.  To separate out the first ([Al/Mg] and [Na/Mg] normal) and second-generation stars ([Al/Mg] and [Na/Mg] rich), we cut the sample at [Mg/Fe]$\leq0.1$ vs. [Mg/Fe]$\leq0.4$ to help the second-generation population stand out more clearly  (the first generation stars and MW halo-like stars possess [Mg/Fe]$\sim0.4$). 
The  [Na/Mg] vs. [Mg/Fe] abundances are shown in Figure~\ref{gcsecfig}.
Chemical abundances from \citet{Pancino17} are also shown, which are not NLTE corrected, hence we report the LTE measurements for our sample as well.
Two GRACES stars, P171457 and P184700, clearly populate the second-generation GC region, given their high [Na/Mg] and low [Mg/Fe] ratios. Both have inner halo-like orbits, and one of them (P184700) is a CEMP star (see Section~\ref{cempgc}).  Two other stars (P180118 and P182221) have lower [Na/Mg] ratios compatible with the first generation stars. 

\citet{Lucey22} also reported two stars compatible with the second-generation using the [Al/Fe] vs. [Mg/Fe] space.  We do not show their sample in Figure~\ref{gcsecfig} since they do not provide LTE Na abundances and the majority of their stars are not very metal-poor.

\subsubsection{Extragalactic globular clusters}\label{extragc}
The chemical abundances of 7 GCs in M31 were compared with GCs in the MW, Sagittarius, Fornax, and the Large Magellanic Clouds by \citet{Sakari15}. They found that the [Mg/Ca] distribution of the extragalactic GCs is wider and mainly negative in the M31 GCs, whereas it peaks at [Mg/Ca]$\sim 0$ in the MW GCs.
The only exception is NGC~2419, which has stars with a wide range of [Mg/Fe] values, and negative values for [Mg/$\alpha$] \citep[\eg][]{Mucciarelli12}. These features, together with the dispersion in the abundances of K and Sc, and its retrograde orbit, have led to speculation that this GC has an extragalactic origin \citep{Cohen12}. 
\citet{Pancino17} collected data from various Galactic and extragalactic GCs. They looked at the distribution of the $\alpha$-elements available, removing the second-generation stars (Al-enhanced and Mg-poor). They reinforce the hypothesis that all the extragalactic GCs have statistically lower mean Mg content than MW GCs.

\begin{figure}

\includegraphics[width=0.5\textwidth]{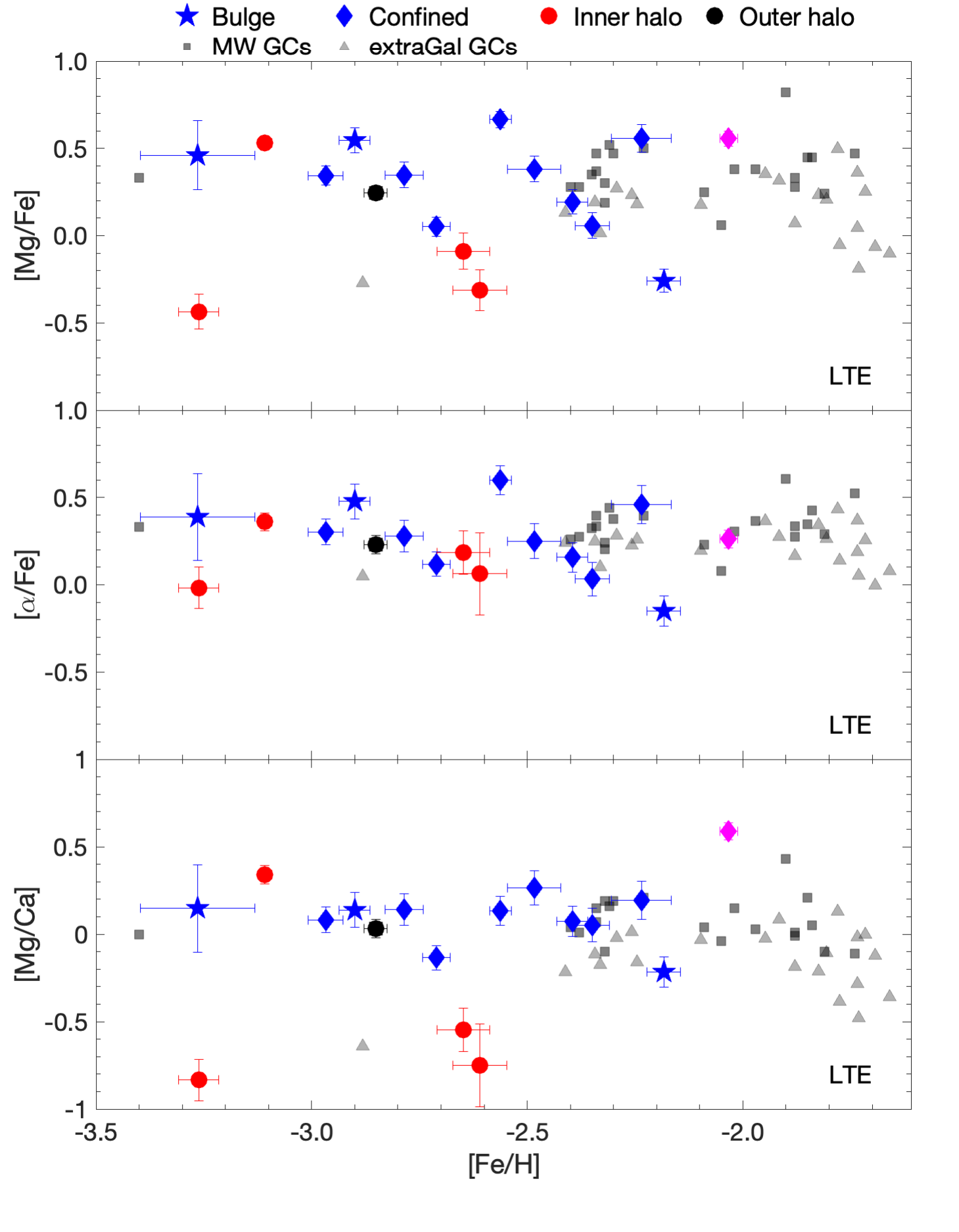}
\caption{Comparison with globular clusters. Top panel: [Mg/Fe] ratio vs. \FeH{} as in Figure~\ref{chemfig} but in LTE. Central panel: [$\alpha$/Fe] vs. \FeH{} in LTE. [$\alpha$/Fe] is measured as the mean between [Mg/Fe] and [Ca/Fe]. Bottom panel: [Mg/Ca] vs. \FeH{} in LTE. The MW GCs compilation (squares) is from \citet{Pritzl05} with the addition of C-19 \citep{Martin22}, while the extragalactic compilation (triangles) is from \citet{Larsen22}. This three panel plot has a similar intent as Figure~6 from \citet{Pancino17} and Figure~8 from \citet{Sakari15}. The scarcity of GCs at \FeH$<-2.5$ is due to the so-called metallicity floor for GC \citep[\eg][]{Beasley19}.}
\label{gc_extra}
\end{figure}

In our GRACES dataset, we find four stars with [Mg/Fe]$<0$ (see Figure~\ref{gcsecfig}).  
Figure~\ref{gc_extra} displays three panels, [Mg/Fe], [$\alpha$/Fe] and [Mg/Ca] ratios as a function of \FeH. 
The three inner halo-like stars (P171457, P182221 and P184700) have [Mg/Ca] in the range $[-0.85,-0.55]$, compatible with an extragalactic GC origin. The one bulge star (P180118) has a [Mg/Ca] $\sim-0.25$, which is low but also in the overlapping region between the Galactic and extragalactic GCs, thus its origin is less clear. 

We highlight that if the inner-halo star P171457 truly formed in an accreted extragalactic GC, then its metallicity (\FeH$_{\rm NLTE}=-3.25\pm0.05$) challenges the current estimates for the metallicity floor for GC \citep[\FeH$\sim-2.8$, \eg][and references therein]{Beasley19}. 
This would not be the first observation that calls into question the metallicity floor threshold; \citet[][]{Martin22} report the discovery of the remnant of the most metal-poor (\FeH$\sim-3.4$) GC known to date, the C$-19$ stellar stream. The chemistry of  C$-19$ indicates a GC origin (e.g., range in Na abundances with no dispersion in Fe).
However, its dynamics as measured from the radial velocity dispersion of its member stars are hotter than expected for a classical GC stream \citep{Errani22,Yuan22}, suggesting perhaps a new formation mechanism for GCs in ancient dwarf galaxies. 
If P171457 is confirmed as a star stripped and accreted from a dissolved GC, this would indicate that more extremely metal-poor GCs formed at early times but that tidal interactions with the MW may have dispersed them in the bulge and the halo.

\subsection{Three CEMP stars: P182221, P183229, and P184700}
\label{cempgc}

Carbon-enhanced metal-poor (CEMP) stars are common amongst the metal-poor stars in the Galactic halo, reaching frequencies of  $\sim30$\% at \FeH$<-2.0$ and up to $80$\% in the UMP regime \citep[\FeH$<-4.0$, \eg see][]{Placco14,Yoon18}, but see \citet{Arentsen22} for caveats. 

According to \citet{Beers05} and later confirmed by \citet{Norris13}, there are various classes of CEMP which differentiate by their Eu and Ba content. Stars are r-process-enhanced if [Eu/Fe]$>1.0$ (CEMP-r), s-process-enhanced if [Ba/Fe]$>1.0$ and [Ba/Eu]$>0.5$ (CEMP-s), mixed if $0.0<$[Ba/Eu]$<0.5$ (CEMP-r/s), and with no overabundance of n-capture elements if [Ba/Fe]$<0$ (CEMP-no). In particular, CEMP-s stars are enhanced due to the contribution of an AGB donor  \citep[\eg][]{Masseron10,Hansen16}, \ie they are or were in a binary system. While CEMP-no are important tracers of spinstars \citep[\eg][]{Meynet06,Meynet10} or faint supernovae \citep[\eg][]{Umeda03,Umeda05,Tominaga14}.

In the PIGS survey, \citet{Arentsen21} discovered 96 new CEMP stars in the inner Galaxy, including 62 with \FeH$<-2.0$.  Previously, it was thought that CEMP stars were not common in the bulge, as only 
one CEMP-s \citet{Koch16} and one CEMP-no \citep{Howes15,Howes16} star had been identified. \citet{Arentsen21} further showed that in the EMP regime (\FeH$<-3.0$) the percentage of CEMP stars is $42_{-13}^{+14}$ percent, in agreement with the Galactic halo populations.

There are two evident C-enhanced stars in this sample, P183229 and P184700, which was first noticed in the low-resolution AAT spectra from the PIGS sample \citep{Arentsen21} and the C$_2$ Swan bands can be seen in our GRACES spectra (\eg see Fig.~\ref{spectraex}).  Both stars are also enhanced in Ba (see Figure~\ref{bafig}), identifying them as CEMP-s stars \citep[\eg][]{Milone12,Lucatello15,Hansen16}.

The chemistry of P184700 is also compatible with second-generation stars in globular clusters (discussed in Sections~\ref{secgen}-\ref{extragc}). 
P184700 has a high [Na/Fe]$_{\rm LTE}=0.83\pm0.10$, and low [Mg/Fe]$_{\rm LTE}=-0.09\pm0.10$.
Some CEMP stars have shown extreme Na enhancements, up to [Na/Fe]$\sim3.0$ on the main sequence, whereas most CEMP stars on the RGB have Na ranging from [Na/Fe]$=0.0-1.5$ \citep{Aoki07,Aoki08}.  These enhancements are thought to depend on properties of the AGB donor \citep[e.g.,][]{Stancliffe09}.
However, a low [Mg/Fe] value is not expected from the CEMP nucleosynthesis and mass transfer models, but can be lower in second-generation GC stars.  
Furthermore, lower [Mg/Fe] ratios are typically found in extragalactic systems, in both field stars and GCs \citep[see Section~\ref{extragc} and][]{Venn04, Pritzl05, Sakari15, Pancino17, Hasselquist21}.

Given the rarity of binarity in very dense environments as shown in various works \citep[\eg][]{Dorazi10,Milone12,Lucatello15}, the possible association of P184700 with globular clusters would be quite a rare event. Further high-resolution spectroscopic follow-up of this star can provide a better insight into its origin.

A third star, P182221, is likely to be (or was) a CEMP-s star.  The large  uncertainty on the C (see Section~\ref{cempsec}), the slightly enhancement in Ba (see Figure~\ref{bafig}), and the stellar parameters suggest (see Figure~\ref{kielfig}) this star is a horizontal branch star that has experienced maximal carbon-depletion once up to the RGB tip \citep{Placco14}. This depletion can reach up to 0.5 dex in [C/Fe]. If so, this would indicate the CEMP-s nature of this star. In favour of this scenario, the GRACES-AAT RV discrepancy (see Figure~\ref{rvgracesferre}, up to $\sim30\kms$) would also suggest that P182221 is in a binary system. Further measurements of the RV are needed to confirm the RV variability.

\subsection{P180956: A very interesting star with planar orbit}\label{planarsec}

\begin{figure}
\includegraphics[width=0.5\textwidth]{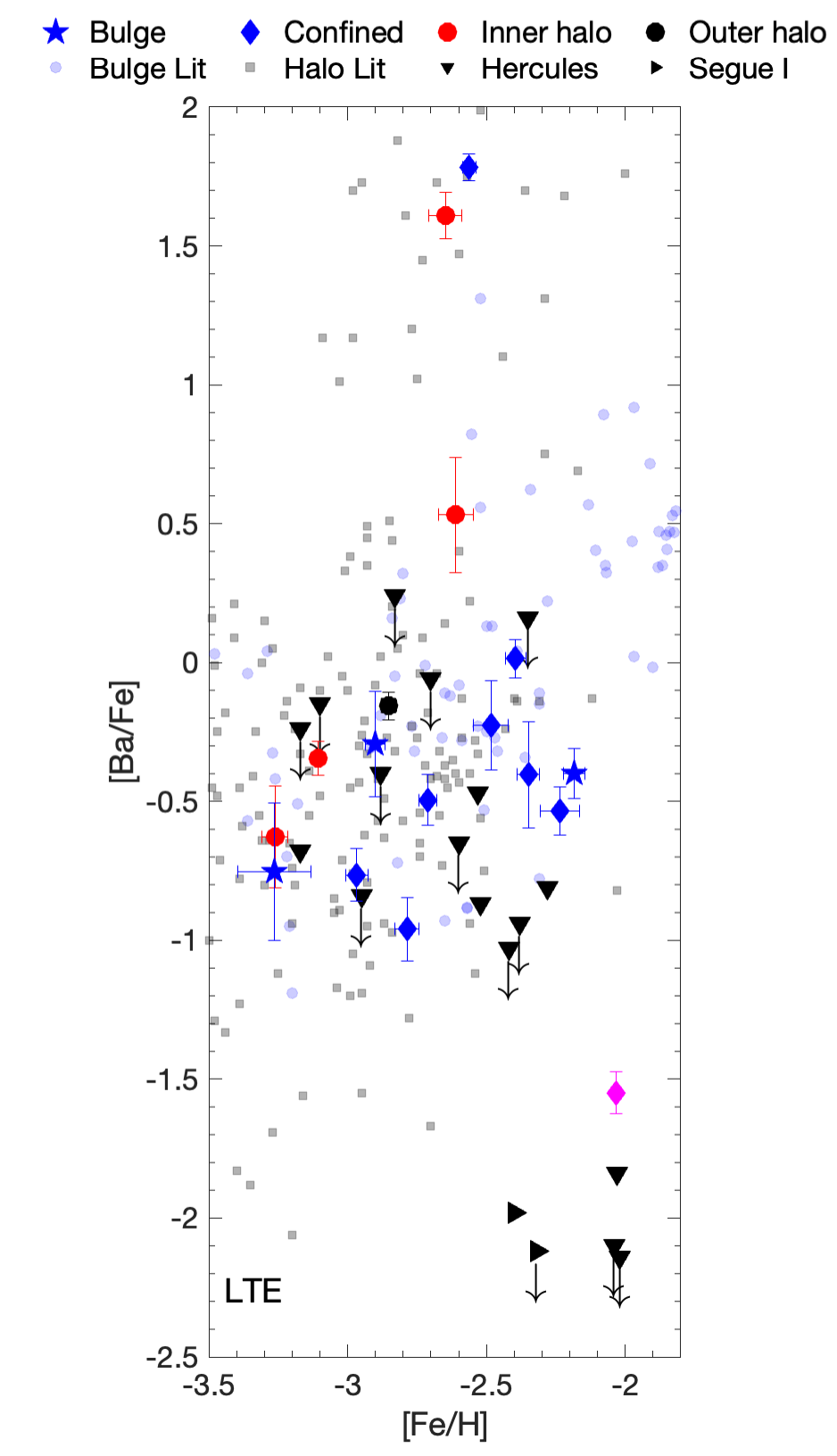}
\caption{[Ba/Fe] vs. \FeH. Ba abundances are not corrected by NLTE effect. The stars in the PIGS/GRACES sample are marked with the same symbols and colours as Figure~\ref{kinefig}. The bulge literature sample (blue circles) is composed by stars from \citet{Howes14,Howes15,Howes16,Koch16,Reggiani20,Lucey22}. Halo literature compilation (grey squares) are from \citet{Aoki13,Yong13,Kielty21}. Hercules stars are from \citet{Koch08,Koch13,Francois16}, while Segue 1 stars are from \citet{Frebel14}. Upper limits on the [Ba/Fe] ratios for Hercules and Segue 1 stars are denoted with an arrow. The low-Ba star (magenta diamond marker) is likely accreted from a dwarf galaxy and its chemistry is very similar to some stars in Hercules and Segue 1.}
\label{badwarf}
\end{figure}

The majority of the stars in the PIGS/GRACES sample that are confined to the MW plane (Z$_{\mathrm{max}}\leq3.5$\kpc) have  small apocentric distances ($<6.5$\kpc), implying that they are confined to the inner region of the MW. The only planar-like star that plunges very far and beyond the Sun is P180956. This star has an apocentre of R$_{\mathrm{apo}} = 12.9 \pm 0.2\kpc$  and a Z$_{\mathrm{max}}=1.81\pm0.04\kpc$. This star at $\FeH\sim-2.0$, also exhibits a peculiar chemistry in the $\alpha$-elements, in the odd-Z elements, and Ba.
The $\alpha$-elements are slightly enriched in Mg, yet challenged in Ca and Ti, such that [Ca/Mg]$\sim -0.6$. The odd-Z elements show that this star is extremely Na-poor  ([Na/Fe]$\sim-1.0$), but also rich in K and Sc ([K/Fe]$\sim0.9$, [Sc/Fe]$\sim0.3$). The heavy element Ba is quite low, where [Ba/Fe]$\sim-1.6$ is amongst the lowest values of all stars known in this metallicity range (see Figure~\ref{badwarf}).  
A similar pattern in the $\alpha$-elements has been seen in a few low-metallicity stars \citep[HE~1424-0241, HE~2323-0256, HE~2139-5432, and HE~1327-2326, see][]{Sitnova19}, which have inner halo orbits with eccentricities $\epsilon\geq0.7$ \citep{Sestito19}. 
A few inner halo stars with similar behaviours in the $\alpha$-elements were also found by \citet{Venn20, Kielty21}, although at slightly lower metallicities ($\FeH\sim-2.5$). 

The peculiar chemistry, \ie very low [Ca/Mg] and [Ba/Fe] ratios, can be explained by the enrichment of one or very few core collapse supernovae \citep[the ``one-shot" model,][]{Frebel12}. The expectation is that strong feedback effects shut off the star formation after a Population III star explodes. The ejected yields would then produce very little Ba, even at higher metallicities ([Fe/H]$>-2$, as found in some stars in Coma Berenices \citep{Frebel12}, Segue 1 \citep{Frebel14}, and Hercules \citep{Koch08,Koch13,Francois16}. Figure~\ref{badwarf} shows the [Ba/Fe] ratio vs. \FeH{} for the PIGS/GRACES sample in comparison with Segue 1 and Hercules stars. 
The ``one-shot" model has, however, been questioned, \eg the linear decrease in [$\alpha$/Fe] with increasing [Fe/H] in Hercules and Com Ber suggests contributions from SNe Ia \citep[][]{Koch08,Waller22}. Regardless its origin, this chemical abundance signature is unique to stars that form in  UFD galaxies, strongly suggesting that P180956 was captured from an UFD galaxy. It has been proposed by several recent publications that VMPs confined to the MW plane might be the relics of the building blocks that formed the proto-MW
\citep[\eg][]{Sestito19,Sestito20,DiMatteo20,Carter21,Cordoni21}. \citet{Sestito21} proposed that a retrograde planar population is an excellent proxy for the early Galactic assembly, while a prograde VMP planar population could trace later accretion events at very low inclination angles. 
In fact, observations of a planar population at very high eccentricities have been found by \citet{Sestito19,Sestito20}, and proposed to have been brought in by one big merger event at early times. 
However, these studies used cosmological simulations or kinematical properties only, and lacked evidence from chemical abundances.
The few planar stars that have been also observed with both high-resolution spectroscopic abundances and detailed Gaia-based kinematics \citep[][ and this work]{Kielty21,Venn20,DiMatteo20} have all pointed to peculiarities in  their chemo-dynamical properties. A thorough high-resolution investigation of a larger sample of VMP planar stars will provide better insight, \eg using the upcoming spectroscopic surveys WEAVE \citep{WEAVE12} and 4MOST \citep{deJong19}.

\subsection{Connection with the Inner Galaxy  Structure and with Aurora}\label{igs}

\citet{Horta21} used SDSS/APOGEE DR16 \citep{Ahumada20} to discover a population of metal-poor stars in the inner region of the MW ($R_{GC}<4\kpc$), which they called the Inner Galaxy Structure (IGS).  This population has different chemo-dynamical properties from the more metal-rich bulge (\FeH$>-0.5$).  
The IGS is composed of high eccentricity stars ($\epsilon>0.6$), with $\FeH >-1.7$ and  $0.3<$[Mg/Fe]$<0.4$, but an unusually negative ratio of [Al/Fe]$<0$ and  positive [Mg/Mn]$>0.3$.
They infer that the IGS host would have had stellar mass of  $\sim5\times10^8\msun$, \ie twice the mass of the Gaia-Enceladus/Sausage system \citep[\eg][]{Belokurov18,Helmi18}. 
As all the stars in this paper are more metal-poor (\FeH$_{\rm LTE}<-2.0$) than the  IGS stars discovered so far ($\FeH >-1.7$), then our sample is not sufficient to search for new members of the IGS.  
Nevertheless, one star P180956 (discussed in Section~\ref{planarsec}) has $\FeH_{\rm NLTE} =-1.8\pm0.1$.  An evaluation of its chemical abundances shows that its [Mg/Fe] may be slightly too large for membership in the IGS, but its low Na is consistent with the expectation of low Al.  The high eccentricity of P180956 would be compatible with the IGS, given the wide range selected by \citet{Horta21}, $e>0.6$. Many accreted structures have been extended towards the VMP regime \citep{Yuan20} applying self-organizing maps algorithms in the 4D space of orbital energy and angular momentum. However, higher signal-to-noise spectral observations and over a wider wavelength region than available to GRACES are necessary to measure the chemical abundance of Al, Mn, and  N, and to test its connection to the IGS.

Very recently, \citet{Belokurov22} used SDSS/APOGEE DR 17 \citep{ApogeeDR17}  and Gaia data to investigate the metal-poor ([Fe/H]$>-1.5$) in-situ population of the Galaxy. They propose that a metal-poor halo could have formed early in the proto-Galaxy. This chaotic early phase of the Milky Way, dubbed Aurora, would produce stars with a large scatter in their chemistry due to the lumpy ISM. This would affect the elements usually used to discern the multi-populations in globular cluster stars (\eg Na, Al, N). According to \citet{Belokurov22}, high-eccentricity structures in the inner Galaxy, such as the IGS, would be part of Aurora. Our sample is more metal-poor than the Aurora’s stars studied so far. Therefore, a firm conclusion on their association with Aurora is beyond the scope of this work.

\section{Conclusions}
We present a chemo-dynamical investigation of 17 stars selected from the low/medium resolution spectroscopic campaign of the Pristine Inner Galaxy survey \citep{Arentsen20b}.  Spectral observations with the Gemini North/GRACES and  {\it Gaia} EDR3 astrometric solutions were used to infer precise chemical abundances, stellar parameters, distances, and orbits. 
Our sample is divided into four dynamical sub-groups; (i) stars confined to the bulge, (ii) stars confined into the MW plane, and two groups that dive further out to the (iii) inner and (iv) outer halo.
The red spectral coverage of GRACES allowed us to determine the chemical abundances for several species, including Fe-peak (Fe, Cr, Ni), $\alpha-$ (Mg, Ca, Ti), odd-Z (Na, K, Sc), and neutron-capture (Ba) elements.  By combining the chemistry and kinematics, we have investigated the properties of this sample of stars to find the following:

\begin{enumerate}
    \item The majority of the stars in this sample have a chemical signature indistinguishable from  that of Milky Way halo stars.  If the bulge and the halo formed at early times from numerous building blocks, that are depositing their stars, their pristine gas, and dark matter into the forming Galaxy, then
    the chemistry of the bulge stars should resemble those in the halo (with the exception that the outer halo where more recent accretion will preferentially deposit their stars). This is additional confirmation of the hierarchical assembly of the Galaxy. 
    \item We do not detect the signatures of PISNe yields \citep[as in][]{Takahashi18}, nor PISNe + CCSNe yields \citep[as in][]{Salvadori19}.
    \item Some of the stars in our sample are chemically compatible with second-generation stars in GCs. This is reinforced through examination of the chemistry of stars in extragalactic GCs, mainly  their negative [Mg/Ca]. 
    \item One possible second-generation GC star, P171457, is extremely metal-poor ($\FeH\sim-3.2$).  If confirmed, this would indicate another dissolved extremely metal-poor GC in the MW, similar to the recently discovered halo stellar stream \citep[C-19,][]{Martin22}.
    \item We confirm the nature of two C-enhanced stars. We find high Ba, indicating they are CEMP-s stars.   One of them, P184700, has the chemical abundances of a second-generation GC star. 
    \item P182221 is likely a CEMP-s star. [C/Fe] measurement from AAT is highly uncertain, while its stellar parameters and and the slightly enhancement in Ba suggest this star past the RGB phase. This would indicate the depletion of C in its atmosphere. Moreover, the high RV discrepancy between GRACES and AAT would also be in favour of the binarity of this star.
    \item P180956 is a very metal-poor ($\FeH\sim-2.0$) star  confined to the MW plane with an apocentric distance of $\sim12\kpc$ and a pericentre $<1\kpc$.  
    Its chemistry includes low [Ba/Fe], [Na/Fe], and [Ca/Mg], which suggest it originated in an UFD galaxy that was polluted by only 1 or a few core collapse supernovae ($\sim20\msun$).  
    Similar stars discovered in the Pristine survey \citep{Sestito20} and followed up with Gemini-North/GRACES and CFHT/ESPaDoNS spectroscopy have been found on planar orbits \citep{Venn20, Kielty21}.
    These may point to one or more very early accretion events, compatible with the building block merger phase \citep[][]{Sestito21}. 
    These chemo-dynamical studies confirm the importance of very metal-poor stars with planar orbits as tracers of the early MW assembly.
\end{enumerate}

This chemo-dynamical investigation of the very metal-poor tail of the inner Galaxy opens a window on the early assembly of the Milky Way. It unveils the variety of the building blocks, from systems chemically similar to globular clusters to ultra faint dwarfs galaxies. Further spectroscopic observations towards bluer regions of the spectra are needed to better characterise the properties of the relics of these ancient and dissolved systems. This is a task easily achievable by the forthcoming Gemini High-Resolution Optical SpecTrograph \citep[GHOST,][]{Pazder20}.

\section*{Acknowledgements}
We acknowledge and respect the l\textschwa\textvbaraccent {k}$^{\rm w}$\textschwa\ng{}\textschwa n peoples on whose traditional territory the University of Victoria stands and the Songhees, Esquimalt and $\ubar{\rm W}$S\'ANE\'C  peoples whose historical relationships with the land continue to this day.

The authors wish to recognize and acknowledge the very significant cultural role and reverence that the summit of Maunakea has always had within the Native Hawaiian community. We are very fortunate to have had the opportunity to conduct observations from this mountain.

We acknowledge the traditional owners of the land on which the Anglo Australian Telescope stands, the Gamilaraay people, and pay our respects to elders past and present.

FS thanks  Tim Beers, Ani Chiti, Anna Frebel, and Ian Roederer for the interesting discussions and feedback about this work at the 2022 JINA-CEE meeting in Notre Dame. The authors thanks Vasily Belokurov for his feedback on Aurora. We thank the anonymous referee for their comments that helped to improve this manuscript.

FS thanks the Dr. Margaret "Marmie" Perkins Hess postdoctoral fellowship for funding his work at the University of Victoria.
KAV thanks the National Sciences and Engineering Research Council of Canada for funding through the Discovery Grants and CREATE programs. AA, NFM, and ZY gratefully acknowledge support from the French National Research Agency (ANR) funded project ``Pristine'' (ANR-18-CE31-0017) and from the European Research Council (ERC) under the European Unions Horizon 2020 research and innovation programme (grant agreement No. 834148). DA acknowledges support from the ERC Starting Grant NEFERTITI H2020/808240. JIGH acknowledges financial support from the Spanish Ministry of Science and Innovation (MICINN) project PID2020-117493GB-I00.
ES acknowledges funding through VIDI grant "Pushing Galactic Archaeology to its limits" (with project number VI.Vidi.193.093) which is funded by the Dutch Research Council (NWO). The authors thanks the International Space Science Institute (ISSI) in Bern, Switzerland, for funding the Team “The Early Milky Way” led by Else Starkenburg.

This work is based on observations obtained with GRACES, as part of the Gemini Large and Long Program, GN-X-LP-102 (where X includes semesters 2019A–2021A). Based on observations obtained at the international Gemini Observatory, a program of NSF’s NOIRLab, which is managed by the Association of Universities for Research in Astronomy (AURA) under a cooperative agreement with the National Science Foundation. On behalf of the Gemini Observatory partnership: the National Science Foundation (United States), National Research Council (Canada), Agencia Nacional de Investigaci\'{o}n y Desarrollo (Chile), Ministerio de Ciencia, Tecnolog\'{i}a e Innovaci\'{o}n (Argentina), Minist\'{e}rio da Ci\^{e}ncia, Tecnologia, Inova\c{c}\~{o}es e Comunica\c{c}\~{o}es (Brazil), and Korea Astronomy and Space Science Institute (Republic of Korea).

Based on observations obtained through the Gemini Remote Access to CFHT ESPaDOnS Spectrograph (GRACES). ESPaDOnS is located at the Canada-France-Hawaii Telescope (CFHT), which is operated by the National Research Council of Canada, the Institut National des Sciences de l’Univers of the Centre National de la Recherche Scientifique of France, and the University of Hawai’i. ESPaDOnS is a collaborative project funded by France (CNRS, MENESR, OMP, LATT), Canada (NSERC), CFHT and ESA. ESPaDOnS was remotely controlled from the international Gemini Observatory, a program of NSF’s NOIRLab, which is managed by the Association of Universities for Research in Astronomy (AURA) under a cooperative agreement with the National Science Foundation on behalf of the Gemini partnership: the National Science Foundation (United States), the National Research Council (Canada), Agencia Nacional de Investigaci\'{o}n y Desarrollo (Chile), Ministerio de Ciencia, Tecnolog\'{i}a e Innovaci\'{o}n (Argentina), Minist\'{e}rio da Ci\^{e}ncia, Tecnologia, Inova\c{c}\~{o}es e Comunica\c{c}\~{o}es (Brazil), and Korea Astronomy and Space Science Institute (Republic of Korea).

Based on observations obtained with MegaPrime/MegaCam, a joint project of CFHT and CEA/DAPNIA, at the Canada-France-Hawaii Telescope (CFHT) which is operated by the National Research Council (NRC) of Canada, the Institut National des Sciences de l'Univers of the Centre National de la Recherche Scientifique of France, and the University of Hawaii.

We thank the Australian Astronomical Observatory, which have made these observations possible.

This work has made use of data from the European Space Agency (ESA) mission
{\it Gaia} (\url{https://www.cosmos.esa.int/gaia}), processed by the {\it Gaia}
Data Processing and Analysis Consortium (DPAC,
\url{https://www.cosmos.esa.int/web/gaia/dpac/consortium}). Funding for the DPAC
has been provided by national institutions, in particular the institutions
participating in the {\it Gaia} Multilateral Agreement.

This research has made use of the SIMBAD database, operated at CDS, Strasbourg, France \citep{Wenger00}. This work made extensive use of \textsc{TOPCAT} \citep{Taylor05}, the \textsc{DUSTMAPS} \textsc{Python} package \citep{Green19}, and \textsc{MATLAB} \citep{MATLAB21}. 

This research was enabled in part by support provided by WestGrid (\url{https://www.westgrid.ca}) and Compute Canada (\url{www.computecanada.ca}).

\section*{Data Availability}
GRACES spectra are available at the Gemini Archive web page \url{https://archive.gemini.edu/searchform}. The data underlying this article are available in the article and in its online supplementary material.

\bibliographystyle{mnras}
\bibliography{graces_bulge.bib}

\begin{thebibliography}{}
\makeatletter
\relax
\def\mn@urlcharsother{\let\do\@makeother \do\$\do\&\do\#\do\^\do\_\do\%\do\~}
\def\mn@doi{\begingroup\mn@urlcharsother \@ifnextchar [ {\mn@doi@}
  {\mn@doi@[]}}
\def\mn@doi@[#1]#2{\def\@tempa{#1}\ifx\@tempa\@empty \href
  {http://dx.doi.org/#2} {doi:#2}\else \href {http://dx.doi.org/#2} {#1}\fi
  \endgroup}
\def\mn@eprint#1#2{\mn@eprint@#1:#2::\@nil}
\def\mn@eprint@arXiv#1{\href {http://arxiv.org/abs/#1} {{\tt arXiv:#1}}}
\def\mn@eprint@dblp#1{\href {http://dblp.uni-trier.de/rec/bibtex/#1.xml}
  {dblp:#1}}
\def\mn@eprint@#1:#2:#3:#4\@nil{\def\@tempa {#1}\def\@tempb {#2}\def\@tempc
  {#3}\ifx \@tempc \@empty \let \@tempc \@tempb \let \@tempb \@tempa \fi \ifx
  \@tempb \@empty \def\@tempb {arXiv}\fi \@ifundefined
  {mn@eprint@\@tempb}{\@tempb:\@tempc}{\expandafter \expandafter \csname
  mn@eprint@\@tempb\endcsname \expandafter{\@tempc}}}

\bibitem[\protect\citeauthoryear{{Abadi}, {Navarro}, {Steinmetz}  \&
  {Eke}}{{Abadi} et~al.}{2003}]{Abadi03}
{Abadi} M.~G.,  {Navarro} J.~F.,  {Steinmetz} M.,   {Eke} V.~R.,  2003, \mn@doi
  [\apj] {10.1086/378316}, \href
  {http://adsabs.harvard.edu/abs/2003ApJ...597...21A} {597, 21}

\bibitem[\protect\citeauthoryear{{Abdurro'uf} et~al.,}{{Abdurro'uf}
  et~al.}{2022}]{ApogeeDR17}
{Abdurro'uf} et~al., 2022, \mn@doi [\apjs] {10.3847/1538-4365/ac4414}, \href
  {https://ui.adsabs.harvard.edu/abs/2022ApJS..259...35A} {259, 35}

\bibitem[\protect\citeauthoryear{{Aguado} et~al.,}{{Aguado}
  et~al.}{2019}]{Aguado19}
{Aguado} D.~S.,  et~al., 2019, \mn@doi [\mnras] {10.1093/mnras/stz2643}, \href
  {https://ui.adsabs.harvard.edu/abs/2019MNRAS.490.2241A} {490, 2241}

\bibitem[\protect\citeauthoryear{{Ahumada} et~al.,}{{Ahumada}
  et~al.}{2020}]{Ahumada20}
{Ahumada} R.,  et~al., 2020, \mn@doi [\apjs] {10.3847/1538-4365/ab929e}, \href
  {https://ui.adsabs.harvard.edu/abs/2020ApJS..249....3A} {249, 3}

\bibitem[\protect\citeauthoryear{{Allende Prieto}, {Beers}, {Wilhelm},
  {Newberg}, {Rockosi}, {Yanny}  \& {Lee}}{{Allende Prieto}
  et~al.}{2006}]{Allende06}
{Allende Prieto} C.,  {Beers} T.~C.,  {Wilhelm} R.,  {Newberg} H.~J.,
  {Rockosi} C.~M.,  {Yanny} B.,   {Lee} Y.~S.,  2006, \mn@doi [\apj]
  {10.1086/498131}, \href
  {https://ui.adsabs.harvard.edu/abs/2006ApJ...636..804A} {636, 804}

\bibitem[\protect\citeauthoryear{{Amarsi}, {Lind}, {Asplund}, {Barklem}  \&
  {Collet}}{{Amarsi} et~al.}{2016}]{Amarsi16}
{Amarsi} A.~M.,  {Lind} K.,  {Asplund} M.,  {Barklem} P.~S.,   {Collet} R.,
  2016, \mn@doi [\mnras] {10.1093/mnras/stw2077}, \href
  {https://ui.adsabs.harvard.edu/abs/2016MNRAS.463.1518A} {463, 1518}

\bibitem[\protect\citeauthoryear{{Anders} et~al.,}{{Anders}
  et~al.}{2019}]{Anders19}
{Anders} F.,  et~al., 2019, \mn@doi [\aap] {10.1051/0004-6361/201935765}, \href
  {https://ui.adsabs.harvard.edu/abs/2019A&A...628A..94A} {628, A94}

\bibitem[\protect\citeauthoryear{{Andrae} et~al.,}{{Andrae}
  et~al.}{2018}]{Andrae18}
{Andrae} R.,  et~al., 2018, \mn@doi [\aap] {10.1051/0004-6361/201732516}, \href
  {https://ui.adsabs.harvard.edu/abs/2018A&A...616A...8A} {616, A8}

\bibitem[\protect\citeauthoryear{{Andrae} et~al.,}{{Andrae}
  et~al.}{2022}]{Andrae22}
{Andrae} R.,  et~al., 2022, arXiv e-prints, \href
  {https://ui.adsabs.harvard.edu/abs/2022arXiv220606138A} {p. arXiv:2206.06138}

\bibitem[\protect\citeauthoryear{{Aoki}, {Beers}, {Christlieb}, {Norris},
  {Ryan}  \& {Tsangarides}}{{Aoki} et~al.}{2007}]{Aoki07}
{Aoki} W.,  {Beers} T.~C.,  {Christlieb} N.,  {Norris} J.~E.,  {Ryan} S.~G.,
  {Tsangarides} S.,  2007, \mn@doi [\apj] {10.1086/509817}, \href
  {https://ui.adsabs.harvard.edu/abs/2007ApJ...655..492A} {655, 492}

\bibitem[\protect\citeauthoryear{{Aoki} et~al.,}{{Aoki} et~al.}{2008}]{Aoki08}
{Aoki} W.,  et~al., 2008, \mn@doi [\apj] {10.1086/533517}, \href
  {https://ui.adsabs.harvard.edu/abs/2008ApJ...678.1351A} {678, 1351}

\bibitem[\protect\citeauthoryear{{Aoki} et~al.,}{{Aoki} et~al.}{2013}]{Aoki13}
{Aoki} W.,  et~al., 2013, \mn@doi [\aj] {10.1088/0004-6256/145/1/13}, \href
  {https://ui.adsabs.harvard.edu/abs/2013AJ....145...13A} {145, 13}

\bibitem[\protect\citeauthoryear{{Aoki}, {Tominaga}, {Beers}, {Honda}  \&
  {Lee}}{{Aoki} et~al.}{2014}]{Aoki14}
{Aoki} W.,  {Tominaga} N.,  {Beers} T.~C.,  {Honda} S.,   {Lee} Y.~S.,  2014,
  \mn@doi [Science] {10.1126/science.1252633}, \href
  {https://ui.adsabs.harvard.edu/abs/2014Sci...345..912A} {345, 912}

\bibitem[\protect\citeauthoryear{{Arentsen} et~al.,}{{Arentsen}
  et~al.}{2020a}]{Arentsen20a}
{Arentsen} A.,  et~al., 2020a, \mn@doi [\mnras] {10.1093/mnrasl/slz156}, \href
  {https://ui.adsabs.harvard.edu/abs/2020MNRAS.491L..11A} {491, L11}

\bibitem[\protect\citeauthoryear{{Arentsen} et~al.,}{{Arentsen}
  et~al.}{2020b}]{Arentsen20b}
{Arentsen} A.,  et~al., 2020b, \mn@doi [\mnras] {10.1093/mnras/staa1661}, \href
  {https://ui.adsabs.harvard.edu/abs/2020MNRAS.496.4964A} {496, 4964}

\bibitem[\protect\citeauthoryear{{Arentsen} et~al.,}{{Arentsen}
  et~al.}{2021}]{Arentsen21}
{Arentsen} A.,  et~al., 2021, \mn@doi [\mnras] {10.1093/mnras/stab1343}, \href
  {https://ui.adsabs.harvard.edu/abs/2021MNRAS.505.1239A} {505, 1239}

\bibitem[\protect\citeauthoryear{{Arentsen}, {Placco}, {Lee}, {Aguado},
  {Martin}, {Starkenburg}  \& {Yoon}}{{Arentsen} et~al.}{2022}]{Arentsen22}
{Arentsen} A.,  {Placco} V.~M.,  {Lee} Y.~S.,  {Aguado} D.~S.,  {Martin} N.~F.,
   {Starkenburg} E.,   {Yoon} J.,  2022, arXiv e-prints, \href
  {https://ui.adsabs.harvard.edu/abs/2022arXiv220604081A} {p. arXiv:2206.04081}

\bibitem[\protect\citeauthoryear{{Asplund}, {Grevesse}, {Sauval}  \&
  {Scott}}{{Asplund} et~al.}{2009}]{Asplund09}
{Asplund} M.,  {Grevesse} N.,  {Sauval} A.~J.,   {Scott} P.,  2009, \mn@doi
  [\araa] {10.1146/annurev.astro.46.060407.145222}, \href
  {https://ui.adsabs.harvard.edu/abs/2009ARA&A..47..481A} {47, 481}

\bibitem[\protect\citeauthoryear{Bailer-Jones}{Bailer-Jones}{2015}]{Bailer15}
Bailer-Jones C. A.~L.,  2015, Publications of the Astronomical Society of the
  Pacific, 127, 994

\bibitem[\protect\citeauthoryear{{Bailer-Jones}, {Rybizki}, {Fouesneau},
  {Mantelet}  \& {Andrae}}{{Bailer-Jones} et~al.}{2018}]{Bailer18}
{Bailer-Jones} C.~A.~L.,  {Rybizki} J.,  {Fouesneau} M.,  {Mantelet} G.,
  {Andrae} R.,  2018, \mn@doi [\aj] {10.3847/1538-3881/aacb21}, \href
  {http://adsabs.harvard.edu/abs/2018AJ....156...58B} {156, 58}

\bibitem[\protect\citeauthoryear{{Bastian} \& {Lardo}}{{Bastian} \&
  {Lardo}}{2018}]{Bastian18}
{Bastian} N.,  {Lardo} C.,  2018, \mn@doi [\araa]
  {10.1146/annurev-astro-081817-051839}, \href
  {https://ui.adsabs.harvard.edu/abs/2018ARA&A..56...83B} {56, 83}

\bibitem[\protect\citeauthoryear{{Beasley}, {Leaman}, {Gallart}, {Larsen},
  {Battaglia}, {Monelli}  \& {Pedreros}}{{Beasley} et~al.}{2019}]{Beasley19}
{Beasley} M.~A.,  {Leaman} R.,  {Gallart} C.,  {Larsen} S.~S.,  {Battaglia} G.,
   {Monelli} M.,   {Pedreros} M.~H.,  2019, \mn@doi [\mnras]
  {10.1093/mnras/stz1349}, \href
  {https://ui.adsabs.harvard.edu/abs/2019MNRAS.487.1986B} {487, 1986}

\bibitem[\protect\citeauthoryear{{Beers} \& {Christlieb}}{{Beers} \&
  {Christlieb}}{2005}]{Beers05}
{Beers} T.~C.,  {Christlieb} N.,  2005, \mn@doi [\araa]
  {10.1146/annurev.astro.42.053102.134057}, \href
  {http://adsabs.harvard.edu/abs/2005ARA%26A..43..531B} {43, 531}

\bibitem[\protect\citeauthoryear{{Belokurov} \& {Kravtsov}}{{Belokurov} \&
  {Kravtsov}}{2022}]{Belokurov22}
{Belokurov} V.,  {Kravtsov} A.,  2022, \mn@doi [\mnras]
  {10.1093/mnras/stac1267}, \href
  {https://ui.adsabs.harvard.edu/abs/2022MNRAS.514..689B} {514, 689}

\bibitem[\protect\citeauthoryear{{Belokurov}, {Erkal}, {Evans}, {Koposov}  \&
  {Deason}}{{Belokurov} et~al.}{2018}]{Belokurov18}
{Belokurov} V.,  {Erkal} D.,  {Evans} N.~W.,  {Koposov} S.~E.,   {Deason}
  A.~J.,  2018, \mn@doi [\mnras] {10.1093/mnras/sty982}, \href
  {http://adsabs.harvard.edu/abs/2018MNRAS.478..611B} {478, 611}

\bibitem[\protect\citeauthoryear{{Bensby} et~al.,}{{Bensby}
  et~al.}{2013}]{Bensby13}
{Bensby} T.,  et~al., 2013, \mn@doi [\aap] {10.1051/0004-6361/201220678}, \href
  {https://ui.adsabs.harvard.edu/abs/2013A&A...549A.147B} {549, A147}

\bibitem[\protect\citeauthoryear{{Bensby} et~al.,}{{Bensby}
  et~al.}{2017}]{Bensby17}
{Bensby} T.,  et~al., 2017, \mn@doi [\aap] {10.1051/0004-6361/201730560}, \href
  {https://ui.adsabs.harvard.edu/abs/2017A&A...605A..89B} {605, A89}

\bibitem[\protect\citeauthoryear{{Bergemann}}{{Bergemann}}{2011}]{Bergemann2011}
{Bergemann} M.,  2011, \mn@doi [\mnras] {10.1111/j.1365-2966.2011.18295.x},
  \href {https://ui.adsabs.harvard.edu/abs/2011MNRAS.413.2184B} {413, 2184}

\bibitem[\protect\citeauthoryear{{Bergemann} \& {Cescutti}}{{Bergemann} \&
  {Cescutti}}{2010}]{Bergemann2010b}
{Bergemann} M.,  {Cescutti} G.,  2010, \mn@doi [\aap]
  {10.1051/0004-6361/201014250}, \href
  {https://ui.adsabs.harvard.edu/abs/2010A&A...522A...9B} {522, A9}

\bibitem[\protect\citeauthoryear{{Bergemann}, {Lind}, {Collet}, {Magic}  \&
  {Asplund}}{{Bergemann} et~al.}{2012}]{Bergemann2012}
{Bergemann} M.,  {Lind} K.,  {Collet} R.,  {Magic} Z.,   {Asplund} M.,  2012,
  \mn@doi [\mnras] {10.1111/j.1365-2966.2012.21687.x}, \href
  {https://ui.adsabs.harvard.edu/abs/2012MNRAS.427...27B} {427, 27}

\bibitem[\protect\citeauthoryear{{Bergemann}, {Collet}, {Amarsi}, {Kovalev},
  {Ruchti}  \& {Magic}}{{Bergemann} et~al.}{2017}]{Bergemann2017}
{Bergemann} M.,  {Collet} R.,  {Amarsi} A.~M.,  {Kovalev} M.,  {Ruchti} G.,
  {Magic} Z.,  2017, \mn@doi [\apj] {10.3847/1538-4357/aa88cb}, \href
  {https://ui.adsabs.harvard.edu/abs/2017ApJ...847...15B} {847, 15}

\bibitem[\protect\citeauthoryear{{Bessell}, {Bloxham}, {Schmidt}, {Keller},
  {Tisserand}  \& {Francis}}{{Bessell} et~al.}{2011}]{Bessell11}
{Bessell} M.,  {Bloxham} G.,  {Schmidt} B.,  {Keller} S.,  {Tisserand} P.,
  {Francis} P.,  2011, \mn@doi [\pasp] {10.1086/660849}, \href
  {https://ui.adsabs.harvard.edu/abs/2011PASP..123..789B} {123, 789}

\bibitem[\protect\citeauthoryear{{Bi{\'e}mont} et~al.,}{{Bi{\'e}mont}
  et~al.}{2011}]{Biemont11}
{Bi{\'e}mont} {\'E}.,  et~al., 2011, \mn@doi [\mnras]
  {10.1111/j.1365-2966.2011.18637.x}, \href
  {https://ui.adsabs.harvard.edu/abs/2011MNRAS.414.3350B} {414, 3350}

\bibitem[\protect\citeauthoryear{{Bland-Hawthorn} \&
  {Gerhard}}{{Bland-Hawthorn} \& {Gerhard}}{2016}]{BlandHawthorn16}
{Bland-Hawthorn} J.,  {Gerhard} O.,  2016, \mn@doi [\araa]
  {10.1146/annurev-astro-081915-023441}, \href
  {http://adsabs.harvard.edu/abs/2016ARA%26A..54..529B} {54, 529}

\bibitem[\protect\citeauthoryear{{Bournaud}}{{Bournaud}}{2016}]{Bournaud16}
{Bournaud} F.,  2016, in {Laurikainen} E.,  {Peletier} R.,   {Gadotti} D.,
  eds,  Astrophysics and Space Science Library Vol. 418, Galactic Bulges.
  p.~355 (\mn@eprint {arXiv} {1503.07660}),
  \mn@doi{10.1007/978-3-319-19378-6\_13}

\bibitem[\protect\citeauthoryear{{Bovy}}{{Bovy}}{2015}]{Bovy15}
{Bovy} J.,  2015, \mn@doi [\apjs] {10.1088/0067-0049/216/2/29}, \href
  {http://adsabs.harvard.edu/abs/2015ApJS..216...29B} {216, 29}

\bibitem[\protect\citeauthoryear{{Bromm}, {Coppi}  \& {Larson}}{{Bromm}
  et~al.}{2002}]{Bromm02}
{Bromm} V.,  {Coppi} P.~S.,   {Larson} R.~B.,  2002, \mn@doi [\apj]
  {10.1086/323947}, \href
  {https://ui.adsabs.harvard.edu/abs/2002ApJ...564...23B} {564, 23}

\bibitem[\protect\citeauthoryear{{Buder} et~al.,}{{Buder}
  et~al.}{2021}]{Buder21}
{Buder} S.,  et~al., 2021, \mn@doi [\mnras] {10.1093/mnras/stab1242}, \href
  {https://ui.adsabs.harvard.edu/abs/2021MNRAS.506..150B} {506, 150}

\bibitem[\protect\citeauthoryear{{Bullock} \& {Johnston}}{{Bullock} \&
  {Johnston}}{2005}]{Bullock2005}
{Bullock} J.~S.,  {Johnston} K.~V.,  2005, \mn@doi [\apj] {10.1086/497422},
  \href {https://ui.adsabs.harvard.edu/abs/2005ApJ...635..931B} {635, 931}

\bibitem[\protect\citeauthoryear{{Carretta}, {D'Orazi}, {Gratton}  \&
  {Lucatello}}{{Carretta} et~al.}{2012}]{Carretta12}
{Carretta} E.,  {D'Orazi} V.,  {Gratton} R.~G.,   {Lucatello} S.,  2012,
  \mn@doi [\aap] {10.1051/0004-6361/201219277}, \href
  {https://ui.adsabs.harvard.edu/abs/2012A&A...543A.117C} {543, A117}

\bibitem[\protect\citeauthoryear{{Carter} et~al.,}{{Carter}
  et~al.}{2021}]{Carter21}
{Carter} C.,  et~al., 2021, \mn@doi [\apj] {10.3847/1538-4357/abcda4}, \href
  {https://ui.adsabs.harvard.edu/abs/2021ApJ...908..208C} {908, 208}

\bibitem[\protect\citeauthoryear{{Casagrande} et~al.,}{{Casagrande}
  et~al.}{2021}]{Casagrande21}
{Casagrande} L.,  et~al., 2021, \mn@doi [\mnras] {10.1093/mnras/stab2304},
  \href {https://ui.adsabs.harvard.edu/abs/2021MNRAS.507.2684C} {507, 2684}

\bibitem[\protect\citeauthoryear{{Chene} et~al.,}{{Chene}
  et~al.}{2014}]{Chene14}
{Chene} A.-N.,  et~al., 2014, in {Navarro} R.,  {Cunningham} C.~R.,   {Barto}
  A.~A.,  eds,  Society of Photo-Optical Instrumentation Engineers (SPIE)
  Conference Series Vol. 9151, Advances in Optical and Mechanical Technologies
  for Telescopes and Instrumentation. p. 915147 (\mn@eprint {arXiv}
  {1409.7448}), \mn@doi{10.1117/12.2057417}

\bibitem[\protect\citeauthoryear{{Choi}, {Dotter}, {Conroy}, {Cantiello},
  {Paxton}  \& {Johnson}}{{Choi} et~al.}{2016}]{Choi16}
{Choi} J.,  {Dotter} A.,  {Conroy} C.,  {Cantiello} M.,  {Paxton} B.,
  {Johnson} B.~D.,  2016, \mn@doi [\apj] {10.3847/0004-637X/823/2/102}, \href
  {http://adsabs.harvard.edu/abs/2016ApJ...823..102C} {823, 102}

\bibitem[\protect\citeauthoryear{{Cohen} \& {Kirby}}{{Cohen} \&
  {Kirby}}{2012}]{Cohen12}
{Cohen} J.~G.,  {Kirby} E.~N.,  2012, \mn@doi [\apj]
  {10.1088/0004-637X/760/1/86}, \href
  {https://ui.adsabs.harvard.edu/abs/2012ApJ...760...86C} {760, 86}

\bibitem[\protect\citeauthoryear{{Cordoni} et~al.,}{{Cordoni}
  et~al.}{2021}]{Cordoni21}
{Cordoni} G.,  et~al., 2021, \mn@doi [\mnras] {10.1093/mnras/staa3417}, \href
  {https://ui.adsabs.harvard.edu/abs/2021MNRAS.503.2539C} {503, 2539}

\bibitem[\protect\citeauthoryear{{D'Orazi}, {Gratton}, {Lucatello}, {Carretta},
  {Bragaglia}  \& {Marino}}{{D'Orazi} et~al.}{2010}]{Dorazi10}
{D'Orazi} V.,  {Gratton} R.,  {Lucatello} S.,  {Carretta} E.,  {Bragaglia} A.,
   {Marino} A.~F.,  2010, \mn@doi [\apjl] {10.1088/2041-8205/719/2/L213}, \href
  {https://ui.adsabs.harvard.edu/abs/2010ApJ...719L.213D} {719, L213}

\bibitem[\protect\citeauthoryear{{Dalton} et~al.,}{{Dalton}
  et~al.}{2012}]{WEAVE12}
{Dalton} G.,  et~al., 2012, {WEAVE: the next generation wide-field spectroscopy
  facility for the William Herschel Telescope}.
p. 84460P, \mn@doi{10.1117/12.925950}

\bibitem[\protect\citeauthoryear{{Dehnen}}{{Dehnen}}{2000}]{Dehnen00}
{Dehnen} W.,  2000, \mn@doi [\aj] {10.1086/301226}, \href
  {https://ui.adsabs.harvard.edu/abs/2000AJ....119..800D} {119, 800}

\bibitem[\protect\citeauthoryear{{Di Matteo}, {Spite}, {Haywood}, {Bonifacio},
  {G{\'o}mez}, {Spite}  \& {Caffau}}{{Di Matteo} et~al.}{2020}]{DiMatteo20}
{Di Matteo} P.,  {Spite} M.,  {Haywood} M.,  {Bonifacio} P.,  {G{\'o}mez} A.,
  {Spite} F.,   {Caffau} E.,  2020, \mn@doi [\aap]
  {10.1051/0004-6361/201937016}, \href
  {https://ui.adsabs.harvard.edu/abs/2020A&A...636A.115D} {636, A115}

\bibitem[\protect\citeauthoryear{{Donati}, {Catala}, {Landstreet}  \&
  {Petit}}{{Donati} et~al.}{2006}]{Donati06}
{Donati} J.~F.,  {Catala} C.,  {Landstreet} J.~D.,   {Petit} P.,  2006, in
  {Casini} R.,  {Lites} B.~W.,  eds,  Astronomical Society of the Pacific
  Conference Series Vol. 358, Solar Polarization 4. p.~362

\bibitem[\protect\citeauthoryear{{Dotter}}{{Dotter}}{2016}]{Dotter16}
{Dotter} A.,  2016, \mn@doi [\apjs] {10.3847/0067-0049/222/1/8}, \href
  {http://adsabs.harvard.edu/abs/2016ApJS..222....8D} {222, 8}

\bibitem[\protect\citeauthoryear{{El-Badry} et~al.,}{{El-Badry}
  et~al.}{2018}]{ElBadry18}
{El-Badry} K.,  et~al., 2018, \mn@doi [\mnras] {10.1093/mnras/sty1864}, \href
  {http://adsabs.harvard.edu/abs/2018MNRAS.480..652E} {480, 652}

\bibitem[\protect\citeauthoryear{{Errani} et~al.,}{{Errani}
  et~al.}{2022}]{Errani22}
{Errani} R.,  et~al., 2022, \mn@doi [\mnras] {10.1093/mnras/stac1516}, \href
  {https://ui.adsabs.harvard.edu/abs/2022MNRAS.514.3532E} {514, 3532}

\bibitem[\protect\citeauthoryear{{Evans} et~al.,}{{Evans}
  et~al.}{2018}]{Evans18}
{Evans} D.~W.,  et~al., 2018, \mn@doi [\aap] {10.1051/0004-6361/201832756},
  \href {https://ui.adsabs.harvard.edu/#abs/2018A&A...616A...4E} {616, A4}

\bibitem[\protect\citeauthoryear{{Falke}, {Tiemann}, {Lisdat}, {Schnatz}  \&
  {Grosche}}{{Falke} et~al.}{2006}]{Falke06}
{Falke} S.,  {Tiemann} E.,  {Lisdat} C.,  {Schnatz} H.,   {Grosche} G.,  2006,
  \mn@doi [\pra] {10.1103/PhysRevA.74.032503}, \href
  {https://ui.adsabs.harvard.edu/abs/2006PhRvA..74c2503F} {74, 032503}

\bibitem[\protect\citeauthoryear{{Fern{\'a}ndez-Trincado}
  et~al.,}{{Fern{\'a}ndez-Trincado} et~al.}{2017}]{Trincado17}
{Fern{\'a}ndez-Trincado} J.~G.,  et~al., 2017, \mn@doi [\apjl]
  {10.3847/2041-8213/aa8032}, \href
  {https://ui.adsabs.harvard.edu/abs/2017ApJ...846L...2F} {846, L2}

\bibitem[\protect\citeauthoryear{{Fran{\c{c}}ois}, {Monaco}, {Bonifacio}, {Moni
  Bidin}, {Geisler}  \& {Sbordone}}{{Fran{\c{c}}ois} et~al.}{2016}]{Francois16}
{Fran{\c{c}}ois} P.,  {Monaco} L.,  {Bonifacio} P.,  {Moni Bidin} C.,
  {Geisler} D.,   {Sbordone} L.,  2016, \mn@doi [\aap]
  {10.1051/0004-6361/201527181}, \href
  {https://ui.adsabs.harvard.edu/abs/2016A&A...588A...7F} {588, A7}

\bibitem[\protect\citeauthoryear{{Frebel} \& {Bromm}}{{Frebel} \&
  {Bromm}}{2012}]{Frebel12}
{Frebel} A.,  {Bromm} V.,  2012, \mn@doi [\apj] {10.1088/0004-637X/759/2/115},
  \href {https://ui.adsabs.harvard.edu/abs/2012ApJ...759..115F} {759, 115}

\bibitem[\protect\citeauthoryear{{Frebel}, {Simon}  \& {Kirby}}{{Frebel}
  et~al.}{2014}]{Frebel14}
{Frebel} A.,  {Simon} J.~D.,   {Kirby} E.~N.,  2014, \mn@doi [\apj]
  {10.1088/0004-637X/786/1/74}, \href
  {https://ui.adsabs.harvard.edu/abs/2014ApJ...786...74F} {786, 74}

\bibitem[\protect\citeauthoryear{{Freeman} \& {Bland-Hawthorn}}{{Freeman} \&
  {Bland-Hawthorn}}{2002}]{Freeman02}
{Freeman} K.,  {Bland-Hawthorn} J.,  2002, \mn@doi [\araa]
  {10.1146/annurev.astro.40.060401.093840}, \href
  {http://adsabs.harvard.edu/abs/2002ARA%26A..40..487F} {40, 487}

\bibitem[\protect\citeauthoryear{{Gaia Collaboration} et~al.,}{{Gaia
  Collaboration} et~al.}{2016}]{Gaia16}
{Gaia Collaboration} et~al., 2016, \mn@doi [\aap]
  {10.1051/0004-6361/201629272}, \href
  {https://ui.adsabs.harvard.edu/abs/2016A%26A...595A...1G} {595, A1}

\bibitem[\protect\citeauthoryear{{Gaia Collaboration} et~al.,}{{Gaia
  Collaboration} et~al.}{2018}]{Gaia18}
{Gaia Collaboration} et~al., 2018, \mn@doi [\aap]
  {10.1051/0004-6361/201833051}, \href
  {https://ui.adsabs.harvard.edu/abs/2018A%26A...616A...1G} {616, A1}

\bibitem[\protect\citeauthoryear{{Gaia Collaboration} et~al.,}{{Gaia
  Collaboration} et~al.}{2021}]{GaiaEDR3}
{Gaia Collaboration} et~al., 2021, \mn@doi [\aap]
  {10.1051/0004-6361/202039657}, \href
  {https://ui.adsabs.harvard.edu/abs/2021A&A...649A...1G} {649, A1}

\bibitem[\protect\citeauthoryear{{Gonz{\'a}lez Hern{\'a}ndez} \&
  {Bonifacio}}{{Gonz{\'a}lez Hern{\'a}ndez} \& {Bonifacio}}{2009}]{Gonzalez09}
{Gonz{\'a}lez Hern{\'a}ndez} J.~I.,  {Bonifacio} P.,  2009, \mn@doi [\aap]
  {10.1051/0004-6361/200810904}, \href
  {https://ui.adsabs.harvard.edu/abs/2009A&A...497..497G} {497, 497}

\bibitem[\protect\citeauthoryear{{Gonz{\'a}lez Hern{\'a}ndez}, {Aguado},
  {Allende Prieto}, {Burgasser}  \& {Rebolo}}{{Gonz{\'a}lez Hern{\'a}ndez}
  et~al.}{2020}]{Ganzalez20}
{Gonz{\'a}lez Hern{\'a}ndez} J.~I.,  {Aguado} D.~S.,  {Allende Prieto} C.,
  {Burgasser} A.~J.,   {Rebolo} R.,  2020, \mn@doi [\apjl]
  {10.3847/2041-8213/ab62ae}, \href
  {https://ui.adsabs.harvard.edu/abs/2020ApJ...889L..13G} {889, L13}

\bibitem[\protect\citeauthoryear{{Gratton}, {Sneden}  \& {Carretta}}{{Gratton}
  et~al.}{2004}]{Gratton04}
{Gratton} R.,  {Sneden} C.,   {Carretta} E.,  2004, \mn@doi [\araa]
  {10.1146/annurev.astro.42.053102.133945}, \href
  {https://ui.adsabs.harvard.edu/abs/2004ARA&A..42..385G} {42, 385}

\bibitem[\protect\citeauthoryear{{Green}, {Schlafly}, {Zucker}, {Speagle}  \&
  {Finkbeiner}}{{Green} et~al.}{2019}]{Green19}
{Green} G.~M.,  {Schlafly} E.,  {Zucker} C.,  {Speagle} J.~S.,   {Finkbeiner}
  D.,  2019, \mn@doi [\apj] {10.3847/1538-4357/ab5362}, \href
  {https://ui.adsabs.harvard.edu/abs/2019ApJ...887...93G} {887, 93}

\bibitem[\protect\citeauthoryear{{Gustafsson}, {Edvardsson}, {Eriksson},
  {J{\o}rgensen}, {Nordlund}  \& {Plez}}{{Gustafsson}
  et~al.}{2008}]{Gustafsson08}
{Gustafsson} B.,  {Edvardsson} B.,  {Eriksson} K.,  {J{\o}rgensen} U.~G.,
  {Nordlund} {\r{A}}.,   {Plez} B.,  2008, \mn@doi [\aap]
  {10.1051/0004-6361:200809724}, \href
  {https://ui.adsabs.harvard.edu/abs/2008A&A...486..951G} {486, 951}

\bibitem[\protect\citeauthoryear{{Hannaford}, {Lowe}, {Grevesse}, {Biemont}  \&
  {Whaling}}{{Hannaford} et~al.}{1982}]{Hannaford82}
{Hannaford} P.,  {Lowe} R.~M.,  {Grevesse} N.,  {Biemont} E.,   {Whaling} W.,
  1982, \mn@doi [\apj] {10.1086/160384}, \href
  {https://ui.adsabs.harvard.edu/abs/1982ApJ...261..736H} {261, 736}

\bibitem[\protect\citeauthoryear{{Hansen}, {Andersen}, {Nordstr{\"o}m},
  {Beers}, {Placco}, {Yoon}  \& {Buchhave}}{{Hansen} et~al.}{2016}]{Hansen16}
{Hansen} T.~T.,  {Andersen} J.,  {Nordstr{\"o}m} B.,  {Beers} T.~C.,  {Placco}
  V.~M.,  {Yoon} J.,   {Buchhave} L.~A.,  2016, \mn@doi [\aap]
  {10.1051/0004-6361/201527409}, \href
  {https://ui.adsabs.harvard.edu/abs/2016A&A...588A...3H} {588, A3}

\bibitem[\protect\citeauthoryear{{Hansen} et~al.,}{{Hansen}
  et~al.}{2018}]{Hansen18}
{Hansen} T.~T.,  et~al., 2018, \mn@doi [\apj] {10.3847/1538-4357/aabacc}, \href
  {https://ui.adsabs.harvard.edu/abs/2018ApJ...858...92H} {858, 92}

\bibitem[\protect\citeauthoryear{{Hasselquist} et~al.,}{{Hasselquist}
  et~al.}{2021}]{Hasselquist21}
{Hasselquist} S.,  et~al., 2021, \mn@doi [\apj] {10.3847/1538-4357/ac25f9},
  \href {https://ui.adsabs.harvard.edu/abs/2021ApJ...923..172H} {923, 172}

\bibitem[\protect\citeauthoryear{{Heger} \& {Woosley}}{{Heger} \&
  {Woosley}}{2002}]{Heger02}
{Heger} A.,  {Woosley} S.~E.,  2002, \mn@doi [\apj] {10.1086/338487}, \href
  {https://ui.adsabs.harvard.edu/abs/2002ApJ...567..532H} {567, 532}

\bibitem[\protect\citeauthoryear{{Heger} \& {Woosley}}{{Heger} \&
  {Woosley}}{2010}]{Heger10}
{Heger} A.,  {Woosley} S.~E.,  2010, \mn@doi [\apj]
  {10.1088/0004-637X/724/1/341}, \href
  {https://ui.adsabs.harvard.edu/abs/2010ApJ...724..341H} {724, 341}

\bibitem[\protect\citeauthoryear{{Helmi}, {Babusiaux}, {Koppelman}, {Massari},
  {Veljanoski}  \& {Brown}}{{Helmi} et~al.}{2018}]{Helmi18}
{Helmi} A.,  {Babusiaux} C.,  {Koppelman} H.~H.,  {Massari} D.,  {Veljanoski}
  J.,   {Brown} A. G.~A.,  2018, \mn@doi [\nat] {10.1038/s41586-018-0625-x},
  \href {https://ui.adsabs.harvard.edu/abs/2018Natur.563...85H} {563, 85}

\bibitem[\protect\citeauthoryear{{Horta} et~al.,}{{Horta}
  et~al.}{2021a}]{Horta21}
{Horta} D.,  et~al., 2021a, \mn@doi [\mnras] {10.1093/mnras/staa2987}, \href
  {https://ui.adsabs.harvard.edu/abs/2021MNRAS.500.1385H} {500, 1385}

\bibitem[\protect\citeauthoryear{{Horta} et~al.,}{{Horta}
  et~al.}{2021b}]{Horta21b}
{Horta} D.,  et~al., 2021b, \mn@doi [\mnras] {10.1093/mnras/staa3598}, \href
  {https://ui.adsabs.harvard.edu/abs/2021MNRAS.500.5462H} {500, 5462}

\bibitem[\protect\citeauthoryear{{Howes} et~al.,}{{Howes}
  et~al.}{2014}]{Howes14}
{Howes} L.~M.,  et~al., 2014, \mn@doi [\mnras] {10.1093/mnras/stu1991}, \href
  {https://ui.adsabs.harvard.edu/abs/2014MNRAS.445.4241H} {445, 4241}

\bibitem[\protect\citeauthoryear{{Howes} et~al.,}{{Howes}
  et~al.}{2015}]{Howes15}
{Howes} L.~M.,  et~al., 2015, \mn@doi [\nat] {10.1038/nature15747}, \href
  {https://ui.adsabs.harvard.edu/abs/2015Natur.527..484H} {527, 484}

\bibitem[\protect\citeauthoryear{{Howes} et~al.,}{{Howes}
  et~al.}{2016}]{Howes16}
{Howes} L.~M.,  et~al., 2016, \mn@doi [\mnras] {10.1093/mnras/stw1004}, \href
  {https://ui.adsabs.harvard.edu/abs/2016MNRAS.460..884H} {460, 884}

\bibitem[\protect\citeauthoryear{{Ishigaki}, {Tominaga}, {Kobayashi}  \&
  {Nomoto}}{{Ishigaki} et~al.}{2018}]{Ishigaki18}
{Ishigaki} M.~N.,  {Tominaga} N.,  {Kobayashi} C.,   {Nomoto} K.,  2018,
  \mn@doi [\apj] {10.3847/1538-4357/aab3de}, \href
  {http://adsabs.harvard.edu/abs/2018ApJ...857...46I} {857, 46}

\bibitem[\protect\citeauthoryear{{Ji}, {Frebel}  \& {Bromm}}{{Ji}
  et~al.}{2015}]{Ji15}
{Ji} A.~P.,  {Frebel} A.,   {Bromm} V.,  2015, \mn@doi [\mnras]
  {10.1093/mnras/stv2052}, \href
  {https://ui.adsabs.harvard.edu/abs/2015MNRAS.454..659J} {454, 659}

\bibitem[\protect\citeauthoryear{{Joggerst}, {Almgren}, {Bell}, {Heger},
  {Whalen}  \& {Woosley}}{{Joggerst} et~al.}{2010}]{Joggerst10}
{Joggerst} C.~C.,  {Almgren} A.,  {Bell} J.,  {Heger} A.,  {Whalen} D.,
  {Woosley} S.~E.,  2010, \mn@doi [\apj] {10.1088/0004-637X/709/1/11}, \href
  {https://ui.adsabs.harvard.edu/abs/2010ApJ...709...11J} {709, 11}

\bibitem[\protect\citeauthoryear{{Johnston}, {Bullock}, {Sharma}, {Font},
  {Robertson}  \& {Leitner}}{{Johnston} et~al.}{2008}]{Johnston2008}
{Johnston} K.~V.,  {Bullock} J.~S.,  {Sharma} S.,  {Font} A.,  {Robertson}
  B.~E.,   {Leitner} S.~N.,  2008, \mn@doi [\apj] {10.1086/592228}, \href
  {https://ui.adsabs.harvard.edu/abs/2008ApJ...689..936J} {689, 936}

\bibitem[\protect\citeauthoryear{{Karlsson}, {Bromm}  \&
  {Bland-Hawthorn}}{{Karlsson} et~al.}{2013}]{Karlsson13}
{Karlsson} T.,  {Bromm} V.,   {Bland-Hawthorn} J.,  2013, \mn@doi [Reviews of
  Modern Physics] {10.1103/RevModPhys.85.809}, \href
  {http://adsabs.harvard.edu/abs/2013RvMP...85..809K} {85, 809}

\bibitem[\protect\citeauthoryear{{Karovicova} et~al.,}{{Karovicova}
  et~al.}{2018}]{Karovicova18}
{Karovicova} I.,  et~al., 2018, \mn@doi [\mnras] {10.1093/mnrasl/sly010}, \href
  {https://ui.adsabs.harvard.edu/abs/2018MNRAS.475L..81K} {475, L81}

\bibitem[\protect\citeauthoryear{{Karovicova}, {White}, {Nordlander},
  {Casagrande}, {Ireland}, {Huber}  \& {Jofr{\'e}}}{{Karovicova}
  et~al.}{2020}]{Karovicova20}
{Karovicova} I.,  {White} T.~R.,  {Nordlander} T.,  {Casagrande} L.,  {Ireland}
  M.,  {Huber} D.,   {Jofr{\'e}} P.,  2020, \mn@doi [\aap]
  {10.1051/0004-6361/202037590}, \href
  {https://ui.adsabs.harvard.edu/abs/2020A&A...640A..25K} {640, A25}

\bibitem[\protect\citeauthoryear{{Kielty} et~al.,}{{Kielty}
  et~al.}{2021}]{Kielty21}
{Kielty} C.~L.,  et~al., 2021, \mn@doi [\mnras] {10.1093/mnras/stab1783}, \href
  {https://ui.adsabs.harvard.edu/abs/2021MNRAS.506.1438K} {506, 1438}

\bibitem[\protect\citeauthoryear{{Kobayashi}, {Karakas}  \&
  {Lugaro}}{{Kobayashi} et~al.}{2020}]{Kobayashi20}
{Kobayashi} C.,  {Karakas} A.~I.,   {Lugaro} M.,  2020, \mn@doi [\apj]
  {10.3847/1538-4357/abae65}, \href
  {https://ui.adsabs.harvard.edu/abs/2020ApJ...900..179K} {900, 179}

\bibitem[\protect\citeauthoryear{{Koch}, {McWilliam}, {Grebel}, {Zucker}  \&
  {Belokurov}}{{Koch} et~al.}{2008}]{Koch08}
{Koch} A.,  {McWilliam} A.,  {Grebel} E.~K.,  {Zucker} D.~B.,   {Belokurov} V.,
   2008, \mn@doi [\apjl] {10.1086/595001}, \href
  {https://ui.adsabs.harvard.edu/abs/2008ApJ...688L..13K} {688, L13}

\bibitem[\protect\citeauthoryear{{Koch}, {Feltzing}, {Ad{\'e}n}  \&
  {Matteucci}}{{Koch} et~al.}{2013}]{Koch13}
{Koch} A.,  {Feltzing} S.,  {Ad{\'e}n} D.,   {Matteucci} F.,  2013, \mn@doi
  [\aap] {10.1051/0004-6361/201220742}, \href
  {https://ui.adsabs.harvard.edu/abs/2013A&A...554A...5K} {554, A5}

\bibitem[\protect\citeauthoryear{{Koch}, {McWilliam}, {Preston}  \&
  {Thompson}}{{Koch} et~al.}{2016}]{Koch16}
{Koch} A.,  {McWilliam} A.,  {Preston} G.~W.,   {Thompson} I.~B.,  2016,
  \mn@doi [\aap] {10.1051/0004-6361/201527413}, \href
  {https://ui.adsabs.harvard.edu/abs/2016A&A...587A.124K} {587, A124}

\bibitem[\protect\citeauthoryear{{Koleva}, {Prugniel}, {Bouchard}  \&
  {Wu}}{{Koleva} et~al.}{2009}]{Koleva09}
{Koleva} M.,  {Prugniel} P.,  {Bouchard} A.,   {Wu} Y.,  2009, \mn@doi [\aap]
  {10.1051/0004-6361/200811467}, \href
  {https://ui.adsabs.harvard.edu/abs/2009A&A...501.1269K} {501, 1269}

\bibitem[\protect\citeauthoryear{{Koppelman}, {Helmi}, {Massari},
  {Price-Whelan}  \& {Starkenburg}}{{Koppelman} et~al.}{2019}]{Koppelman19}
{Koppelman} H.~H.,  {Helmi} A.,  {Massari} D.,  {Price-Whelan} A.~M.,
  {Starkenburg} T.~K.,  2019, \mn@doi [\aap] {10.1051/0004-6361/201936738},
  \href {https://ui.adsabs.harvard.edu/abs/2019A&A...631L...9K} {631, L9}

\bibitem[\protect\citeauthoryear{Kramida, {Yu.~Ralchenko}, Reader  \& {and NIST
  ASD Team}}{Kramida et~al.}{2021}]{NIST_ASD}
Kramida A.,  {Yu.~Ralchenko} Reader J.,   {and NIST ASD Team} 2021, {NIST
  Atomic Spectra Database (ver. 5.9), [Online]. Available:
  {\tt{https://physics.nist.gov/asd}} [2022, March 23]. National Institute of
  Standards and Technology, Gaithersburg, MD.}

\bibitem[\protect\citeauthoryear{{Kruijssen}}{{Kruijssen}}{2015}]{Kruijssen15}
{Kruijssen} J.~M.~D.,  2015, \mn@doi [\mnras] {10.1093/mnras/stv2026}, \href
  {https://ui.adsabs.harvard.edu/abs/2015MNRAS.454.1658K} {454, 1658}

\bibitem[\protect\citeauthoryear{{Lardo} et~al.,}{{Lardo}
  et~al.}{2021}]{Lardo21}
{Lardo} C.,  et~al., 2021, \mn@doi [\mnras] {10.1093/mnras/stab2847}, \href
  {https://ui.adsabs.harvard.edu/abs/2021MNRAS.508.3068L} {508, 3068}

\bibitem[\protect\citeauthoryear{{Larsen}, {Eitner}, {Magg}, {Bergemann},
  {Moltzer}, {Brodie}, {Romanowsky}  \& {Strader}}{{Larsen}
  et~al.}{2022}]{Larsen22}
{Larsen} S.~S.,  {Eitner} P.,  {Magg} E.,  {Bergemann} M.,  {Moltzer} C.~A.~S.,
   {Brodie} J.~P.,  {Romanowsky} A.~J.,   {Strader} J.,  2022, \mn@doi [\aap]
  {10.1051/0004-6361/202142243}, \href
  {https://ui.adsabs.harvard.edu/abs/2022A&A...660A..88L} {660, A88}

\bibitem[\protect\citeauthoryear{{Lawler}, {Guzman}, {Wood}, {Sneden}  \&
  {Cowan}}{{Lawler} et~al.}{2013}]{Lawler13}
{Lawler} J.~E.,  {Guzman} A.,  {Wood} M.~P.,  {Sneden} C.,   {Cowan} J.~J.,
  2013, \mn@doi [\apjs] {10.1088/0067-0049/205/2/11}, \href
  {https://ui.adsabs.harvard.edu/abs/2013ApJS..205...11L} {205, 11}

\bibitem[\protect\citeauthoryear{{Lawler}, {Hala}, {Sneden}, {Nave}, {Wood}  \&
  {Cowan}}{{Lawler} et~al.}{2019}]{Lawler19}
{Lawler} J.~E.,  {Hala} {Sneden} C.,  {Nave} G.,  {Wood} M.~P.,   {Cowan}
  J.~J.,  2019, \mn@doi [\apjs] {10.3847/1538-4365/ab08ef}, \href
  {https://ui.adsabs.harvard.edu/abs/2019ApJS..241...21L} {241, 21}

\bibitem[\protect\citeauthoryear{{Lind}, {Bergemann}  \& {Asplund}}{{Lind}
  et~al.}{2012}]{Lind2012}
{Lind} K.,  {Bergemann} M.,   {Asplund} M.,  2012, \mn@doi [\mnras]
  {10.1111/j.1365-2966.2012.21686.x}, \href
  {https://ui.adsabs.harvard.edu/abs/2012MNRAS.427...50L} {427, 50}

\bibitem[\protect\citeauthoryear{{Lindegren} et~al.,}{{Lindegren}
  et~al.}{2018}]{Lindegren18}
{Lindegren} L.,  et~al., 2018, \mn@doi [\aap] {10.1051/0004-6361/201832727},
  \href {https://ui.adsabs.harvard.edu/abs/2018A&A...616A...2L} {616, A2}

\bibitem[\protect\citeauthoryear{{Lindegren} et~al.,}{{Lindegren}
  et~al.}{2021}]{Lindegren21}
{Lindegren} L.,  et~al., 2021, \mn@doi [\aap] {10.1051/0004-6361/202039709},
  \href {https://ui.adsabs.harvard.edu/abs/2021A&A...649A...2L} {649, A2}

\bibitem[\protect\citeauthoryear{{Lucatello}, {Sollima}, {Gratton},
  {Vesperini}, {D'Orazi}, {Carretta}  \& {Bragaglia}}{{Lucatello}
  et~al.}{2015}]{Lucatello15}
{Lucatello} S.,  {Sollima} A.,  {Gratton} R.,  {Vesperini} E.,  {D'Orazi} V.,
  {Carretta} E.,   {Bragaglia} A.,  2015, \mn@doi [\aap]
  {10.1051/0004-6361/201526957}, \href
  {https://ui.adsabs.harvard.edu/abs/2015A&A...584A..52L} {584, A52}

\bibitem[\protect\citeauthoryear{{Lucchesi} et~al.,}{{Lucchesi}
  et~al.}{2022}]{Lucchesi22}
{Lucchesi} R.,  et~al., 2022, \mn@doi [\mnras] {10.1093/mnras/stab3721}, \href
  {https://ui.adsabs.harvard.edu/abs/2022MNRAS.511.1004L} {511, 1004}

\bibitem[\protect\citeauthoryear{{Lucey} et~al.,}{{Lucey}
  et~al.}{2019}]{Lucey19}
{Lucey} M.,  et~al., 2019, \mn@doi [\mnras] {10.1093/mnras/stz1847}, \href
  {https://ui.adsabs.harvard.edu/abs/2019MNRAS.488.2283L} {488, 2283}

\bibitem[\protect\citeauthoryear{{Lucey} et~al.,}{{Lucey}
  et~al.}{2021}]{Lucey21}
{Lucey} M.,  et~al., 2021, \mn@doi [\mnras] {10.1093/mnras/stab003}, \href
  {https://ui.adsabs.harvard.edu/abs/2021MNRAS.501.5981L} {501, 5981}

\bibitem[\protect\citeauthoryear{{Lucey} et~al.,}{{Lucey}
  et~al.}{2022}]{Lucey22}
{Lucey} M.,  et~al., 2022, \mn@doi [\mnras] {10.1093/mnras/stab2878}, \href
  {https://ui.adsabs.harvard.edu/abs/2022MNRAS.509..122L} {509, 122}

\bibitem[\protect\citeauthoryear{MATLAB}{MATLAB}{2021}]{MATLAB21}
MATLAB 2021, version 9.11.0 (R2021b).
The MathWorks Inc., Natick, Massachusetts

\bibitem[\protect\citeauthoryear{{Marigo}, {Girardi}, {Bressan}, {Groenewegen},
  {Silva}  \& {Granato}}{{Marigo} et~al.}{2008}]{Marigo08}
{Marigo} P.,  {Girardi} L.,  {Bressan} A.,  {Groenewegen} M.~A.~T.,  {Silva}
  L.,   {Granato} G.~L.,  2008, \mn@doi [\aap] {10.1051/0004-6361:20078467},
  \href {http://adsabs.harvard.edu/abs/2008A%26A...482..883M} {482, 883}

\bibitem[\protect\citeauthoryear{{Martell}, {Smolinski}, {Beers}  \&
  {Grebel}}{{Martell} et~al.}{2011}]{Martell11}
{Martell} S.~L.,  {Smolinski} J.~P.,  {Beers} T.~C.,   {Grebel} E.~K.,  2011,
  \mn@doi [\aap] {10.1051/0004-6361/201117644}, \href
  {https://ui.adsabs.harvard.edu/abs/2011A&A...534A.136M} {534, A136}

\bibitem[\protect\citeauthoryear{{Martin} et~al.,}{{Martin}
  et~al.}{2022}]{Martin22}
{Martin} N.~F.,  et~al., 2022, \mn@doi [\nat] {10.1038/s41586-021-04162-2},
  \href {https://ui.adsabs.harvard.edu/abs/2022Natur.601...45M} {601, 45}

\bibitem[\protect\citeauthoryear{{Martioli}, {Teeple}, {Manset}, {Devost},
  {Withington}, {Venne}  \& {Tannock}}{{Martioli} et~al.}{2012}]{Martioli12}
{Martioli} E.,  {Teeple} D.,  {Manset} N.,  {Devost} D.,  {Withington} K.,
  {Venne} A.,   {Tannock} M.,  2012, in {Radziwill} N.~M.,  {Chiozzi} G.,  eds,
   Society of Photo-Optical Instrumentation Engineers (SPIE) Conference Series
  Vol. 8451, Software and Cyberinfrastructure for Astronomy II. p. 84512B,
  \mn@doi{10.1117/12.926627}

\bibitem[\protect\citeauthoryear{{Mashonkina}, {Jablonka}, {Pakhomov},
  {Sitnova}  \& {North}}{{Mashonkina} et~al.}{2017}]{Mashonkina17}
{Mashonkina} L.,  {Jablonka} P.,  {Pakhomov} Y.,  {Sitnova} T.,   {North} P.,
  2017, \mn@doi [\aap] {10.1051/0004-6361/201730779}, \href
  {https://ui.adsabs.harvard.edu/abs/2017A&A...604A.129M} {604, A129}

\bibitem[\protect\citeauthoryear{{Masseron}, {Johnson}, {Plez}, {van Eck},
  {Primas}, {Goriely}  \& {Jorissen}}{{Masseron} et~al.}{2010}]{Masseron10}
{Masseron} T.,  {Johnson} J.~A.,  {Plez} B.,  {van Eck} S.,  {Primas} F.,
  {Goriely} S.,   {Jorissen} A.,  2010, \mn@doi [\aap]
  {10.1051/0004-6361/200911744}, \href
  {https://ui.adsabs.harvard.edu/abs/2010A&A...509A..93M} {509, A93}

\bibitem[\protect\citeauthoryear{{Meynet}, {Ekstr{\"o}m}  \& {Maeder}}{{Meynet}
  et~al.}{2006}]{Meynet06}
{Meynet} G.,  {Ekstr{\"o}m} S.,   {Maeder} A.,  2006, \mn@doi [\aap]
  {10.1051/0004-6361:20053070}, \href
  {https://ui.adsabs.harvard.edu/abs/2006A&A...447..623M} {447, 623}

\bibitem[\protect\citeauthoryear{{Meynet}, {Hirschi}, {Ekstrom}, {Maeder},
  {Georgy}, {Eggenberger}  \& {Chiappini}}{{Meynet} et~al.}{2010}]{Meynet10}
{Meynet} G.,  {Hirschi} R.,  {Ekstrom} S.,  {Maeder} A.,  {Georgy} C.,
  {Eggenberger} P.,   {Chiappini} C.,  2010, \mn@doi [\aap]
  {10.1051/0004-6361/200913377}, \href
  {https://ui.adsabs.harvard.edu/abs/2010A&A...521A..30M} {521, A30}

\bibitem[\protect\citeauthoryear{{Milone} et~al.,}{{Milone}
  et~al.}{2012}]{Milone12}
{Milone} A.~P.,  et~al., 2012, \mn@doi [\aap] {10.1051/0004-6361/201016384},
  \href {https://ui.adsabs.harvard.edu/abs/2012A&A...540A..16M} {540, A16}

\bibitem[\protect\citeauthoryear{{Miyamoto} \& {Nagai}}{{Miyamoto} \&
  {Nagai}}{1975}]{MiyamotoNagai}
{Miyamoto} M.,  {Nagai} R.,  1975, \pasj, \href
  {https://ui.adsabs.harvard.edu/abs/1975PASJ...27..533M} {27, 533}

\bibitem[\protect\citeauthoryear{{Monari}, {Famaey}, {Siebert}, {Grand},
  {Kawata}  \& {Boily}}{{Monari} et~al.}{2016}]{Monari16}
{Monari} G.,  {Famaey} B.,  {Siebert} A.,  {Grand} R. J.~J.,  {Kawata} D.,
  {Boily} C.,  2016, \mn@doi [\mnras] {10.1093/mnras/stw1564}, \href
  {https://ui.adsabs.harvard.edu/abs/2016MNRAS.461.3835M} {461, 3835}

\bibitem[\protect\citeauthoryear{{Monty}, {Venn}, {Lane}, {Lokhorst}  \&
  {Yong}}{{Monty} et~al.}{2020}]{Monty20}
{Monty} S.,  {Venn} K.~A.,  {Lane} J. M.~M.,  {Lokhorst} D.,   {Yong} D.,
  2020, \mn@doi [\mnras] {10.1093/mnras/staa1995}, \href
  {https://ui.adsabs.harvard.edu/abs/2020MNRAS.497.1236M} {497, 1236}

\bibitem[\protect\citeauthoryear{{Mucciarelli}, {Bellazzini}, {Ibata}, {Merle},
  {Chapman}, {Dalessandro}  \& {Sollima}}{{Mucciarelli}
  et~al.}{2012}]{Mucciarelli12}
{Mucciarelli} A.,  {Bellazzini} M.,  {Ibata} R.,  {Merle} T.,  {Chapman} S.~C.,
   {Dalessandro} E.,   {Sollima} A.,  2012, \mn@doi [\mnras]
  {10.1111/j.1365-2966.2012.21847.x}, \href
  {https://ui.adsabs.harvard.edu/abs/2012MNRAS.426.2889M} {426, 2889}

\bibitem[\protect\citeauthoryear{{Mucciarelli}, {Bellazzini}  \&
  {Massari}}{{Mucciarelli} et~al.}{2021}]{Mucciarelli21}
{Mucciarelli} A.,  {Bellazzini} M.,   {Massari} D.,  2021, \mn@doi [\aap]
  {10.1051/0004-6361/202140979}, \href
  {https://ui.adsabs.harvard.edu/abs/2021A&A...653A..90M} {653, A90}

\bibitem[\protect\citeauthoryear{{Myeong}, {Vasiliev}, {Iorio}, {Evans}  \&
  {Belokurov}}{{Myeong} et~al.}{2019}]{Myeong19}
{Myeong} G.~C.,  {Vasiliev} E.,  {Iorio} G.,  {Evans} N.~W.,   {Belokurov} V.,
  2019, \mn@doi [\mnras] {10.1093/mnras/stz1770}, \href
  {https://ui.adsabs.harvard.edu/abs/2019MNRAS.488.1235M} {488, 1235}

\bibitem[\protect\citeauthoryear{{Navarro}, {Frenk}  \& {White}}{{Navarro}
  et~al.}{1997}]{NavarroFrenkWhite97}
{Navarro} J.~F.,  {Frenk} C.~S.,   {White} S.~D.~M.,  1997, \mn@doi [\apj]
  {10.1086/304888}, \href {http://adsabs.harvard.edu/abs/1997ApJ...490..493N}
  {490, 493}

\bibitem[\protect\citeauthoryear{{Ness} et~al.,}{{Ness}
  et~al.}{2013a}]{Ness13a}
{Ness} M.,  et~al., 2013a, \mn@doi [\mnras] {10.1093/mnras/sts629}, \href
  {https://ui.adsabs.harvard.edu/abs/2013MNRAS.430..836N} {430, 836}

\bibitem[\protect\citeauthoryear{{Ness} et~al.,}{{Ness}
  et~al.}{2013b}]{Ness13b}
{Ness} M.,  et~al., 2013b, \mn@doi [\mnras] {10.1093/mnras/stt533}, \href
  {https://ui.adsabs.harvard.edu/abs/2013MNRAS.432.2092N} {432, 2092}

\bibitem[\protect\citeauthoryear{{Ness}, {Debattista}, {Bensby}, {Feltzing},
  {Ro{\v{s}}kar}, {Cole}, {Johnson}  \& {Freeman}}{{Ness}
  et~al.}{2014}]{Ness2014}
{Ness} M.,  {Debattista} V.~P.,  {Bensby} T.,  {Feltzing} S.,  {Ro{\v{s}}kar}
  R.,  {Cole} D.~R.,  {Johnson} J.~A.,   {Freeman} K.,  2014, \mn@doi [\apjl]
  {10.1088/2041-8205/787/2/L19}, \href
  {https://ui.adsabs.harvard.edu/abs/2014ApJ...787L..19N} {787, L19}

\bibitem[\protect\citeauthoryear{{Nomoto}, {Kobayashi}  \& {Tominaga}}{{Nomoto}
  et~al.}{2013}]{Nomoto13}
{Nomoto} K.,  {Kobayashi} C.,   {Tominaga} N.,  2013, \mn@doi [\araa]
  {10.1146/annurev-astro-082812-140956}, \href
  {https://ui.adsabs.harvard.edu/abs/2013ARA&A..51..457N} {51, 457}

\bibitem[\protect\citeauthoryear{{Norris} et~al.,}{{Norris}
  et~al.}{2013}]{Norris13}
{Norris} J.~E.,  et~al., 2013, \mn@doi [\apj] {10.1088/0004-637X/762/1/28},
  \href {https://ui.adsabs.harvard.edu/abs/2013ApJ...762...28N} {762, 28}

\bibitem[\protect\citeauthoryear{{Norris}, {Yong}, {Venn}, {Gilmore},
  {Casagrande}  \& {Dotter}}{{Norris} et~al.}{2017}]{Norris17}
{Norris} J.~E.,  {Yong} D.,  {Venn} K.~A.,  {Gilmore} G.,  {Casagrande} L.,
  {Dotter} A.,  2017, \mn@doi [\apjs] {10.3847/1538-4365/aa755e}, \href
  {https://ui.adsabs.harvard.edu/abs/2017ApJS..230...28N} {230, 28}

\bibitem[\protect\citeauthoryear{{Omukai} \& {Palla}}{{Omukai} \&
  {Palla}}{2001}]{Omukai01}
{Omukai} K.,  {Palla} F.,  2001, \mn@doi [\apjl] {10.1086/324410}, \href
  {https://ui.adsabs.harvard.edu/abs/2001ApJ...561L..55O} {561, L55}

\bibitem[\protect\citeauthoryear{{Pancino} et~al.,}{{Pancino}
  et~al.}{2017}]{Pancino17}
{Pancino} E.,  et~al., 2017, \mn@doi [\aap] {10.1051/0004-6361/201730474},
  \href {https://ui.adsabs.harvard.edu/abs/2017A&A...601A.112P} {601, A112}

\bibitem[\protect\citeauthoryear{{Pazder}, {Fournier}, {Pawluczyk}  \& {van
  Kooten}}{{Pazder} et~al.}{2014}]{Pazder14}
{Pazder} J.,  {Fournier} P.,  {Pawluczyk} R.,   {van Kooten} M.,  2014, in
  {Navarro} R.,  {Cunningham} C.~R.,   {Barto} A.~A.,  eds,  Society of
  Photo-Optical Instrumentation Engineers (SPIE) Conference Series Vol. 9151,
  Advances in Optical and Mechanical Technologies for Telescopes and
  Instrumentation. p. 915124, \mn@doi{10.1117/12.2057327}

\bibitem[\protect\citeauthoryear{{Pazder} et~al.,}{{Pazder}
  et~al.}{2020}]{Pazder20}
{Pazder} J.,  et~al., 2020, in Society of Photo-Optical Instrumentation
  Engineers (SPIE) Conference Series. p. 1144743, \mn@doi{10.1117/12.2561985}

\bibitem[\protect\citeauthoryear{{Placco}, {Frebel}, {Beers}  \&
  {Stancliffe}}{{Placco} et~al.}{2014}]{Placco14}
{Placco} V.~M.,  {Frebel} A.,  {Beers} T.~C.,   {Stancliffe} R.~J.,  2014,
  \mn@doi [\apj] {10.1088/0004-637X/797/1/21}, \href
  {https://ui.adsabs.harvard.edu/abs/2014ApJ...797...21P} {797, 21}

\bibitem[\protect\citeauthoryear{{Placco}, {Sneden}, {Roederer}, {Lawler}, {Den
  Hartog}, {Hejazi}, {Maas}  \& {Bernath}}{{Placco} et~al.}{2021}]{Placco21}
{Placco} V.~M.,  {Sneden} C.,  {Roederer} I.~U.,  {Lawler} J.~E.,  {Den Hartog}
  E.~A.,  {Hejazi} N.,  {Maas} Z.,   {Bernath} P.,  2021, \mn@doi [Research
  Notes of the American Astronomical Society] {10.3847/2515-5172/abf651}, \href
  {https://ui.adsabs.harvard.edu/abs/2021RNAAS...5...92P} {5, 92}

\bibitem[\protect\citeauthoryear{{Plez}}{{Plez}}{2012}]{Plez12}
{Plez} B.,  2012, {Turbospectrum: Code for spectral synthesis} (\mn@eprint
  {ascl} {1205.004})

\bibitem[\protect\citeauthoryear{{Pritzl}, {Venn}  \& {Irwin}}{{Pritzl}
  et~al.}{2005}]{Pritzl05}
{Pritzl} B.~J.,  {Venn} K.~A.,   {Irwin} M.,  2005, \mn@doi [\aj]
  {10.1086/432911}, \href
  {https://ui.adsabs.harvard.edu/abs/2005AJ....130.2140P} {130, 2140}

\bibitem[\protect\citeauthoryear{{Reggiani}, {Schlaufman}, {Casey}  \&
  {Ji}}{{Reggiani} et~al.}{2020}]{Reggiani20}
{Reggiani} H.,  {Schlaufman} K.~C.,  {Casey} A.~R.,   {Ji} A.~P.,  2020,
  \mn@doi [\aj] {10.3847/1538-3881/aba948}, \href
  {https://ui.adsabs.harvard.edu/abs/2020AJ....160..173R} {160, 173}

\bibitem[\protect\citeauthoryear{{Sakari}, {Venn}, {Mackey}, {Shetrone},
  {Dotter}, {Ferguson}  \& {Huxor}}{{Sakari} et~al.}{2015}]{Sakari15}
{Sakari} C.~M.,  {Venn} K.~A.,  {Mackey} D.,  {Shetrone} M.~D.,  {Dotter} A.,
  {Ferguson} A. M.~N.,   {Huxor} A.,  2015, \mn@doi [\mnras]
  {10.1093/mnras/stv020}, \href
  {https://ui.adsabs.harvard.edu/abs/2015MNRAS.448.1314S} {448, 1314}

\bibitem[\protect\citeauthoryear{{Salvadori}, {Ferrara}, {Schneider},
  {Scannapieco}  \& {Kawata}}{{Salvadori} et~al.}{2010}]{Salvadori10}
{Salvadori} S.,  {Ferrara} A.,  {Schneider} R.,  {Scannapieco} E.,   {Kawata}
  D.,  2010, \mn@doi [\mnras] {10.1111/j.1745-3933.2009.00772.x}, \href
  {http://adsabs.harvard.edu/abs/2010MNRAS.401L...5S} {401, L5}

\bibitem[\protect\citeauthoryear{{Salvadori}, {Bonifacio}, {Caffau}, {Korotin},
  {Andreevsky}, {Spite}  \& {Sk{\'u}lad{\'o}ttir}}{{Salvadori}
  et~al.}{2019}]{Salvadori19}
{Salvadori} S.,  {Bonifacio} P.,  {Caffau} E.,  {Korotin} S.,  {Andreevsky} S.,
   {Spite} M.,   {Sk{\'u}lad{\'o}ttir} {\'A}.,  2019, \mn@doi [\mnras]
  {10.1093/mnras/stz1464}, \href
  {https://ui.adsabs.harvard.edu/abs/2019MNRAS.487.4261S} {487, 4261}

\bibitem[\protect\citeauthoryear{{Santistevan}, {Wetzel}, {Sanderson},
  {El-Badry}, {Samuel}  \& {Faucher-Gigu{\`e}re}}{{Santistevan}
  et~al.}{2021}]{Santistevan21}
{Santistevan} I.~B.,  {Wetzel} A.,  {Sanderson} R.~E.,  {El-Badry} K.,
  {Samuel} J.,   {Faucher-Gigu{\`e}re} C.-A.,  2021, \mn@doi [\mnras]
  {10.1093/mnras/stab1345}, \href
  {https://ui.adsabs.harvard.edu/abs/2021MNRAS.505..921S} {505, 921}

\bibitem[\protect\citeauthoryear{{Schiavon} et~al.,}{{Schiavon}
  et~al.}{2017}]{Schiavon17}
{Schiavon} R.~P.,  et~al., 2017, \mn@doi [\mnras] {10.1093/mnras/stw3093},
  \href {https://ui.adsabs.harvard.edu/abs/2017MNRAS.466.1010S} {466, 1010}

\bibitem[\protect\citeauthoryear{{Schlaufman} \& {Casey}}{{Schlaufman} \&
  {Casey}}{2014}]{Schlaufman14}
{Schlaufman} K.~C.,  {Casey} A.~R.,  2014, \mn@doi [\apj]
  {10.1088/0004-637X/797/1/13}, \href
  {https://ui.adsabs.harvard.edu/abs/2014ApJ...797...13S} {797, 13}

\bibitem[\protect\citeauthoryear{{Schultheis}, {Rich}, {Origlia}, {Ryde},
  {Nandakumar}, {Thorsbro}  \& {Neumayer}}{{Schultheis}
  et~al.}{2019}]{Schultheis19}
{Schultheis} M.,  {Rich} R.~M.,  {Origlia} L.,  {Ryde} N.,  {Nandakumar} G.,
  {Thorsbro} B.,   {Neumayer} N.,  2019, \mn@doi [\aap]
  {10.1051/0004-6361/201935772}, \href
  {https://ui.adsabs.harvard.edu/abs/2019A&A...627A.152S} {627, A152}

\bibitem[\protect\citeauthoryear{{Schultz} \& {Wiemer}}{{Schultz} \&
  {Wiemer}}{1975}]{Schultz75}
{Schultz} G.~V.,  {Wiemer} W.,  1975, \aap, \href
  {https://ui.adsabs.harvard.edu/abs/1975A&A....43..133S} {43, 133}

\bibitem[\protect\citeauthoryear{{Sestito} et~al.,}{{Sestito}
  et~al.}{2019}]{Sestito19}
{Sestito} F.,  et~al., 2019, \mn@doi [\mnras] {10.1093/mnras/stz043}, \href
  {http://adsabs.harvard.edu/abs/2019MNRAS.484.2166S} {484, 2166}

\bibitem[\protect\citeauthoryear{{Sestito} et~al.,}{{Sestito}
  et~al.}{2020}]{Sestito20}
{Sestito} F.,  et~al., 2020, \mn@doi [\mnras] {10.1093/mnrasl/slaa022}, \href
  {https://ui.adsabs.harvard.edu/abs/2020MNRAS.497L...7S} {497, L7}

\bibitem[\protect\citeauthoryear{{Sestito} et~al.,}{{Sestito}
  et~al.}{2021}]{Sestito21}
{Sestito} F.,  et~al., 2021, \mn@doi [\mnras] {10.1093/mnras/staa3479}, \href
  {https://ui.adsabs.harvard.edu/abs/2021MNRAS.500.3750S} {500, 3750}

\bibitem[\protect\citeauthoryear{{Shapiro}, {Genzel}  \& {F{\"o}rster
  Schreiber}}{{Shapiro} et~al.}{2010}]{Shapiro10}
{Shapiro} K.~L.,  {Genzel} R.,   {F{\"o}rster Schreiber} N.~M.,  2010, \mn@doi
  [\mnras] {10.1111/j.1745-3933.2010.00810.x}, \href
  {https://ui.adsabs.harvard.edu/abs/2010MNRAS.403L..36S} {403, L36}

\bibitem[\protect\citeauthoryear{{Sitnova}, {Mashonkina}, {Ezzeddine}  \&
  {Frebel}}{{Sitnova} et~al.}{2019}]{Sitnova19}
{Sitnova} T.~M.,  {Mashonkina} L.~I.,  {Ezzeddine} R.,   {Frebel} A.,  2019,
  \mn@doi [\mnras] {10.1093/mnras/stz626}, \href
  {https://ui.adsabs.harvard.edu/abs/2019MNRAS.485.3527S} {485, 3527}

\bibitem[\protect\citeauthoryear{{Sneden}}{{Sneden}}{1973}]{Sneden73}
{Sneden} C.~A.,  1973, PhD thesis, THE UNIVERSITY OF TEXAS AT AUSTIN.

\bibitem[\protect\citeauthoryear{{Sneden}, {Cowan}, {Kobayashi}, {Pignatari},
  {Lawler}, {Den Hartog}  \& {Wood}}{{Sneden} et~al.}{2016}]{Sneden16}
{Sneden} C.,  {Cowan} J.~J.,  {Kobayashi} C.,  {Pignatari} M.,  {Lawler} J.~E.,
   {Den Hartog} E.~A.,   {Wood} M.~P.,  2016, \mn@doi [\apj]
  {10.3847/0004-637X/817/1/53}, \href
  {https://ui.adsabs.harvard.edu/abs/2016ApJ...817...53S} {817, 53}

\bibitem[\protect\citeauthoryear{{Sobeck}, {Lawler}  \& {Sneden}}{{Sobeck}
  et~al.}{2007}]{Sobeck07}
{Sobeck} J.~S.,  {Lawler} J.~E.,   {Sneden} C.,  2007, \mn@doi [\apj]
  {10.1086/519987}, \href
  {https://ui.adsabs.harvard.edu/abs/2007ApJ...667.1267S} {667, 1267}

\bibitem[\protect\citeauthoryear{{Sobeck} et~al.,}{{Sobeck}
  et~al.}{2011}]{Sobeck11}
{Sobeck} J.~S.,  et~al., 2011, \mn@doi [\aj] {10.1088/0004-6256/141/6/175},
  \href {https://ui.adsabs.harvard.edu/abs/2011AJ....141..175S} {141, 175}

\bibitem[\protect\citeauthoryear{{Stacy}, {Greif}  \& {Bromm}}{{Stacy}
  et~al.}{2010}]{Stacy10}
{Stacy} A.,  {Greif} T.~H.,   {Bromm} V.,  2010, \mn@doi [\mnras]
  {10.1111/j.1365-2966.2009.16113.x}, \href
  {https://ui.adsabs.harvard.edu/abs/2010MNRAS.403...45S} {403, 45}

\bibitem[\protect\citeauthoryear{{Stancliffe}}{{Stancliffe}}{2009}]{Stancliffe09}
{Stancliffe} R.~J.,  2009, \mn@doi [\mnras] {10.1111/j.1365-2966.2009.14394.x},
  \href {https://ui.adsabs.harvard.edu/abs/2009MNRAS.394.1051S} {394, 1051}

\bibitem[\protect\citeauthoryear{{Starkenburg}, {Oman}, {Navarro}, {Crain},
  {Fattahi}, {Frenk}, {Sawala}  \& {Schaye}}{{Starkenburg}
  et~al.}{2017a}]{Starkenburg17a}
{Starkenburg} E.,  {Oman} K.~A.,  {Navarro} J.~F.,  {Crain} R.~A.,  {Fattahi}
  A.,  {Frenk} C.~S.,  {Sawala} T.,   {Schaye} J.,  2017a, \mn@doi [\mnras]
  {10.1093/mnras/stw2873}, \href
  {http://adsabs.harvard.edu/abs/2017MNRAS.465.2212S} {465, 2212}

\bibitem[\protect\citeauthoryear{{Starkenburg} et~al.,}{{Starkenburg}
  et~al.}{2017b}]{Starkenburg17b}
{Starkenburg} E.,  et~al., 2017b, \mn@doi [\mnras] {10.1093/mnras/stx1068},
  \href {http://adsabs.harvard.edu/abs/2017MNRAS.471.2587S} {471, 2587}

\bibitem[\protect\citeauthoryear{{Starkenburg} et~al.,}{{Starkenburg}
  et~al.}{2019}]{Starkenburg19}
{Starkenburg} E.,  et~al., 2019, \mn@doi [\mnras] {10.1093/mnras/stz2935},
  \href {https://ui.adsabs.harvard.edu/abs/2019MNRAS.490.5757S} {490, 5757}

\bibitem[\protect\citeauthoryear{{Takahashi}, {Yoshida}  \&
  {Umeda}}{{Takahashi} et~al.}{2018}]{Takahashi18}
{Takahashi} K.,  {Yoshida} T.,   {Umeda} H.,  2018, \mn@doi [\apj]
  {10.3847/1538-4357/aab95f}, \href
  {https://ui.adsabs.harvard.edu/abs/2018ApJ...857..111T} {857, 111}

\bibitem[\protect\citeauthoryear{{Taylor}}{{Taylor}}{2005}]{Taylor05}
{Taylor} M.~B.,  2005, in {Shopbell} P.,  {Britton} M.,   {Ebert} R.,  eds,
  Astronomical Society of the Pacific Conference Series Vol. 347, Astronomical
  Data Analysis Software and Systems XIV. p.~29

\bibitem[\protect\citeauthoryear{{Timmes}, {Woosley}  \& {Weaver}}{{Timmes}
  et~al.}{1995}]{Timmes95}
{Timmes} F.~X.,  {Woosley} S.~E.,   {Weaver} T.~A.,  1995, \mn@doi [\apjs]
  {10.1086/192172}, \href
  {https://ui.adsabs.harvard.edu/abs/1995ApJS...98..617T} {98, 617}

\bibitem[\protect\citeauthoryear{{Tissera}, {White}  \&
  {Scannapieco}}{{Tissera} et~al.}{2012}]{Tissera12}
{Tissera} P.~B.,  {White} S. D.~M.,   {Scannapieco} C.,  2012, \mn@doi [\mnras]
  {10.1111/j.1365-2966.2011.20028.x}, \href
  {https://ui.adsabs.harvard.edu/abs/2012MNRAS.420..255T} {420, 255}

\bibitem[\protect\citeauthoryear{{Tody}}{{Tody}}{1986}]{Tody86}
{Tody} D.,  1986, in {Crawford} D.~L.,  ed.,  Society of Photo-Optical
  Instrumentation Engineers (SPIE) Conference Series Vol. 627, Society of
  Photo-Optical Instrumentation Engineers (SPIE). p.~733,
  \mn@doi{10.1117/12.968154}

\bibitem[\protect\citeauthoryear{{Tody}}{{Tody}}{1993}]{Tody93}
{Tody} D.,  1993, in {Hanisch} R.~J.,  {Brissenden} R.~J.~V.,   {Barnes} J.,
  eds,  Astronomical Society of the Pacific Conference Series Vol. 52,
  Astronomical Data Analysis Software and Systems II. p.~173

\bibitem[\protect\citeauthoryear{{Tolstoy}, {Hill}  \& {Tosi}}{{Tolstoy}
  et~al.}{2009}]{Tolstoy09}
{Tolstoy} E.,  {Hill} V.,   {Tosi} M.,  2009, \mn@doi [\araa]
  {10.1146/annurev-astro-082708-101650}, \href
  {https://ui.adsabs.harvard.edu/abs/2009ARA&A..47..371T} {47, 371}

\bibitem[\protect\citeauthoryear{{Tominaga}, {Iwamoto}  \& {Nomoto}}{{Tominaga}
  et~al.}{2014}]{Tominaga14}
{Tominaga} N.,  {Iwamoto} N.,   {Nomoto} K.,  2014, \mn@doi [\apj]
  {10.1088/0004-637X/785/2/98}, \href
  {https://ui.adsabs.harvard.edu/abs/2014ApJ...785...98T} {785, 98}

\bibitem[\protect\citeauthoryear{{Trubko}, {Gregoire}, {Holmgren}  \&
  {Cronin}}{{Trubko} et~al.}{2017}]{Trubko17}
{Trubko} R.,  {Gregoire} M.~D.,  {Holmgren} W.~F.,   {Cronin} A.~D.,  2017,
  \mn@doi [\pra] {10.1103/PhysRevA.95.052507}, \href
  {https://ui.adsabs.harvard.edu/abs/2017PhRvA..95e2507T} {95, 052507}

\bibitem[\protect\citeauthoryear{{Tumlinson}}{{Tumlinson}}{2010}]{Tumlinson10}
{Tumlinson} J.,  2010, \mn@doi [The Astrophysical Journal]
  {10.1088/0004-637X/708/2/1398}, \href
  {https://ui.adsabs.harvard.edu/abs/2010ApJ...708.1398T} {708, 1398}

\bibitem[\protect\citeauthoryear{{Umeda} \& {Nomoto}}{{Umeda} \&
  {Nomoto}}{2003}]{Umeda03}
{Umeda} H.,  {Nomoto} K.,  2003, \mn@doi [\nat] {10.1038/nature01571}, \href
  {https://ui.adsabs.harvard.edu/abs/2003Natur.422..871U} {422, 871}

\bibitem[\protect\citeauthoryear{{Umeda} \& {Nomoto}}{{Umeda} \&
  {Nomoto}}{2005}]{Umeda05}
{Umeda} H.,  {Nomoto} K.,  2005, \mn@doi [\apj] {10.1086/426097}, \href
  {https://ui.adsabs.harvard.edu/abs/2005ApJ...619..427U} {619, 427}

\bibitem[\protect\citeauthoryear{{Venn}, {Irwin}, {Shetrone}, {Tout}, {Hill}
  \& {Tolstoy}}{{Venn} et~al.}{2004}]{Venn04}
{Venn} K.~A.,  {Irwin} M.,  {Shetrone} M.~D.,  {Tout} C.~A.,  {Hill} V.,
  {Tolstoy} E.,  2004, \mn@doi [\aj] {10.1086/422734}, \href
  {https://ui.adsabs.harvard.edu/abs/2004AJ....128.1177V} {128, 1177}

\bibitem[\protect\citeauthoryear{{Venn} et~al.,}{{Venn} et~al.}{2020}]{Venn20}
{Venn} K.~A.,  et~al., 2020, \mn@doi [\mnras] {10.1093/mnras/stz3546}, \href
  {https://ui.adsabs.harvard.edu/abs/2020MNRAS.492.3241V} {492, 3241}

\bibitem[\protect\citeauthoryear{{Waller}, {Venn}, {Sestito}, {Jensen},
  {Kielty}, {Hayes}, {McConnachie}  \& {Navarro}}{{Waller}
  et~al.}{2022}]{Waller22}
{Waller} F.,  {Venn} K.,  {Sestito} F.,  {Jensen} J.,  {Kielty} C.,  {Hayes}
  C.,  {McConnachie} A.,   {Navarro} J.,  2022, arXiv e-prints, \href
  {https://ui.adsabs.harvard.edu/abs/2022arXiv220807948W} {p. arXiv:2208.07948}

\bibitem[\protect\citeauthoryear{{Wenger} et~al.,}{{Wenger}
  et~al.}{2000}]{Wenger00}
{Wenger} M.,  et~al., 2000, \mn@doi [\aaps] {10.1051/aas:2000332}, \href
  {https://ui.adsabs.harvard.edu/abs/2000A&AS..143....9W} {143, 9}

\bibitem[\protect\citeauthoryear{{Wise}, {Turk}, {Norman}  \& {Abel}}{{Wise}
  et~al.}{2012}]{Wise12}
{Wise} J.~H.,  {Turk} M.~J.,  {Norman} M.~L.,   {Abel} T.,  2012, \mn@doi
  [\apj] {10.1088/0004-637X/745/1/50}, \href
  {https://ui.adsabs.harvard.edu/abs/2012ApJ...745...50W} {745, 50}

\bibitem[\protect\citeauthoryear{{Wolf} et~al.,}{{Wolf} et~al.}{2018}]{Wolf18}
{Wolf} C.,  et~al., 2018, \mn@doi [\pasa] {10.1017/pasa.2018.5}, \href
  {http://adsabs.harvard.edu/abs/2018PASA...35...10W} {35, e010}

\bibitem[\protect\citeauthoryear{{Wood}, {Lawler}, {Sneden}  \& {Cowan}}{{Wood}
  et~al.}{2013}]{Wood13}
{Wood} M.~P.,  {Lawler} J.~E.,  {Sneden} C.,   {Cowan} J.~J.,  2013, \mn@doi
  [\apjs] {10.1088/0067-0049/208/2/27}, \href
  {https://ui.adsabs.harvard.edu/abs/2013ApJS..208...27W} {208, 27}

\bibitem[\protect\citeauthoryear{{Wood}, {Lawler}, {Sneden}  \& {Cowan}}{{Wood}
  et~al.}{2014}]{Wood14}
{Wood} M.~P.,  {Lawler} J.~E.,  {Sneden} C.,   {Cowan} J.~J.,  2014, \mn@doi
  [\apjs] {10.1088/0067-0049/211/2/20}, \href
  {https://ui.adsabs.harvard.edu/abs/2014ApJS..211...20W} {211, 20}

\bibitem[\protect\citeauthoryear{{Yong} et~al.,}{{Yong} et~al.}{2013}]{Yong13}
{Yong} D.,  et~al., 2013, \mn@doi [\apj] {10.1088/0004-637X/762/1/26}, \href
  {http://adsabs.harvard.edu/abs/2013ApJ...762...26Y} {762, 26}

\bibitem[\protect\citeauthoryear{{Yoon} et~al.,}{{Yoon} et~al.}{2018}]{Yoon18}
{Yoon} J.,  et~al., 2018, \mn@doi [\apj] {10.3847/1538-4357/aaccea}, \href
  {https://ui.adsabs.harvard.edu/abs/2018ApJ...861..146Y} {861, 146}

\bibitem[\protect\citeauthoryear{{Youakim} et~al.,}{{Youakim}
  et~al.}{2017}]{Youakim17}
{Youakim} K.,  et~al., 2017, \mn@doi [\mnras] {10.1093/mnras/stx2005}, \href
  {https://ui.adsabs.harvard.edu/abs/2017MNRAS.472.2963Y} {472, 2963}

\bibitem[\protect\citeauthoryear{{Yuan} et~al.,}{{Yuan} et~al.}{2020}]{Yuan20}
{Yuan} Z.,  et~al., 2020, \mn@doi [\apj] {10.3847/1538-4357/ab6ef7}, \href
  {https://ui.adsabs.harvard.edu/abs/2020ApJ...891...39Y} {891, 39}

\bibitem[\protect\citeauthoryear{{Yuan} et~al.,}{{Yuan} et~al.}{2022}]{Yuan22}
{Yuan} Z.,  et~al., 2022, \mn@doi [\mnras] {10.1093/mnras/stac1399}, \href
  {https://ui.adsabs.harvard.edu/abs/2022MNRAS.514.1664Y} {514, 1664}

\bibitem[\protect\citeauthoryear{{de Jong} et~al.,}{{de Jong}
  et~al.}{2019}]{deJong19}
{de Jong} R.~S.,  et~al., 2019, \mn@doi [The Messenger]
  {10.18727/0722-6691/5117}, \href
  {https://ui.adsabs.harvard.edu/abs/2019Msngr.175....3D} {175, 3}

\makeatother
\end{thebibliography}

\appendix

\section{Orbital parameters as a function of the distance grid}\label{apporb}
\begin{figure*}
\begin{center}
\includegraphics[width=\textwidth]{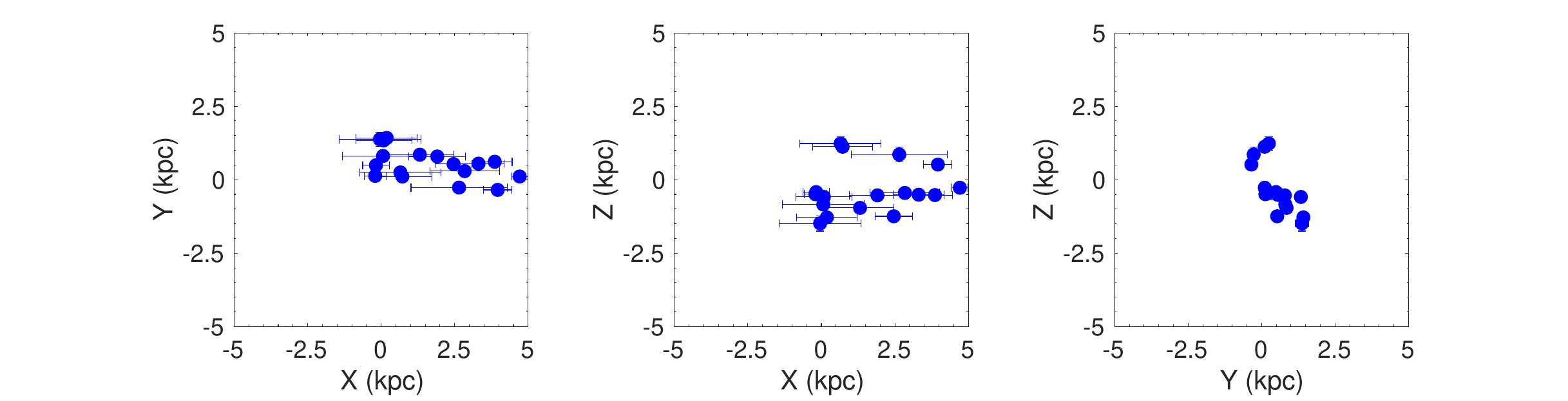}
\caption{Spatial distribution in the Galactic cartesian frame of the targets observed with GRACES. The Galactic centre is at the position (0,0,0) \kpc, while the Sun is at (8.3,0,0) \kpc.}
\label{spacedist}
\end{center}
\end{figure*}

\begin{figure*}
\begin{center}
\includegraphics[width=\textwidth]{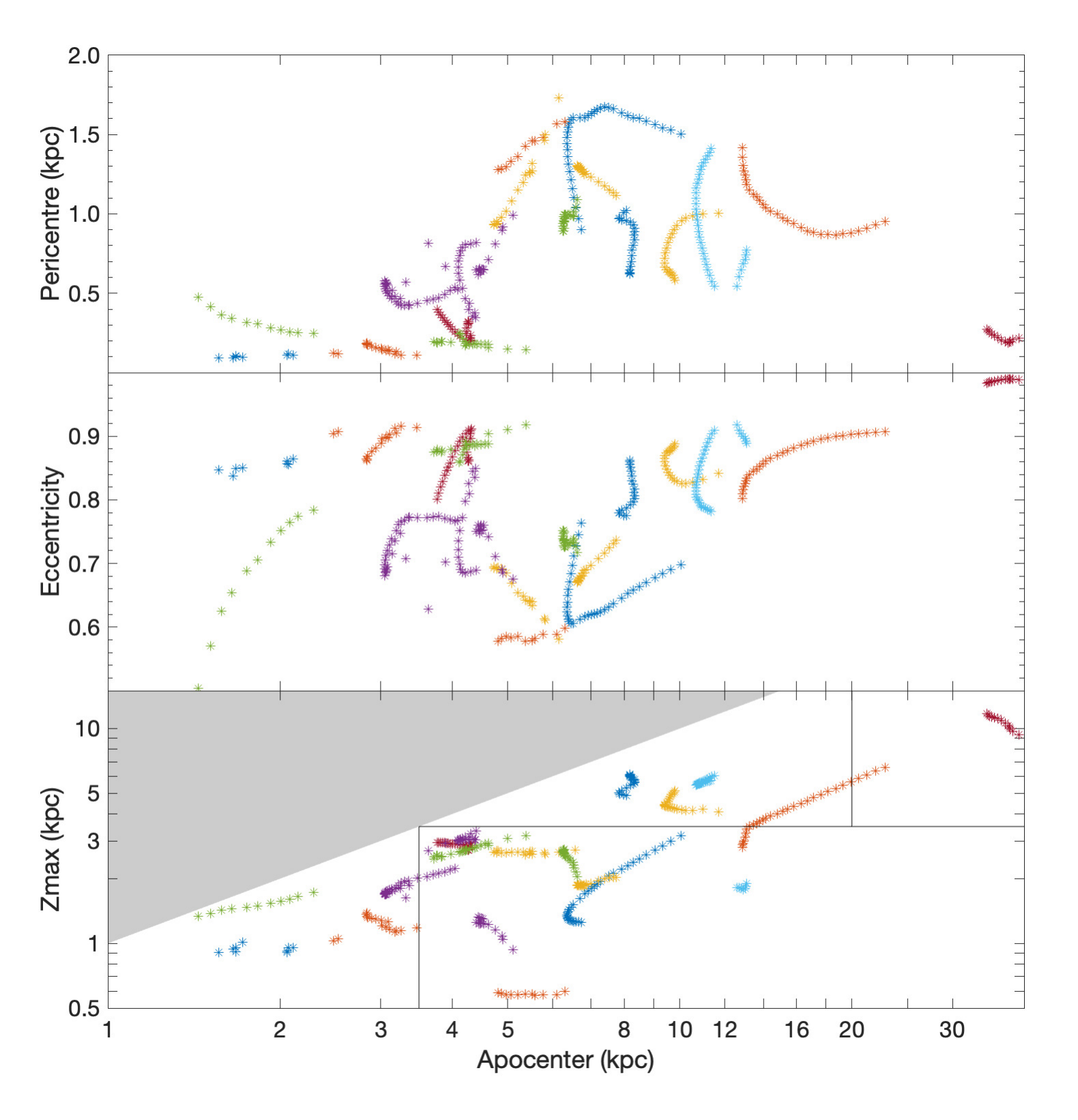}
\caption{Orbital parameters of the targets observed with GRACES. The GRACES sample is divided into 4 dynamical groups according to their $r_{\rm apo}$ $Z_{\rm max}$ as defined in Section~\ref{orbsec}. The grey shaded area denotes the forbidden region in which  $Z_{\rm max}$>$r_{\rm apo}$. The the orbital parameters for each star are inferred at each point of the distance grid. The distance grid span $1\sigma$ around the maximum of the distance probability distribution function with a step of $0.1\kpc$. Each colour represents a star.}
\label{kinefig_grid}
\end{center}
\end{figure*}

The spatial distribution in Galactic Cartesian coordinates is shown in the three panels of Figure~\ref{spacedist}.
As described in Section~\ref{orbsec}, a grid of distances with a step of $0.1\kpc$ within $\pm 1\sigma$ from the maximum of the distance PDF has been created for each star. Then, the orbits has been computed for each step of the grid and varying the other parameters (\eg RV, proper motions, coordinates etc.) with a Monte Carlo. 
The orbital parameters for each step of the distance grid are reported in Figure~\ref{kinefig_grid}.

\section{Comparison with the stellar parameters from  {\it Gaia} DR3}\label{appdr3}
\begin{figure}

\includegraphics[width=0.51\textwidth]{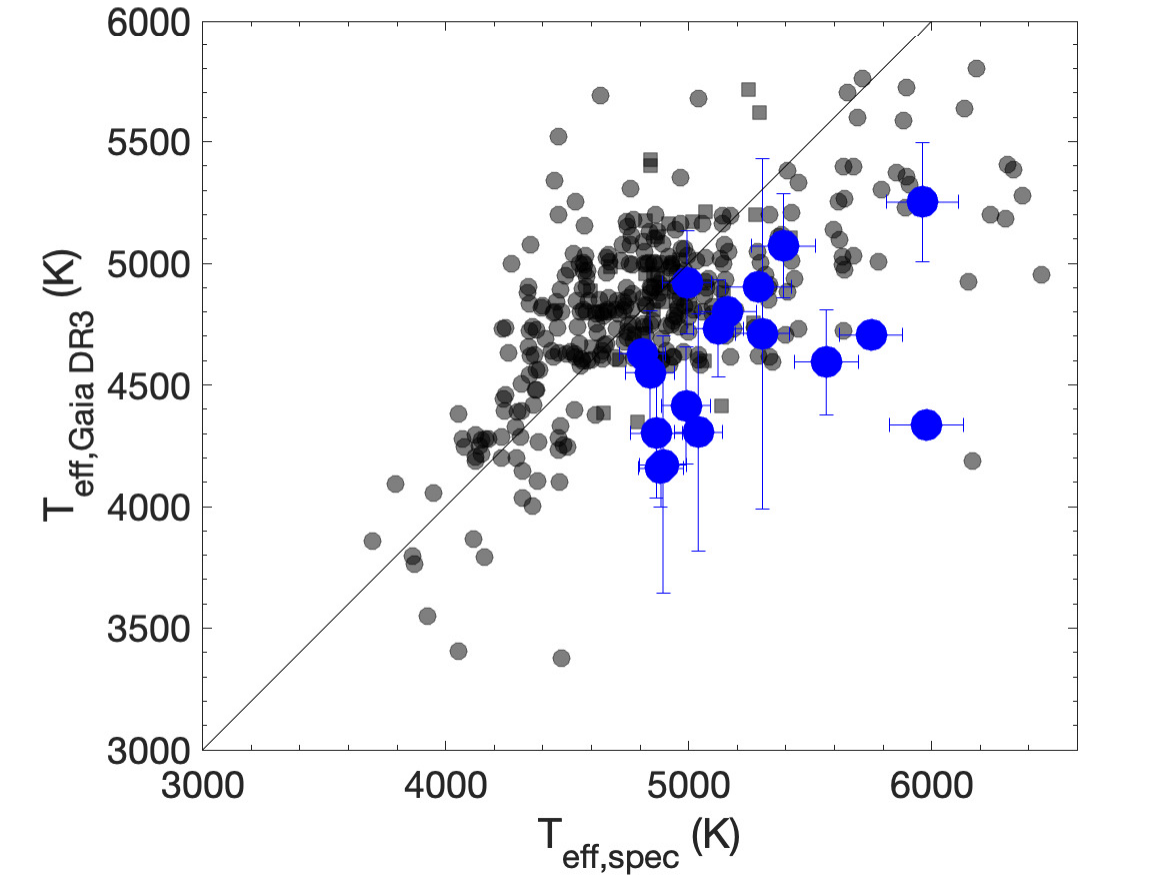}\\
\includegraphics[width=0.5\textwidth]{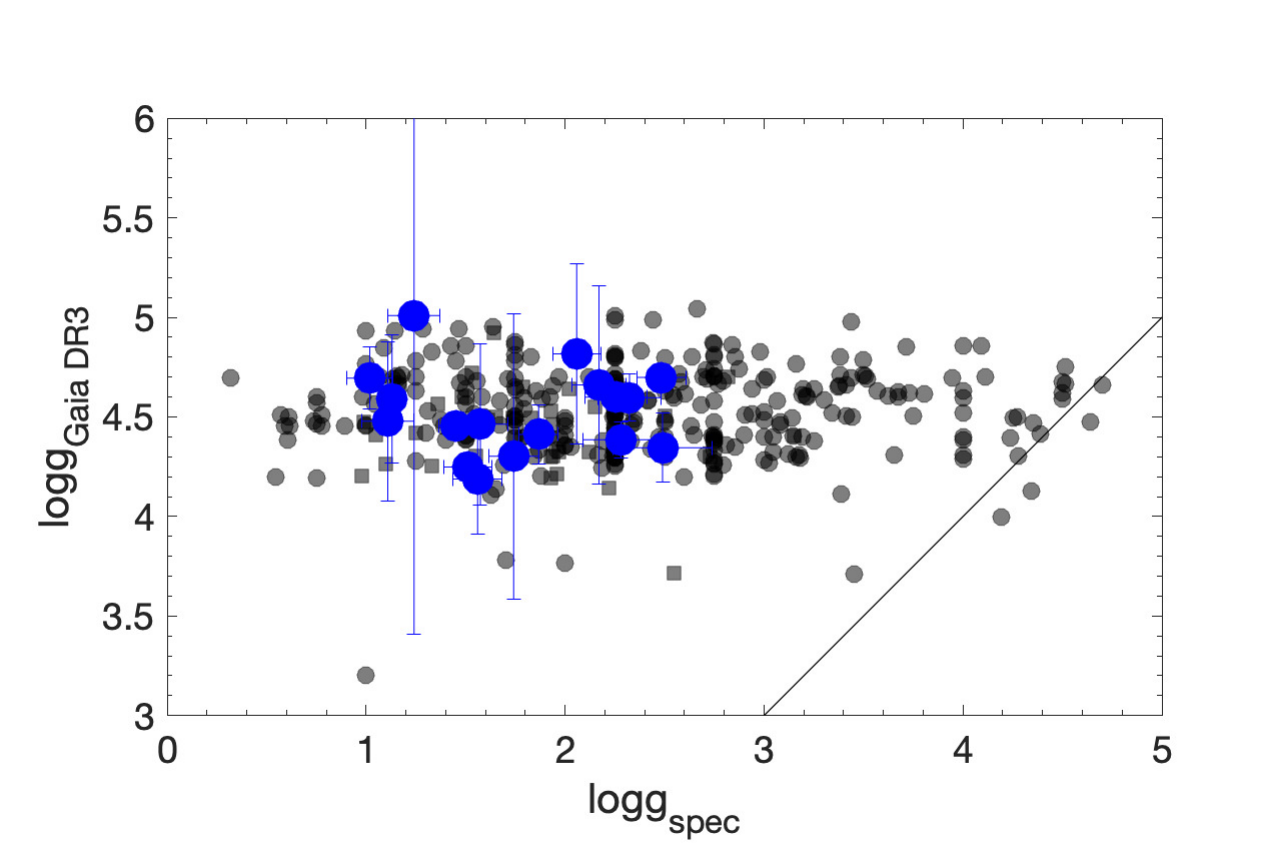}\\
\includegraphics[width=0.51\textwidth]{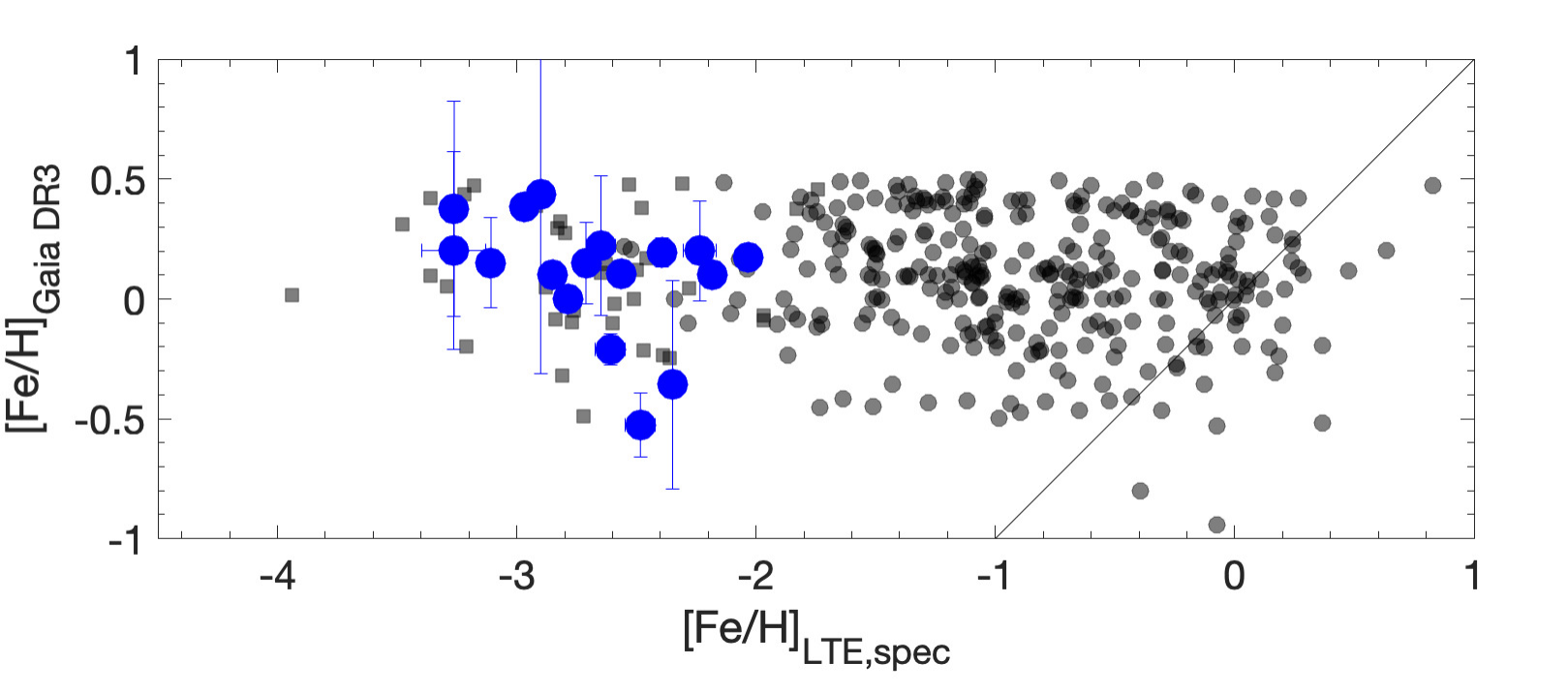}
\caption{Comparison of the stellar parameters from  {\it Gaia} DR3. Top panel: Effective temperature. Central panel: Surface gravity. Bottom panel: Metallicity. Blue circles are the stars from this work, black shaded circles and squares are stars from the COMBS \citep{Lucey22} and EMBLA \citep{Howes14,Howes15,Howes16} surveys, respectively. Black lines are the 1:1 comparisons. The agreement between the stellar parameters from {\it Gaia} DR3 and the spectroscopic ones are very poor as expected from the warnings by \citet{Andrae22}.}
\label{gaiadr3_params}
\end{figure}

It is tempting to use the new  {\it Gaia} DR3 \citep{Andrae22,Gaia16} catalogue of stellar parameters and metallicities; however, \citet{Andrae22} warn on the quality of the stellar parameters for stars with poor parallax measurements (\ie $\varpi/\sigma_{\varpi}<20$) and farther than 2\kpc. This is exactly the regime of the metal-poor stars towards the bulge. As expected, we find the {\it Gaia} DR3 photometric results for these stars to be quite poor.  Figure~\ref{gaiadr3_params} shows the temperatures, gravities, and metallicities from this work and from the COMBS \citep{Lucey22} and EMBLA \citep{Howes14,Howes15,Howes16} surveys compared withe  {\it Gaia} DR3 photometric results.   {\it Gaia} DR3 temperatures are typically lower than the values used in spectroscopic works (top panel of Figure~\ref{gaiadr3_params}).  The  {\it Gaia} DR3 surface gravities have a flat distribution around logg$\sim4.5$ dex, while the sample spans a range of 5 dex (central panel of Figure~\ref{gaiadr3_params}). The  {\it Gaia} DR3 metallicities are distributed around   \FeH$\sim0.0$, while the stars are from  super-solar to extremely metal-poor, spanning a range of 5 dex (bottom panel of Figure~\ref{gaiadr3_params}).

\bsp	
\label{lastpage}
\end{document}